\documentclass{aa}
\usepackage[varg]{txfonts}
\usepackage{graphicx}
\usepackage{placeins}
\bibpunct{(}{)}{;}{a}{}{,} 

\begin{document}


\title{Semi-analytic calculations for extended mid-infrared emission associated with FU Ori-type objects}


\titlerunning{Semi-Analytic Calculations for Extended Mid-IR Emission associated with FUors}


\author{
Michihiro Takami\inst{1}
\and
Pin-Gao Gu\inst{1}
\and
Gilles Otten\inst{1}
\and
Christian Delacroix\inst{2}
\and
Sheng-Yuan Liu\inst{1}
\and
Shiang-Yu Wang\inst{1}
\and
Jennifer L. Karr\inst{1}
}


\authorrunning{Takami et al.}

\offprints{M. Takami, \email{hiro@asiaa.sinica.edu.tw}}

\institute{Institute of Astronomy and Astrophysics, Academia Sinica, 
11F of Astronomy-Mathematics Building,
No.1, Sec. 4, Roosevelt Rd., Taipei 10617, Taiwan, R.O.C.
\and
STAR Institute, Universit\'e de Li\`ege, All\'ee du Six Ao\^ut 19c, B-4000 Li\`ege, Belgium
}

\date{
Accepted 29 August 2023
}


\abstract {}
{Near-infrared imaging polarimetry at high-angular resolutions has revealed an intriguing distribution of circumstellar dust toward FU Ori-type objects (FUors). These dust grains are probably associated with either an accretion disk or an infalling envelope. Follow-up observations in the mid-infrared would lead us to a better understanding of the hierarchy of the mass accretion processes onto FUors (that is envelope and disk accretion), which hold keys for understanding the mechanism of their accretion outbursts and the growth of low-mass young stellar objects (YSOs) in general.
}
{
We developed a semi-analytic method to estimate the mid-infrared intensity distributions using the observed polarized intensity (PI) distributions in the $H$ band ($\lambda$=1.65 $\mu$m). 
This new method allows
us to estimate the intensity levels with an order-of-magnitude accuracy, assuming that the emission is a combination of scattered and thermal emission from circumstellar dust grains
illuminated and heated by a central source, but the radiation heating through the inner edge of the dust disk is negligible due to the obscuration by an optically
thick compact disk. We have derived intensity distributions for two FUors, FU Ori and V1735 Cyg, at three wavelengths ($\lambda$=3.5, 4.8, and 12 $\mu$m) for various cases, with a star or a flat compact self-luminous disk as an illuminating source; an optically thick disk or an optically thin envelope for circumstellar dust grains; and three different dust models.
The calculations were carried out for typical aspect ratios of the disk surface and the envelope $z/r$ of $\sim$0.1, $\sim$0.2, and $\sim$0.4.
}
{We have been able to obtain self-consistent results for many cases and regions, in particular when the viewing angle of the 
disk or envelope is zero (face-on).
Our calculations suggest that the mid-infrared extended emission at the above wavelengths is dominated by the single scattering process. The contribution of thermal emission is negligible unless we add an additional heating mechanism such as adiabatic heating in spiral structures and/or fragments.
The uncertain nature of the central illuminating source, the distribution of circumstellar dust grains
and the optical properties of dust grains yield uncertainties in the intensity levels on orders of magnitude,
for example, 20-800, for the aspect ratio of the disk or the envelope of $\sim$0.2 and $\lambda$=3-13 $\mu$m.
}
{The new method we have developed is useful for estimating the detectability of the extended mid-infrared emission. Observations with the forthcoming extremely large telescopes, with a telescope diameter of 24-39 m, would yield a breakthrough for the above research topic at angular resolutions comparable to the existing near-infrared observations. The new semi-analytic method is complementary to full radiative transfer simulations, which offer more accurate calculations but only with specific dynamical models and significant computational time.}

\keywords{
Methods: analytical
--
Stars: individual (FU Ori, V1735 Cyg)
--
Stars: protostars
---
Infrared: stars
}
\maketitle


\section{Introduction} \label{sec:intro}

The FU Orionis objects (hereafter FUors) are a class of young stellar objects (YSOs) that undergo the most active and violent accretion outbursts. During each burst, the accretion rate rapidly increases by a factor of $\sim$1000, and remains high for several decades or more. Such outbursts have been observed toward about ten stars to date. Astronomers have also identified another dozen YSOs with optical or near-infrared
spectra similar to FUors, distinct from many other YSOs, but whose outbursts have never been observed ("FUor candidates" or "FUor-like stars"). It has been suggested that many low-mass YSOs \citep[and also some high-mass YSOs; see, e.g., ][]{Caratti17} experience FUor outbursts, but  we miss most of them because of the small chance of capturing the events. Readers can refer to
\citet{Hartmann96} and \citet{Audard14} for reviews, for example.

Near-infrared ($\lambda$$\sim$2 $\mu$m) imaging polarimetry
at high-angular resolutions (0\farcs05-0\farcs1) reveals complicated circumstellar structures associated with some FUors \citep{Liu16,Takami18,Laws20}. These were observed via scattering from circumstellar dust grains illuminated by the central source. While the scattered light is significantly fainter than the central source, its large polarization relative to the central object allows us to observe circumstellar structures as close as 0\farcs1-0\farcs2 to the central source.

The observed circumstellar structures include arms similar to those of spirals and the elongated structures that may be associated with gas streams.
\citet{Liu16} and \citet{Takami18} attributed them to gravitationally unstable disks and trails of clump ejections in such disks. These authors qualitatively reproduced these structures using dynamical simulations \citep{Takami18} combined with radiative transfer simulations for near-infrared scattered light \citep{Liu16}. Therefore, many YSOs may actually experience gravitational instabilities in their disks during their lifetimes, as well as the FUor outbursts. Gravitational fragmentation induced by these instabilities may also induce the formation of planets and brown dwarfs at large orbital radii, 
the presence of which the conventional planet formation models cannot simply explain \citep[e.g.,][]{Boss03,Nayaksin10,Vorobyov13,Stamatellos15}.

In contrast, \citet{Laws20} attributed the observed structures surrounding FU Ori, the archetypical FUor, to an infalling envelope. 
This explanation is corroborated by infrared spectral energy distributions and millimeter emissions, which indicate the presence of massive circumstellar envelopes toward some FUors \citep[e.g.,][]{Sandell01,Gramajo14}. Furthermore, \citet{Laws20}  argue that the observed structures are similar to those of infalling gas toward some normal YSOs, observed using Atacama Large Millimeter/submillimeter Array (ALMA) \citep{Yen14,Yen19}. If this is the case, the structures seen in the near-infrared images would provide valuable clues for understanding how the circumstellar disk is fed from the envelope, leading to accretion outbursts \citep[e.g.,][]{Hartmann96}.

Imaging observations in the mid-infrared ($\lambda$$\gtrsim$3 $\mu$m) would be useful for discriminating between these two cases, therefore leading us to understand mass accretion onto FUors better, and perhaps protostellar evolution and planet formation in a general context. As the dust opacity is smaller at longer wavelengths (Sect. \ref{sec:dust}), such observations would be powerful for searching for circumstellar structures embedded in a dusty environment, in particular if a disk is embedded in an envelope seen in the near-infrared. In contrast, if the distribution of mid-infrared emission is similar to that in the near-infrared, this would imply that both are associated with the surface of an optically thick disk (Sect. \ref{sec:eq:thick}).
While the observations at longer wavelengths degrade the diffraction-limited angular resolution, the use of next-generation extremely large telescopes such as
the Extremely Large Telescope (ELT, with a 39-m telescope diameter), the Giant Magellan Telescope (GMT, 25-m), and the Thirty Meter Telescope (TMT, 30-m) will overcome this problem. For example, the ELT will offer a diffraction-limited angular resolution of 20 mas at $\lambda$=3.5 $\mu$m, improving the angular resolution by a factor of $\sim$2 compared with near-infrared imaging polarimetry to date made at 8-10 m telescopes. 

For this study, we developed equations to calculate approximate distributions of mid-infrared emission using existing near-infrared imaging polarimetry. In contrast to extensive numerical simulations, with a combination of dynamical and radiative transfer simulations \citep[e.g.,][]{Liu16,Dong16}, our semi-analytic approach allowed us to easily calculate the emission distributions in various cases, that is, when the extended emission is associated with a disk or an envelope; when the central illuminating source is a star or a compact self-luminous disk; when the radiation from the star to a dusty inner disk edge is obscured by a compact optically thick ionized disk;
and with different dust models.
Furthermore, we have been able to simultaneously investigate whether individual cases yield self-consistent calculations, for example, whether a combination of the compact self-luminous disk and an extended envelope is consistent with the observed near-infrared polarized intensity (PI) distributions.
In summary, the semi-analytic approach developed in this paper will be complementary to detailed radiative transfer simulations using the density distributions provided by dynamical simulations. 

The rest of this paper is organized as follows.
In Sect. \ref{sec:dust}, we describe the dust models. We used their optical properties to derive some approximations, as explored in the next section.
In Sect. \ref{sec:eq}, we describe how we derived equations to calculate the mid-infrared intensity distributions using the observed PI distributions in the $H$ band ($\lambda$=1.65 $\mu$m) and the spectral energy distributions of the central source.
In Sect. \ref{sec:application}, we explain how we applied the calculations to two 
FUors: FU Ori and V1735 Cyg.
In Sect. \ref{sec:sim}, we explore how we verified our calculations using monochromatic Monte-Carlo scattering simulations. 
We summarize our work and discuss it in Sect. \ref{sec:sum}.


\section{Dust models}  \label{sec:dust}

We used three of the dust models used in HO-CHUNK\footnote{https://gemelli.colorado.edu/~bwhitney/codes/}, one of the radiative transfer codes for dusty circumstellar environments associated with YSOs \citep{Whitney03b,Whitney03a,Whitney04,Whitney13}.
We summarize these dust models below.
Each dust model consists of a number of homogeneous spherical particles with ``astronomical silicates" \citep{Draine84} and graphite, with certain size distributions adjusted to reproduce various observations. For one of the models, dust particles are associated with ice coating. 

(1) `Dust1' (labeled as {\it `kmhnew\_extrap'} in HO-CHUNK) --- This model is based on the dust size distribution of the interstellar medium ($R_V$=3.1) measured by \citet{Kim94}. The parameter $R_V=A_V/(A_B-A_V)$ is the optical total-to-selective extinction ratio (where $A_V$ and $A_B$ are extinctions at $\lambda$=0.55 and 0.45 $\mu$m, respectively) often used to represent how large the dust grains are. 
A larger $R_V$ implies larger grain sizes.

(2) `Dust2' ({\it `r400\_ice095\_extrap'}) --- This dust model is for the interstellar medium as well but the outer 5 \% of the radius of the individual grains are covered with water ice. To approximate dust properties in nearby star formation regions, \citet{Whitney03a} produced dust size distribution that fits an extinction curve generated by \citet{Cardelli89}. This size distribution is similar to `Dust1' but it yields $R_V=4$, slightly larger than that for `Dust1', as measured in the Taurus Molecular Clouds \citep{Whittet01}. A 5 \% thickness of the ice coating was selected to best match the polarization observations of the background stars \citep{Whitney03a}.

(3)  `Dust3' ({\it `ww04'}) --- This dust model with larger dust grains was developed by \citet{Cotera01} to reproduce the smaller dependence of wavelength on grain opacities observed at the surface of the HH 30 disk in the near-infrared. We note that this model may not be appropriate for all circumstellar disks associated with YSOs. \citet{Takami14} analyzed near-infrared PI distributions for a few more disks, and suggested that their grain sizes are significantly smaller that this model, even smaller than those for the `Dust1' model.

\begin{figure*}[ht!]
\centering
\includegraphics[width=18cm]{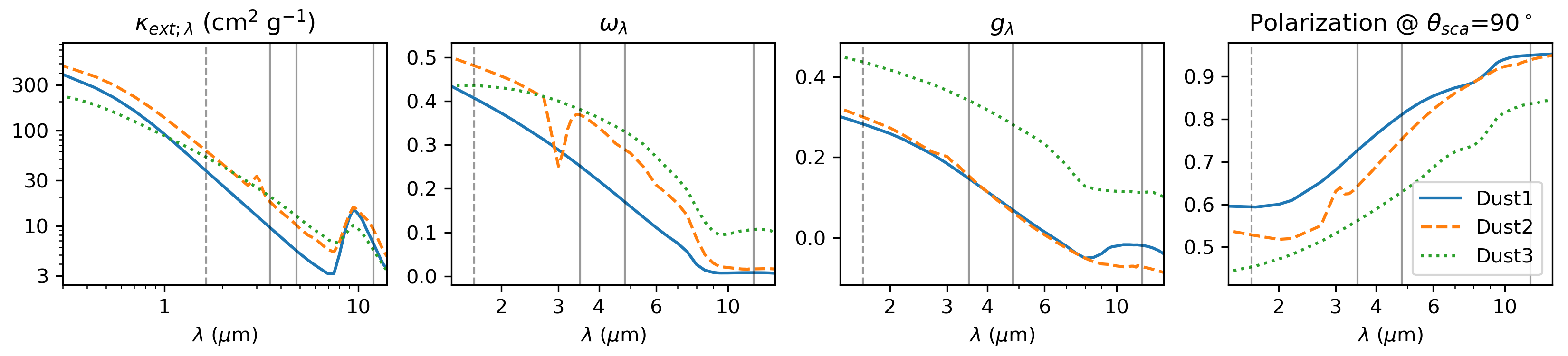}
\caption{Optical properties for three dust models. From left to right: the opacity per unit gas mass, assuming a gas-to-dust mass ratio of 100 
($\kappa_{\mathrm{ext};\lambda}$); 
the scattering albedo
$\omega_\mathrm{\lambda}$;
the forward throwing parameter $g_\mathrm{\lambda}$; and the polarization for the scattering angle of 90$^\circ$.
The gray vertical solid lines indicate three representative wavelengths for our calculations (3.5, 4.8, 12 $\mu$m; see text for details). 
The gray vertical dashed line indicates the wavelength for $H$ band, for which we used imaging polarimetry to model the images for longer wavelengths.
The dust opacity  $\kappa_{\mathrm{ext};\lambda}$ is shown for the optical to mid-infrared wavelengths, while the remaining parameters for scattered and thermal emission are shown for near- to mid-infrared. The variations of optical properties at $\lambda$$\sim$3 and $\sim$10 $\mu$m are due to water ice and silicate, respectively.
\label{fig:dust_general}
}
\end{figure*}

HO-CHUNK uses the parameter files for these dust models with the extinction and scattering cross sections $C_\mathrm{ext}$ and $C_\mathrm{sca}$; the opacity $\kappa_\mathrm{ext}$; the forward throwing parameter $g$; and a degree of polarization for the scattering angle equal to 90$^\circ$, for wavelengths from $\lambda$=0.01 $\mu$m~to 3.6 cm. The authors calculated the optical parameters for these dust models based on the Mie theory and the geometrical optic algorithm \citep[see][for details]{Wood02b}. In Fig. \ref{fig:dust_general} we show some key optical parameters at the wavelength range of our interest.

The nature of the dust grains that are responsible for near-infrared scattered light toward FUors is not clear. Therefore, we used each of the above dust models to calculate the intensity distributions at the target wavelengths, and regard the different results as an uncertainty. 
One may regard these dust models as unrealistic for the following points: (1) the nature of "astronomical silicate" is not clear \citep[e.g.,][]{Greenberg96}; (2) the actual carbon dust may be amorphous carbon rather than graphite \citep[and references therein]{Jager98}; and (3) the actual particles are probably aggregates rather than spherical particles \citep[e.g.,][]{Henning09,Tazaki16}. Even so, the dust models we used yield a good start for our initial study due to the availability of the dust parameters needed for this study, and the fact that these models reproduced the observations described above. 

We selected three representative wavelengths ($\lambda$=3.5, 4.8, and 12 $\mu$m) for our calculations based on the applicability for calculations (up to $\sim$15 $\mu$m; see Sect. \ref{sec:application:ill}); a range of the optical properties for dust grains shown in Fig. \ref{fig:dust_general}; and the observability from the ground through the atmospheric windows. We note that the Mid-Infrared Instrument (MIRI) on the James Webb Space Telescope offers coronagraphic observations at $\lambda$$>$12 $\mu$m but with too large inner working angles ($\gtrsim$0.5 arcsecond) for our targets (see Sect. \ref{sec:application}).



\section{Equations to derive intensity distributions} \label{sec:eq}

\subsection{Overview} \label{sec:eq:overview}

In this section we derive the equations to calculate the intensity distributions for three wavelengths ($\lambda$=3.5, 4.8, and 12 $\mu$m) using the PI distribution observed in $H$ band ($\lambda$=1.65 $\mu$m). We derived the equations for scattered light and thermal emission assuming the following origins for the extended emission:
(1) an optically thick and geometrically thin disk (Sect. \ref{sec:eq:thick}); and 
(2) an optically thin remnant envelope (Sect. \ref{sec:eq:thin}).
Optically thick and geometrically thin disks have actually been seen in many edge-on disks associated with normal YSOs \citep[e.g.,][for a review]{Watson07_PPV}. We note that, however, some numerical simulations suggest that the aspect ratio of the disk surface ($z/r$) can reach up to 0.8-0.9 for FUor disks \citep{Liu16,Dong16}.

The targets associated with an optically
thick envelope are beyond the scope of this study. Such YSOs exhibit an outflow cavity illuminated by the central source at optical and/or near-infrared wavelengths \citep[e.g.,][]{Padgett99}.  Such a reflection nebula is actually observed toward one of the FUor-like stars \citep{Kospal08}. We limit the applicability of our study to classical FUors, which do not show evidence for such optically thick envelopes with outflow cavities \citep{Liu16,Takami18,Laws20}.

%
%
We assume that, at the radii of our interests ($r$$>$10 au), the disk or the envelope is heated by the radiation from the central illuminating source. In practice, adiabatic heating or spiral shocks would significantly heat up gravitationally unstable disks, enhancing mid-infrared emission \citep[e.g.,][]{Ilee11,Vorobyov20,Vorobyov22}.

The viewing angle of the disk or the envelope, which affects the observed emission distribution, is uncertain, as the observed near-infrared intensity distributions are so complicated that they do not show a circular or elliptical distribution as observed for many disks associated with normal YSOs \citep[][--- see also Sects. \ref{sec:intro} and \ref{sec:application}]{Liu16,Takami18,Laws20}. High-resolution millimeter observations using ALMA can be useful for such measurements in some cases, in particular for disks associated with pre-main sequence stars \citep[e.g.,][]{Long18}. However, the millimeter emission associated with FUors is often very compact, comparable to the angular resolutions of the ALMA observations \citep[e.g.,][]{Cieza18,Kospal21}, perhaps due to emission in the inner disk region enhanced by viscous heating \citep{Takami19}.
As many of the near-infrared images for FUors do not show any evidence for large viewing angles, we assumed their viewing angles $i$ to be $\lesssim$45$^\circ$ throughout our calculations.

In Sect. \ref{sec:eq:source}, we extend the equations adding the effect below. First, the central illuminating source may be either a compact self-luminous accretion disk \citep[e.g.,][for reviews]{Hartmann96,Audard14} or a star \citep[e.g.,][]{Herbig03,Elbakyan19}. We derived equations for both cases, simultaneously correcting the fluxes and intensities for foreground extinction.
In Sect. \ref{sec:eq:check} we derive the equations to check self-consistencies of the calculations.
%
In Table 1, we summarize major parameters used in the following subsections.

\begin{table*}
\caption{Major parameters. \label{tbl:params}}
{\footnotesize
\begin{tabular}{ll}
\hline\hline
$A_V$                           & Visual extinction \\
$a_\lambda$, $a_{\lambda; \theta_\mathrm{sca}(x,y,i)}$
                                        & See Eqs. (\ref{eq:a})(\ref{eq:a_const})\\
$B_{\lambda,T}$         & Blackbody function \\
$d$                                     & Distance to the target \\
$F_{\mathrm{obs};\lambda}$
                                        & Observed flux for the central illuminating source at the wavelength $\lambda$\\
$F_\lambda(x,y)$                & Flux that the extended disk or envelope receives from the central source at the wavelength $\lambda$\\
$f_\lambda$                     & A factor to correct foreground extinction for the wavelength $\lambda$ (see Eq. \ref{eq:f_lambda})\\
$I_{a; \lambda}(x,y)$           & Intensity distribution for the extended disk at the wavelength $\lambda$ (via single scattering) \\
$I_{b; \lambda}(x,y)$   & Intensity distribution for the extended disk at the wavelength $\lambda$ (via double scattering) \\
$I_{c; \lambda}(x,y)$   & Intensity distribution for the extended disk at the wavelength $\lambda$ (thermal emission from the surface layer) \\
$I_{d; \lambda}(x,y)$   & Intensity distribution for the extended disk at the wavelength $\lambda$ (thermal emission from the disk interior) \\
$I_{a'; \lambda}(x,y)$  & Intensity distribution for the extended envelope at the wavelength $\lambda$ (via single scattering) \\
$I_{c'; \lambda}(x,y)$  & Intensity distribution for the extended envelope at the wavelength $\lambda$ (thermal emission) \\
$i$                                     & Viewing angle of the extended disk or envelope \\
$P_1 (\lambda,\theta_\mathrm{sca})$             & Scattering phase function for the scattering angle $\theta_\mathrm{sca}$ at the wavelength $\lambda$\\
$P_2 (\lambda,\theta_\mathrm{sca})$             & Product of the scattering phase function and polarization for the scattering angle $\theta_\mathrm{sca}$ at the wavelength $\lambda$\\
$P_2(H)$                        & Representative constant for $P_2 (\lambda; \theta_\mathrm{sca})$ for the $H$ band ($\lambda$=1.65 $\mu$m) \\
$(PI)_{\mathrm{obs};H}(x,y)$
                                        & Observed PI distribution for the $H$ band\\
$r$                                     & Distance to the central source projected to the midplane (=$\sqrt{x^2+y^2}$) \\
$T_\mathrm{d} (x,y)$    & Temperature at the disk interior \\
$T_\mathrm{e} (x,y)$    & Temperature in the envelope \\
$T_\mathrm{s} (x,y)$            & Temperature at the surface layer of the disk \\
$x,y$                           & Coordinate in the midplane \\
$z$                                     & Coordinate perpendicular to the midplane \\
$z_\mathrm{disk}(x,y)$  & Location of the disk surface \\
$z_\mathrm{env}(x,y)$   & Typical height of the envelope \\
$\alpha(x,y)$                   & Sine of the radial gazing angle of the disk surface $\beta (x,y)$\\
$\beta(x,y)$                    & Radial grazing angle of the disk surface with respect to the incident light (see Fig. \ref{fig:concept})\\
$\overline{\gamma}$             & Typical grazing angle of the flat compact disk from the disk surface or the envelope \\
$\delta(x,y)$                   & Radial inclination of the disk surface with respect to the midplane (see Fig. \ref{fig:concept})\\
$\theta_\mathrm{sca} (x,y,i)$
                                        & Scattering angle of the light from the central illuminating source (see Fig. \ref{fig:concept}) \\
$\kappa_{\mathrm{ext};\lambda}$
                                        & Dust opacity for the wavelength $\lambda$ \\
$\lambda$                               & Wavelength \\
$\Sigma_\mathrm{env}(x,y)$
                                        & Surface density for the extended envelope \\
$\tau_{l,\lambda} (x,y)$        & Optical thickness of the extended envelope at line of sight of the observations\\
$\tau_{\mathrm{r},\lambda} (x,y)$
                                        & Optical thickness of the extended envelope from the central source in the radial direction \\     
$\omega_\mathrm{\lambda}$               & Scattering albedo for the wavelength $\lambda$ \\
\hline \\
\end{tabular} \\
}
\end{table*}


\subsection{Emission from an optically thick and geometrically thin disk}  \label{sec:eq:thick}

\subsubsection{Overview} \label{sec:eq:thick:overview}

\begin{figure}[ht!]
\centering
\includegraphics[width=5cm]{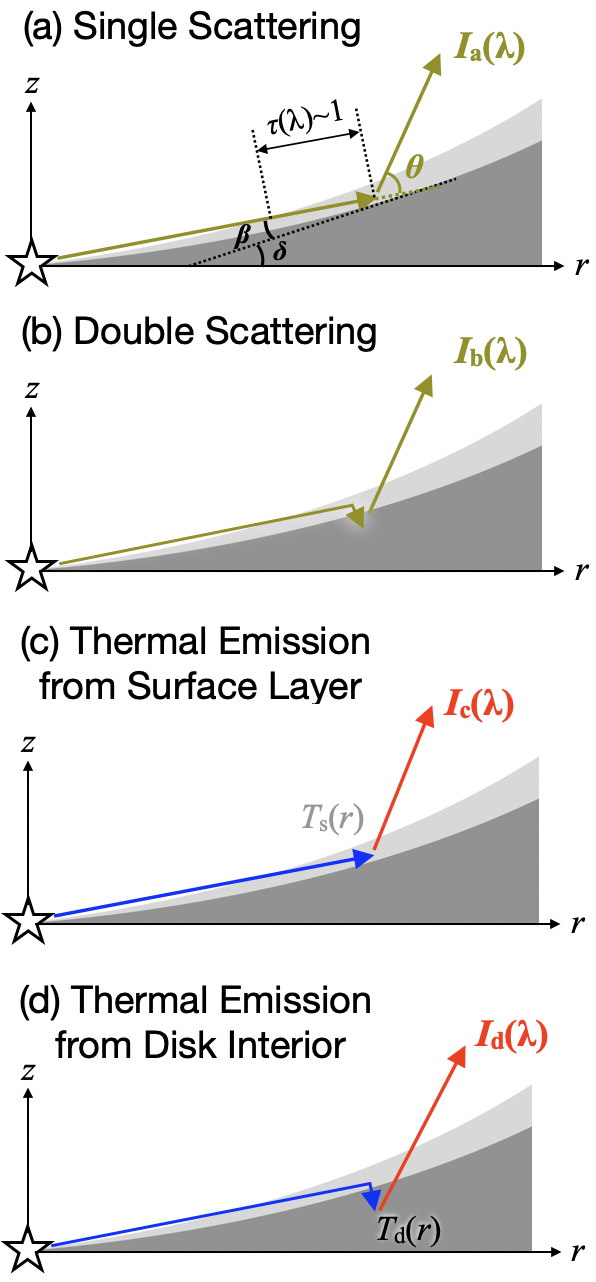}
\caption{A simplified disk geometry with a surface layer and the disk interior. The panels (a)-(d) explain the four emission components we discuss in Sect. \ref{sec:eq:thick:I_a}-\ref{sec:eq:thick:I_d}, respectively. The green arrows indicate scattering of the stellar light at the wavelength of the observations. The blue arrows indicate the stellar light responsible for heating the surface layer and the disk interior. The red arrows indicate thermal emission from the surface layer and the disk interior. In the top panel,
$\beta$ is the radial grazing angle of the disk surface with respect to the incident light from the central illuminating source;
$\delta$ is the radial inclination angle of the disk surface with respect to the midplane;
and $\theta$ is the scattering angle.
\label{fig:concept}}
\end{figure}

We extended the approximation developed by \citet{Chiang97} and \citet{Chiang01}, and discussed in \citet{Dullemond07_PPV}. A disk consists of a geometrically thin surface layer and an optically thick disk interior, as shown in light and dark gray in Fig. \ref{fig:concept}. Throughout the paper, $z$ is the coordinate perpendicular to the disk midplane; $(x,y)$ is the coordinate parallel to the disk midplane; and $r$ is the distance to the central source projected to the disk midplane (=$\sqrt{x^2+y^2}$).

The optical thickness of the surface layer is about 1 along the light path to the central source, and significantly below 1 across the disk surface. Such a disk geometry was verified using realistic radiative transfer simulations with conventional models for circumstellar disks \citep[e.g.,][]{Cotera01,Takami13}, and multiwavelength imaging observations for some edge-on disks \citep[e.g.,][]{Cotera01,Grosso03}. 
As shown in these studies, the disk tends to have a large gradient in the optical thickness across its surface, and as a result, the location of the surface layer is relatively independent of wavelength.
Throughout the paper, we approximated that the location of the disk surface is the same at all the wavelengths (optical to the mid-infrared; Sect. \ref{sec:application:ill}) for our calculations.
\citet{Chiang97,Chiang99} and \citet{Chiang01} successfully reproduced the observed spectral energy distributions (SEDs) for some YSOs with disks using this approximation.

For the following subsections we derived the equations for the following four emission components: (a) direct scattered light of the star via the surface layer (single scattering; Sect. \ref{sec:eq:thick:I_a}); (b) light scattered from the surface layer toward the disk interior, and re-scattered in the disk interior toward the outside (double scattering; Sect. \ref{sec:eq:thick:I_b}); (c) thermal emission from the surface layer (Sect. \ref{sec:eq:thick:I_c}); and (d) thermal emission from the disk interior (Sect. \ref{sec:eq:thick:I_d}). Fig. \ref{fig:concept} shows the schematic views for these components.

Full numerical radiative transfer simulations for optically thick dust disks often include the presence of an inner disk edge that can receive stellar light. This physical process would significantly increase the temperature of the entire disk \citep[e.g.,][]{Whitney03a,Robitaille11}. We assumed that this radiative heating is negligible for the reasons below. It has been suggested that the FUors are associated with an optically thick compact ionized disk inside the dust disk \citep[e.g.,][]{Hartmann96,Zhu08}. Such a disk can block stellar light toward the inner wall of the dusty disk.

\subsubsection{Single scattering} \label{sec:eq:thick:I_a}

The intensity is approximately described as:
%
\begin{eqnarray}
I_{a; \lambda}(x,y) &\sim& \frac{F_\lambda(x,y)~\alpha(x,y)~\omega_\mathrm{\lambda} P_1(\lambda,\theta_{_\mathrm{sca} (x,y,i)})}{\mathrm{cos}~i} ,
\label{eq:thick:I_a0} \\
\alpha(x,y) &=& \mathrm{sin}~\beta (x,y),
\label{eq:alpha:definition}
\end{eqnarray}
%
where
$F_\lambda (x,y)$ is the flux density of illumination by the central source for a unit area perpendicular to the direction of the incident light; 
$\omega_\mathrm{\lambda}$ is the scattering albedo;
$i$ is the viewing angle of the disk;
$P_1(\lambda; \theta_\mathrm{sca  (x,y,i)})$ is the scattering phase function of the dust grains;
$\theta_\mathrm{sca} (x,y,i)$ is the scattering angle;
and $\beta (x,y)$ is the radial grazing angle of the disk surface with respect to the incident light from the central illuminating source.
Eq. (\ref{eq:thick:I_a0}) implies that the light observed via single scattering is a product of
the light that the surface layer received from the star $F_\lambda \alpha$,
the efficiency of scattering $\omega_\mathrm{\lambda} P_1$,
and an enhancement for the observations due to the viewing angle 1/cos $i$.

To accurately estimate the last term, we need to use the viewing angle of the disk surface with respect to line of sight of the observations rather than $i$. The alternative use of 1/cos $i$ allowed us to reasonably simplify the calculations, but it is accurate only if the disk surface is nearly parallel to the disk midplane. In Sects. \ref{sec:eq:check:disk} and \ref{sec:application}, we investigate how this simplification affects the accuracies of our calculations.

As described in Sect \ref{sec:eq:overview}, we assumed the viewing angle of the disk $i$ to be $\lesssim$45$^\circ$. Therefore, the $1$/cos $i$ term also yielded changes in the intensities of $\sim$30 \%; this is significantly smaller than other uncertainties and systematic errors discussed later.

For the polarized intensity in $H$ band, Eq. (\ref{eq:thick:I_a0}) is revised to:
%
\begin{equation}
(PI)_{a;H}(x,y) \sim \frac{F_H(x,y)~\alpha(x,y)~\omega_H P_2(H; \theta_\mathrm{sca (x,y,i)})}{\mathrm{cos}~i } ,
\label{eq:thick:PI_a}
\end{equation}
%
where $P_2$ is the scattering phase function of the dust grains multiplied by the degree of polarization.
For the polarization intensity, this single scattering component dominates the observations. The emission via multiple scattering is significantly fainter for the reason given below, as demonstrated using numerical simulations by \citet{Takami13}. For small grains, for which photons are fairly isotropically scattered, the scattered photons are polarized with a variety of position angles, canceling each other out. For large grains, which scatter most of the photons forward (that is with small $\theta_\mathrm{sca}$), the polarization of the individual photons is significantly reduced after the first scattering \citep[see, e.g., Fig. 6 in][]{Takami13}. \citet{Takami13} executed radiative transfer calculations using more realistic disk models, which yielded an upper limit of the contribution of multiple scattering of 10 \% of the observed PI distribution.

Therefore, we replaced $(PI)_{a;H}(x,y)$ in Eq. (\ref{eq:thick:PI_a}) by $(PI)_{\mathrm{obs};H}(x,y)$, which is the observed PI distribution in $H$ band, and derived the following equation:
%
\begin{equation}
(PI)_{\mathrm{obs};H}(x,y) \sim \frac{F_H(x,y)~\alpha(x,y)~\omega_H P_2(H; \theta_\mathrm{sca,(x,y,i)})}{\mathrm{cos}~i } .
\label{eq:thick:PI_obs}
\end{equation}
%
%
Using Eqs. (\ref{eq:thick:I_a0}) and (\ref{eq:thick:PI_obs}) we derived:
%
\begin{equation}
I_{a; \lambda}(x,y) 
\sim
a_{\lambda; \theta_\mathrm{sca} (x,y,i)}
\frac{F_\lambda (x,y)}{F_H (x,y)}  \frac{\omega_\mathrm{\lambda}}{\omega_H}
(PI)_{\mathrm{obs};H}(x,y),
\label{eq:thick:I_a2}
\end{equation}
%
where
%
\begin{equation}
a_{\lambda; \theta_\mathrm{sca} (x,y,i)} = \frac{P_1(\lambda; \theta_\mathrm{sca (x,y,i)}) }{P_2(H; \theta_\mathrm{sca (x,y,i)})}.
\label{eq:a}
\end{equation}

The scattering angle $\theta_\mathrm{sca}$ depends on the viewing angle $i$, which is uncertain as discussed above. Furthermore, this angle is not exactly the same at the individual positions ($x$,$y$) of the disk: these are smaller and larger at the near and far sides of the disk, respectively. To avoid complexities, we used a representative constant for $a_{\lambda; \theta_\mathrm{sca} (x,y,i)}$ defined below for each dust model:
%
\begin{equation}
a_{\lambda} = \mathrm{exp} [0.5 (
\mathrm{log}~a_{\lambda;\mathrm{min}}
+
\mathrm{log}~a_{\lambda;\mathrm{max}}
)],
\label{eq:a_const}
\end{equation}
 where $a_{\lambda;\mathrm{min}}$ and $a_{\lambda;\mathrm{max}}$ are the minimum and maximum values measured at $\theta_\mathrm{sca}$=45$^\circ$--135$^\circ$.
Replacing $a_{\lambda; \theta_\mathrm{sca} (x,y,i)}$ in Eq. (\ref{eq:thick:I_a2}) by $a_{\lambda}$ in Eq. (\ref{eq:a_const}), we derived:
 %
 \begin{equation}
I_{a; \lambda}(x,y) \sim
a_{\lambda}
\frac{F_\lambda (x,y)}{F_H (x,y)} 
\frac{\omega_\mathrm{\lambda}}{\omega_H}
(PI)_{\mathrm{obs};H}(x,y).
\label{eq:thick:I_a}
%
\end{equation}

\begin{figure*}[ht!]
\centering
\includegraphics[width=16cm]{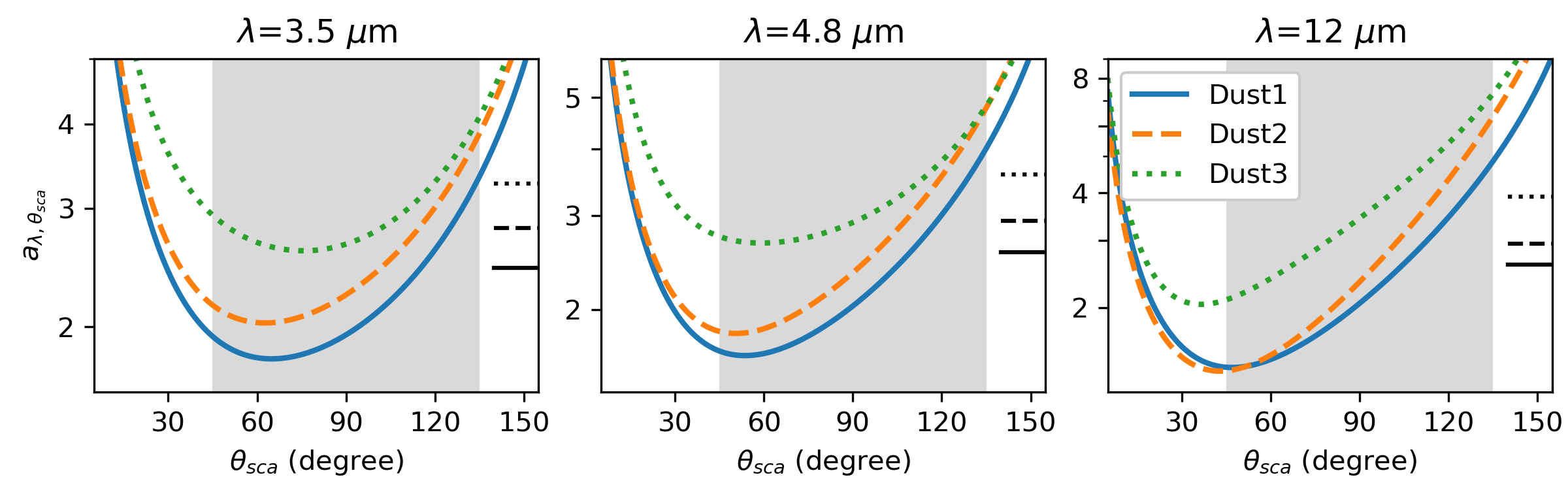}
\caption{$a_{\lambda; \theta_\mathrm{sca}}$ for the three target wavelengths and dust models. The black solid, dashed and dotted lines at the right side of the each panel are $a_\lambda$ derived using Eq. (\ref{eq:a_const}). The gray boxes show the range of the angles used to determine $a_\lambda$ (see the text for details).
\label{fig:phase1}
}
\end{figure*}

Fig. \ref{fig:phase1} shows $a_{\lambda; \theta_\mathrm{sca}}$ for individual dust models at 3.5, 4.8 and 12 $\mu$m. We derived the phase function $(I/F_0)_{\lambda; \theta_\mathrm{sca}}$ using the Henyey-Greenstein approximation. To derive $P_2(H; \theta_\mathrm{sca})$, we used the approximation described below. We first derived the degree of polarization at $\theta_\mathrm{sca}$=90$^\circ$~using the dust parameter files for HO-CHUNK, interpolating the value for the target wavelengths using \texttt{scipy.interpolate.interp1d}. We then scaled the sine function using this value to derive the degree of polarization, and multiplied it by the Henyey-Greenstein function.
The same approximation is also used for the HO-CHUNK radiative transfer code \citep[; Section 2]{Whitney03a}.
\citet{Takami13} executed accurate calculations for the degree of polarization using the Mie theory, and showed that this approximation works well (see their Fig. 6).

In Fig. \ref{fig:phase1} the parameter $a_{\lambda; \theta_\mathrm{sca}}$ 
varies around the representative constant $a_\lambda$ by a factor of 1.3-1.4, 1.3-1.6, and 1.9-2.2 for $\lambda$=3.5, 4.8 and 12 $\mu$m, respectively, at scattering angles of 45$^\circ$-135$^\circ$. We regard these factors as possible systematic errors in the calculated intensity distributions.

The range of scattering angles to derive $a_\lambda$ was selected based on the viewing angle of the disk ($i$$\lesssim$45$^\circ$) discussed in Sect. \ref{sec:eq:overview}. In practice, the height of the disk surface from the midplane makes the scattering angles smaller. Fig. \ref{fig:phase1} shows that $a_{\lambda;\mathrm{min}}$ and $a_{\lambda;\mathrm{max}}$ derived above are actually valid for scattering angles as small as 10$^\circ$-20$^\circ$. Therefore, the adopted approximation would be valid unless the disk does not include extremely small scattering angles in the near side of the disk.

\subsubsection{Double scattering} \label{sec:eq:thick:I_b}

We estimated the intensity distribution via double scattering on small grains using the following equation: 
%
\begin{equation}
I_{b; \lambda}(x,y) \sim
 \frac{F_\lambda(x,y) \alpha(x,y) \omega_\mathrm{\lambda} }{2}
 \frac{\omega_\mathrm{\lambda}}{4 \pi}
  \frac{1}{\mathrm{cos}~i}
 =
  \frac{F_\lambda (x,y) \alpha(x,y) \omega_\mathrm{\lambda}^2}{8 \pi~\mathrm{cos}~i}
\label{eq:thick:I_b1}
.\end{equation}
%
The first, second, and third terms in the middle of the equation correspond to the light scattered at the surface layer into the disk interior, the scattering in the disk interior toward the outside, and an enhancement due to the viewing angle, respectively. The first term implies that half of the scattered light at the surface layer goes to the disk interior. For the second term, we approximated that the scattering occurs isotropically. As described in Sect.  \ref{sec:eq:thick:I_a}, the validity of this approximation was demonstrated by numerical simulations for polarized intensity by \citet{Takami13}.

\begin{figure}[ht!]
\centering
\includegraphics[width=6.5cm]{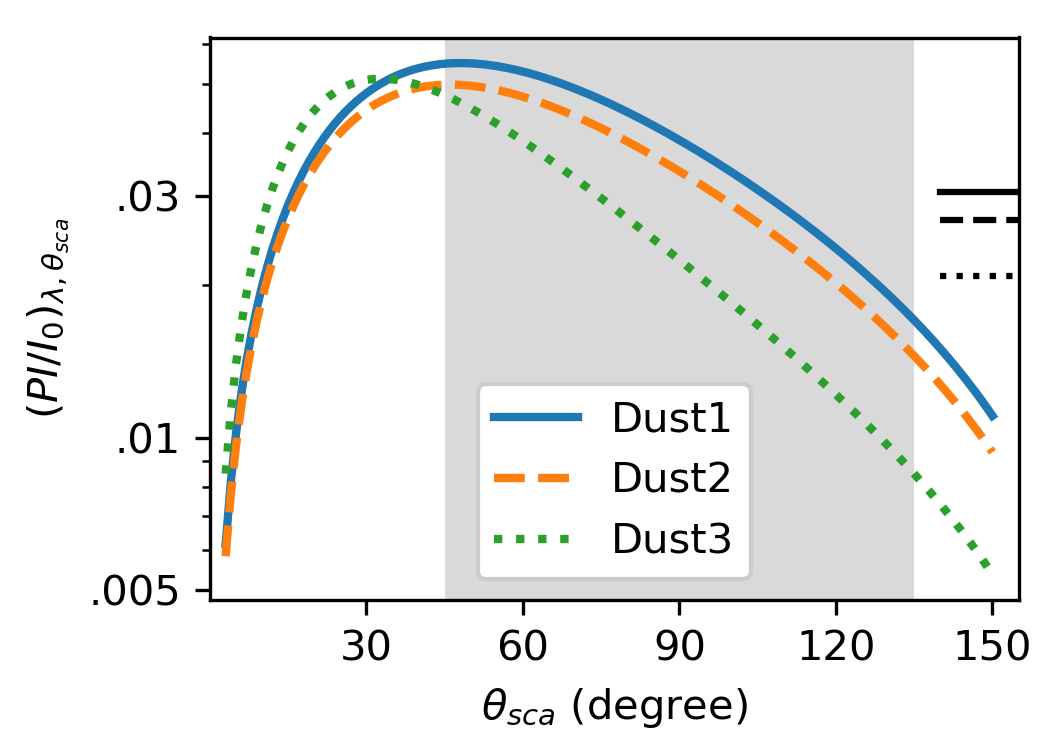}
\caption{$P_2(H; \theta_\mathrm{sca})$ for different dust models. The black solid, dashed and dotted lines at the right side of the panel are the representative constants $P_2(H)$ derived using Eq. (\ref{eq:thick:PI_per_I0_const}). The gray boxes show the range of the angles used to determine $P_2(H)$ (see the text for details).
\label{fig:phase2}
}
\end{figure}

We derived $\alpha(x,y)$ using Eq. (\ref{eq:thick:PI_obs}) as:
%
\begin{equation}
\alpha(x,y) \sim \frac{\mathrm{cos}~i}{\omega_H~F_H(x,y)~P_2 (H;\theta_\mathrm{sca (x,y,i)})}  (PI)_{\mathrm{obs};H}(x,y).
\label{eq:alpha1}
\end{equation}
%
%
Fig. \ref{fig:phase2} shows $P_2 (H; \theta_\mathrm{sca})$ for the three dust models. As for Sect. \ref{sec:eq:thick:I_a}, we derived the representative constants as:
%
\begin{equation}
P_2(H) = \mathrm{exp} \left\{0.5 \left(
\mathrm{log} [P_2(H)_\mathrm{min}]
+
\mathrm{log} [P_2(H)_\mathrm{max}]
\right)
\right\},
\label{eq:thick:PI_per_I0_const}
\end{equation}
where $P_2(H)_\mathrm{min}$ and $P_2(H)_\mathrm{max}$ are the minimum and maximum values measured at $\theta_\mathrm{sca}$=45$^\circ$-135$^\circ$. In this range of scattering angles, $P_2(H; \theta_\mathrm{sca})$ varies by up to a factor of $\sim$2 in terms of the representative constant $P_2(H)$.
As for $a_{\lambda;\mathrm{min}}$ and $a_{\lambda;\mathrm{max}}$ in Sect. \ref{sec:eq:thick:I_a}, $P_2(H)_\mathrm{min}$ and $P_2(H)_\mathrm{max}$ derived in this range of scattering angle is valid for small scattering angles caused by the height of the disk surface, down to $\theta_\mathrm{scat}$=3$^\circ$-8$^\circ$.

Replacing $P_2(H; \theta_\mathrm{sca})$ in Eq. (\ref{eq:alpha1}) by $P_2(H)$, we derived:
%
%
\begin{equation}
\alpha(x,y) \sim \frac{\mathrm{cos}~i}{\omega_H F_H (x,y) P_2(H)}  (PI)_{\mathrm{obs};H}(x,y).
\label{eq:alpha}
\end{equation}
%
Substituting this equation to (\ref{eq:thick:I_b1}), we derived:
%
\begin{eqnarray}
I_{b; \lambda}(x,y)
&\sim&
\frac{F_\lambda (x,y)}{F_H (x,y)}
\frac{\omega_\mathrm{\lambda}^2}{8 \pi \omega_H P_2(H)}
(PI)_{\mathrm{obs};H}(x,y) \nonumber \\
&\sim& 
\frac{\omega_\mathrm{\lambda}}{8 \pi a_\lambda P_2(H)}
I_{a; \lambda}(x,y).
\label{eq:thick:I_b}
\end{eqnarray}
%
Eq. (\ref{eq:thick:I_b}) shows that $I_{b; \lambda}(x,y)$ is identical to $I_{a; \lambda}(x,y)$ but with a different intensity level.

For scattering with large grains, for which forward scattering dominates, the scattered light would be propagated mainly along the disk surface, and faint in the other directions including those toward the disk interior. As a result, the contribution of $I_{b; \lambda}(x,y)$ is smaller than the case with small grains. We therefore used $I_{b; \lambda}(x,y)$ described in Eq. in (\ref{eq:thick:I_b}) as an upper limit.

\subsubsection{Thermal emission from the surface layer} \label{sec:eq:thick:I_c}

The observed intensity would be described as follows:
%
\begin{eqnarray}
I_{c; \lambda}(x,y) &\sim& \frac{\tau_\lambda (x,y) (1-\omega_\mathrm{\lambda}) B_{\lambda,T_\mathrm{s}(x,y)} }{\mathrm{cos}~i} \nonumber \\
                              &\sim& \frac{\alpha(x,y) (1-\omega_\mathrm{\lambda}) B_{\lambda,T_\mathrm{s}(x,y)}}{\mathrm{cos}~i} 
\label{eq:thick:I_c1}
\end{eqnarray}
%
where $\tau_\lambda (x,y)$ is the optical thickness of the surface layer perpendicular to the disk plane;
$B_{\lambda,T}$ is the blackbody function;
and $T_\mathrm{s}(x,y)$ is the temperature of the surface layer. 
Substituting Eq. (\ref{eq:alpha}) to (\ref{eq:thick:I_c1}), we derived:
%
\begin{eqnarray}
I_{c; \lambda}(x,y)  &\sim &
\frac{1-\omega_\mathrm{\lambda}}{\omega_H F_H (x,y) P_2(H)}  B_{\lambda,T_\mathrm{s}(x,y)} (PI)_{\mathrm{obs};H}(x,y).
\label{eq:thick:I_c}
\end{eqnarray}
%

The temperature $T_\mathrm{s}(x,y)$ is derived using the following equation for the thermal budget of the surface layer:
%
\begin{equation}
\alpha(x,y) \int (1-\omega_\mathrm{\lambda}) F_\lambda (x,y) d\lambda
\sim
4 \pi \int (1-\omega_\mathrm{\lambda}) B_{\lambda,T_\mathrm{s}(x,y)} d\lambda.
\label{eq:T_s1}
\end{equation}
%
The left and right sides of the equation correspond to the total energies radiating in and out, respectively.
Substituting Eq. (\ref{eq:alpha}), we derived:
%
\begin{eqnarray}
\int (1-\omega_\mathrm{\lambda}) B_{\lambda,T_\mathrm{s}(x,y)} d\lambda \hspace{5cm} \nonumber \\
\sim
\frac{ \mathrm{cos}~i}{4 \pi \omega_H F_H(x,y) P_2(H)} \left[\int (1-\omega_\mathrm{\lambda}) F_\lambda (x,y)  d\lambda \right] \nonumber \\ \nonumber \\
\times (PI)_{\mathrm{obs};H}(x,y). \hspace{1.5cm}
\label{eq:T_s}
\end{eqnarray}
%

\subsubsection{Thermal emission from the disk interior} \label{sec:eq:thick:I_d}
The observed intensity would be described as follows:
%
\begin{equation}
I_{d; \lambda}(x,y) \sim \frac{B_{\lambda,T_\mathrm{d}(x,y)}}{\mathrm{cos}~i},
\label{eq:thick:I_d1}
\end{equation}
%
where $T_\mathrm{d}(x,y)$ is the temperature of the disk interior just below the surface layer.

To derive $T_\mathrm{d}(x,y)$, we considered the thermal budget as follows: half of radiation reaching to the surface layer is scattered or reemitted toward the disk interior, while the remain half escapes in the opposite direction \citep{Chiang97,Chiang01,Dullemond07_PPV}. We approximated that all the former radiation is absorbed in the disk interior. The thermal radiation from the disk interior escapes only outside of the disk.
Therefore:
%
\begin{equation}
\frac{\alpha(x,y)}{2} \int F_\lambda (x,y) d\lambda
\sim
\pi \int (1-\omega_\mathrm{\lambda})  B_{\lambda,T_\mathrm{d}(x,y)} d\lambda,
\label{eq:T_d1}
\end{equation}
%
where the left and right sides of the equation correspond to total energies radiating in and out, respectively.
Substituting Eq. (\ref{eq:alpha}), we derived:
%
\begin{eqnarray}
\int (1-\omega_\mathrm{\lambda}) B_{\lambda,T_\mathrm{d}(x,y)} d\lambda \hspace{5.5cm} \nonumber \\
\sim
\frac{ \mathrm{cos}~i}{2 \pi \omega_H F_H (x,y) P_2(H)} \left[\int F_\lambda (x,y) d\lambda \right]  (PI)_{\mathrm{obs};H}(x,y). \hspace{0.5cm}
\label{eq:T_d}
\end{eqnarray}
%

In practice, some radiation from the disk surface toward the disk interior is not absorbed but scattered toward the outside of the disk, implying that the right side of Eq. (\ref{eq:T_d}) should be smaller to some extent than described. As a result, the temperature, and therefore $I_{d; \lambda}$ derived using Eq. (\ref{eq:thick:I_d1}) were overestimated to some extent.

\subsection{Emission from an optically thin envelope} \label{sec:eq:thin}

We considered radiative transfer processes similar to (a) and (c) shown in Fig. \ref{fig:concept} and discussed in Sects \ref{sec:eq:thick:I_a} and \ref{sec:eq:thick:I_c}.
If the envelope is geometrically thin, the emission via single scattering is described as:
%
\begin{equation}
I_{a'; \lambda}(x,y) \sim \frac{F_\lambda (x,y)~\kappa_{\mathrm{ext};\lambda} \Sigma_\mathrm{env}(x,y)~\omega_\mathrm{\lambda} P_1(\lambda; \theta_\mathrm{sca (x,y,i)})}{\mathrm{cos}~i} ,
\label{eq:thin:I_a1}
\end{equation}
%
where $\kappa_{\mathrm{ext};\lambda}$ is the dust opacity; and $\Sigma_\mathrm{env}(x,y)$ is the column density of dust associated with unit area of the envelope.
This equation is the same as Eq. (\ref{eq:thick:I_a0}) for the surface layer of the disk, but $\alpha (x,y)$ is replaced by $\kappa_{\mathrm{ext};\lambda} \Sigma_\mathrm{env}(x,y)$. 
In reality, the radiation field at the individual positions in the envelope $F_\lambda (x,y)$ is also a function of $z$. We used a representative value over $z$ (hereafter $z_\mathrm{env}$) to reasonably simplify the calculations (Sect. \ref{sec:eq:source}).

Applying calculations similar to those in Sects. \ref{sec:eq:thick:I_a} and \ref{sec:eq:thick:I_b}, we derived:
%
\begin{eqnarray}
(PI)_{\mathrm{obs};H}(x,y) \hspace{6cm} \nonumber \\
\sim
\frac{F_H (x,y)~\kappa_{\mathrm{ext};H}  \Sigma_\mathrm{env}(x,y)~\omega_H P_2(H; \theta_\mathrm{sca (x,y,i)} )}{\mathrm{cos}~i } ,
\label{eq:thin:PI_a}
\end{eqnarray}

\begin{eqnarray}
\Sigma_\mathrm{env}(x,y) \sim
\frac{\mathrm{cos}~i}{F_H (x,y)~\kappa_{\mathrm{ext};H}~\omega_H P_2(H; \theta_\mathrm{sca (x,y,i)})} 
(PI)_{\mathrm{obs};H}(x,y),
\label{eq:Sigma}
\end{eqnarray}

\begin{equation}
I_{a'; \lambda}(x,y) 
\sim
a_\lambda
\frac{F_\lambda (x,y)}{F_H (x,y)} \frac{\kappa_{\mathrm{ext};\lambda}}{\kappa_{\mathrm{ext};H}} \frac{\omega_\mathrm{\lambda}}{\omega_H}
(PI)_{\mathrm{obs};H}(x,y)
\hspace{0.3cm}
\label{eq:thin:I_a}
\end{equation}
%
%
%

The thermal emission from the optically thin envelope is described as:
%
\begin{equation}
I_{c'; \lambda}(x,y) = \frac{\kappa_{\mathrm{ext};\lambda} \Sigma_\mathrm{env}(x,y) (1-\omega_\mathrm{\lambda}) B_{\lambda,T_\mathrm{e}(x,y)}}{\mathrm{cos}~i}.
\label{eq:thin:I_c1}
\end{equation}
%
Substituting Eq. (\ref{eq:Sigma}) we derived:
%
\begin{eqnarray}
I_{c'; \lambda}(x,y)  &\sim &
\frac{1-\omega_\mathrm{\lambda}}{ \omega_H F_H (x,y) P_2 (H)}
\frac{\kappa_{\mathrm{ext};\lambda}}{\kappa_{\mathrm{ext};H}}
\nonumber \\
&& ~~~~~~~~~~~~~~~~~~~~~~~~\times
B_{\lambda,T_\mathrm{e}(x,y)} (PI)_{\mathrm{obs};H}(x,y).
\label{eq:thin:I_c}
\end{eqnarray}
%
%
We derived the temperature $T_\mathrm{e}(x,y)$ using the following equation:
%
\begin{equation}
4 \pi \int (1-\omega_\mathrm{\lambda}) B_{\lambda,T_\mathrm{e}(x,y)} d\lambda \sim \int (1-\omega_\mathrm{\lambda}) F_\lambda (x,y)  d\lambda.
\label{eq:T_e}
\end{equation}
%
The left and right sides of the equation correspond to the energy radiating out and in, respectively.

If the envelope is geometrically thick, one may want to remove the 1/cos $i$ term from the above equations. This does not affect Eqs. (\ref{eq:thin:I_a})(\ref{eq:thin:I_c})(\ref{eq:T_e}), as the term has been canceled out. Furthermore, the term was canceled out when we used Eq. (\ref{eq:Sigma}) in Sect. \ref{sec:eq:check:env}. Therefore, we can use the same equations for a geometrically thick envelope as well.

\subsection{Radiation from the illuminating source}  \label{sec:eq:source}

\subsubsection{Overview} \label{sec:eq:source:overview}

In Sects. \ref{sec:eq:thick} and \ref{sec:eq:thin}, we derived the equations for the individual emission components as functions of $F_\lambda (x,y)$, that is radiation field by the central illuminating source at individual positions. 
In this subsection, we will replace them using the observed flux for the central source, considering the fact that
the illuminating source is either a star or a flat compact self-luminous disk.

If the illuminating source is a star, the radiation of which is isotropic, the radiation field $F_\lambda$ is approximately described for a geometrically thin disk or envelope as follows:
%
\begin{equation}
F_\lambda (x,y) \sim  \frac{d^2}{r^2}  f_\lambda^{-1} F_{\mathrm{obs},\lambda},
\label{eq:F_star}
\end{equation}
%
where $F_{\mathrm{obs},\lambda}$ is the observed flux from the central illuminating source;
$d$ is the distance from the observer to the target;
and $f_\lambda$ is a factor to correct the foreground extinction, which is defined as:
%
\begin{equation}
f_\lambda = \mathrm{exp} (-0.921 A_\lambda),
\label{eq:f_lambda1}
\end{equation}
%
where $A_\lambda$ is the extinction at the wavelength $\lambda$ \citep{Spitzer78}.
Throughout the paper, we derived $A_\lambda / A_V$ using the opacity for the `Dust2' model for molecular clouds (Sect. \ref{sec:dust}). As a result, Eq. (\ref{eq:f_lambda1}) could alternatively be described as follows:
%
\begin{equation}
f_\lambda = \mathrm{exp} \left[ -0.921 A_V \left( \frac{\kappa_{\mathrm{ext},\lambda}}{\kappa_{\mathrm{ext},V}}  \right)_\mathrm{Dust2} \right],
\label{eq:f_lambda}
\end{equation}
where $(\kappa_{\mathrm{ext},\lambda}/\kappa_{\mathrm{ext},V})_\mathrm{Dust2}$ is the dust opacity normalized to that of the $V$ band ($\lambda$=0.55 $\mu$m) based on the `Dust2' model.
%
We assumed that there is no extinction between the central illuminating source and the disk or the envelope as it is likely that the strong FUor winds blow up dust grains in these regions \citep[][]{Takami19}.

If the illuminating source is a flat compact disk, Eq. (\ref{eq:F_star}) is replaced with:
%
\begin{equation}
F_\lambda (x,y) \sim  \frac{d^2}{r^2} 
\left( \frac{\gamma}{\mathrm{cos}~i} \right)
f_\lambda^{-1} F_{\mathrm{obs},\lambda},
\label{eq:F_disk}
\end{equation}
%
where $\gamma$ is the sine of the grazing angle of the flat inner disk from the observer and the extended disk or envelope, therefore
%
\begin{equation}
\gamma = \mathrm{sin~[tan^{-1}} (z/r)]. 
\label{eq:gamma}
\end{equation}
%
In reality, $\gamma$ varies with position on the disk surface and in the envelope, but inclusion of this variation would make the calculations too complicated to derive the intensity distributions in a straightforward manner. 
Therefore, we used constant values for our calculations (0.1, 0.2, and 0.4; Sect \ref{sec:application}; hereafter $\overline{\gamma}$), and investigate how this approximation affects the accuracy of our calculations.

The term 1/cos $i$ in Eq. (\ref{eq:F_disk}) is to correct the observed flux for the viewing angle of the flat compact disk.
Due to the assumed flat nature ($z$$\sim$0) of the compact disk, the term 1/cos $i$ in this equation is free from the systematic errors discussed in Sect. \ref{sec:eq:thick:I_a}. To discriminate between these two cases, we bracketed 1/cos $i$ with $\gamma$ for those originating from Eq. (\ref{eq:F_disk}).

\citet{Liu16} showed the SEDs observed for a few FUors, with the stellar continuum that the authors derived based on their model fitting of the entire SED using their radiative transfer simulations. For their simulations, the central illuminating source is a star, however, the infrared excess over the stellar continuum is apparent at $\lambda \gtrsim 3$ $\mu$m~as the inner disk is heated by stellar radiation. Therefore we regarded the illuminating source as an inner disk at $\lambda \ge 3$ $\mu$m.

\subsubsection{Revised equations for intensities and temperatures} \label{sec:eq:source:neweqs}

We revised Eqs. (\ref{eq:thick:I_a})(\ref{eq:thick:I_c})(\ref{eq:T_s})(\ref{eq:thick:I_d1})(\ref{eq:T_d})(\ref{eq:thin:I_a})(\ref{eq:thin:I_c})(\ref{eq:T_e}) by substituting (\ref{eq:F_star}) and (\ref{eq:F_disk}). We also applied foreground extinction to the observed polarized intensity distributions $(PI)_{\mathrm{obs};H}(x,y)$;  the derived intensity distributions $I_{a; \lambda}(x,y)$, $I_{c; \lambda}(x,y)$, $I_{d; \lambda}(x,y)$, $I_{a'; \lambda}(x,y)$, and $I_{c'; \lambda}(x,y)$; and the observed flux of the central source $F_{\mathrm{obs},\lambda}$.
We corrected these parameters for foreground extinction by multiplying $f_\lambda$ defined in Eq. (\ref{eq:f_lambda}).

For the case where the illuminating source is a star at $\lambda$$<$3 $\mu$m, we substituted Eq. (\ref{eq:F_star}) to $F_H(x,y)$ for all of the above equations. We also substituted Eq. (\ref{eq:F_star}) to $F_\lambda(x,y)$ in Eqs. (\ref{eq:T_s})(\ref{eq:T_d})(\ref{eq:T_e}) as the heating of the extended disk or envelope is dominated by illumination at $\lambda$$<$3 $\mu$m~(Sect. \ref{sec:application:ill}). In contrast, we substituted Eq. (\ref{eq:F_disk}), that of a flat compact disk, to (\ref{eq:thick:I_a})(\ref{eq:thin:I_a}) to derive the intensity distributions for single scattered light in mid-infrared wavelengths. As a result, the revised equations are:
%
 \begin{equation}
\frac{I_{a; \lambda}(x,y)}{F_{\mathrm{obs},\lambda}}
\sim
a_{\lambda}
\left( \frac{\overline{\gamma}}{\mathrm{cos}~i} \right)
\frac{\omega_\mathrm{\lambda}}{\omega_H}
\frac{(PI)_{\mathrm{obs};H}(x,y)}{F_{\mathrm{obs},H}},
\label{eq:thick:I_a:star}
\end{equation}
\begin{eqnarray}
I_{c; \lambda}(x,y)  &\sim &
\frac{f_\lambda}{P_2(H)}
\frac{r^2}{d^2}
\frac{1-\omega_\mathrm{\lambda}}{\omega_H}  \nonumber \\
&& ~~~~~~~~\times B_{\lambda,T_\mathrm{s}(x,y)}
\frac{(PI)_{\mathrm{obs};H}(x,y)}{F_{\mathrm{obs},H}},
\label{eq:thick:I_c:star}
\end{eqnarray}
\begin{eqnarray}
I_{d; \lambda}(x,y) &\sim& f_\lambda \frac{B_{\lambda,T_\mathrm{d}(x,y)}}{\mathrm{cos}~i}
\label{eq:thick:I_d} \\
\frac{I_{a'; \lambda}(x,y)}{F_{\mathrm{obs},\lambda}}
&\sim&
a_\lambda
\left( \frac{\overline{\gamma}}{\mathrm{cos}~i} \right)
\frac{\kappa_{\mathrm{ext};\lambda}}{\kappa_{\mathrm{ext};H}} \frac{\omega_\mathrm{\lambda}}{\omega_H}
\frac{(PI)_{\mathrm{obs};H}(x,y)}{F_{\mathrm{obs},H}},
\label{eq:thin:I_a:star}
\end{eqnarray}
\begin{eqnarray}
I_{c'; \lambda}(x,y)  &\sim &
\frac{f_\lambda}{P_2(H)}
\frac{r^2}{d^2}
\frac{\kappa_{\mathrm{ext};\lambda}}{\kappa_{\mathrm{ext};H}}
\frac{1-\omega_\mathrm{\lambda}}{\omega_H} \nonumber \\
&& \hspace{2cm} \times B_{\lambda,T_\mathrm{e}(x,y)}
\frac{(PI)_{\mathrm{obs};H}(x,y)}{F_{\mathrm{obs},H}},
\label{eq:thin:I_c:star}
\end{eqnarray}
\begin{eqnarray}
\int (1-\omega_\mathrm{\lambda}) B_{\lambda,T_\mathrm{s}(x,y)} d\lambda \hspace{5cm} \nonumber \\
\sim
\frac{ \mathrm{cos}~i}{4 \pi \omega_H P_2(H)} \left[\int (1-\omega_\mathrm{\lambda}) f_\lambda^{-1} F_{\mathrm{obs},\lambda}  d\lambda \right] \frac{(PI)_{\mathrm{obs};H}(x,y)}{F_{\mathrm{obs},H}},
\label{eq:T_s:star}
\end{eqnarray}
\begin{eqnarray}
\int (1-\omega_\mathrm{\lambda}) B_{\lambda,T_\mathrm{d}(x,y)} d\lambda \hspace{5.5cm} \nonumber \\
\sim
\frac{ \mathrm{cos}~i}{2 \pi \omega_H P_2(H)} \left[\int f_\lambda^{-1} F_{\mathrm{obs},\lambda}  d\lambda \right]
\frac{(PI)_{\mathrm{obs};H}(x,y)}{F_{\mathrm{obs},H}},
\label{eq:T_d:star}
\end{eqnarray}
\begin{equation}
\int (1-\omega_\mathrm{\lambda}) B_{\lambda,T_\mathrm{e}(x,y)} d\lambda \sim \frac{d^2}{4 \pi r^2} \int (1-\omega_\mathrm{\lambda}) f_\lambda^{-1} F_{\mathrm{obs},\lambda}  d\lambda.
\label{eq:T_e:star}
\end{equation}

If the central source is a self-luminous disk,
we alternatively substituted Eq. (\ref{eq:F_disk}) to $F_H(x,y)$ and $F_\lambda(x,y)$ for all equations. As a result,
we replaced Eqs. (\ref{eq:thick:I_a:star})(\ref{eq:thick:I_c:star})(\ref{eq:thin:I_a:star})(\ref{eq:thin:I_c:star})(\ref{eq:T_e:star}) with the following Equations:
 \begin{equation}
\frac{I_{a; \lambda}(x,y)}{F_{\mathrm{obs},\lambda}} \sim
a_{\lambda}
\frac{\omega_\mathrm{\lambda}}{\omega_H}
\frac{(PI)_{\mathrm{obs};H}(x,y)}{F_{\mathrm{obs},H}},
\label{eq:thick:I_a:disk}
\end{equation}
\begin{eqnarray}
I_{c; \lambda}(x,y)  &\sim &
\frac{f_\lambda}{P_2(H)}
\frac{r^2}{d^2}
\frac{1-\omega_\mathrm{\lambda}}{\omega_H} B_{\lambda,T_\mathrm{s}(x,y)}
\nonumber \\
&& \hspace{2cm} \times 
\left( \frac{\overline{\gamma}}{\mathrm{cos}~i} \right)^{-1}
\frac{(PI)_{\mathrm{obs};H}(x,y)}{F_{\mathrm{obs},H}},
\label{eq:thick:I_c:disk}
\end{eqnarray}
\begin{eqnarray}
\frac{I_{a'; \lambda}(x,y)}{F_{\mathrm{obs},\lambda}}
&\sim&
a_\lambda
\frac{\kappa_{\mathrm{ext};\lambda}}{\kappa_{\mathrm{ext};H}} \frac{\omega_\mathrm{\lambda}}{\omega_H}
\frac{(PI)_{\mathrm{obs};H}(x,y)}{F_{\mathrm{obs},H}},
\label{eq:thin:I_a:disk}
\end{eqnarray}
\begin{eqnarray}
I_{c'; \lambda}(x,y)  &\sim &
\frac{f_\lambda}{P_2(H)}
\frac{r^2}{d^2}
\frac{\kappa_{\mathrm{ext};\lambda}}{\kappa_{\mathrm{ext};H}}
\frac{1-\omega_\mathrm{\lambda}}{\omega_H}
B_{\lambda,T_\mathrm{e}(x,y)}
\nonumber \\
&& \hspace{2cm} \times 
\left( \frac{\overline{\gamma}}{\mathrm{cos}~i} \right)^{-1}
\frac{(PI)_{\mathrm{obs};H}(x,y)}{F_{\mathrm{obs},H}},
\label{eq:thin:I_c:disk}
\end{eqnarray}
\begin{equation}
\int (1-\omega_\mathrm{\lambda}) B_{\lambda,T_\mathrm{e}(x,y)} d\lambda \sim
\frac{d^2}{4 \pi r^2}
\left( \frac{\overline{\gamma}}{\mathrm{cos}~i} \right)
\int (1-\omega_\mathrm{\lambda}) f_\lambda^{-1} F_{\mathrm{obs},\lambda}  d\lambda.
\label{eq:T_e:disk}
\end{equation}

%
We used these equations together with (\ref{eq:a_const})(\ref{eq:thick:PI_per_I0_const})(\ref{eq:thick:I_b})
for $a_{\lambda}$, $(P_2)_H$, and $I_{b; \lambda}(x,y)$, respectively,
to derive the intensity distributions.
As shown in Eqs. (\ref{eq:thick:I_a:star})(\ref{eq:thin:I_a:star})(\ref{eq:thick:I_a:disk})(\ref{eq:thin:I_a:disk}), 
$I_{a; \lambda}(x,y)$
and
$I_{a'; \lambda}(x,y)$
(therefore $I_{b; \lambda}(x,y)$ as well; see Sect. \ref{sec:eq:thick:I_b}) are identical to $PI_{\mathrm{obs};H}(x,y)$ but with different intensities levels under the given approximations of scattering angles made in Sects. \ref{sec:eq:thick:I_a} and \ref{sec:eq:thick:I_b}.


\subsection{Checking self-consistencies} \label{sec:eq:check}

In this subsection we derive the equations to check self-consistencies of calculations for an extended disk (Sect. \ref{sec:eq:check:disk}) and an envelope (\ref{sec:eq:check:env}).

\subsubsection{Surface geometry of the extended disk} \label{sec:eq:check:disk}

We summarize below the approximations which can potentially cause significant systematic errors, among those we used in previous subsections:

\begin{itemize}
\item[(A)] We multipled the intensities by 1/cos $i$, where $i$ is the viewing angle of the disk midplane, for the enhancement of the intensity, 
ignoring the inclination of the disk surface from the midplane at individual positions (Sect. \ref{sec:eq:thick:I_a}).

\item[(B)] We derived Eqs. (\ref{eq:F_star}) and (\ref{eq:F_disk}), omitting the term for the height of the disk surface from the midplane or the vertical distribution of the envelope (Sect. \ref{sec:eq:source}).

\item[(C)] We used a representative constant $\overline{\gamma}$, without including its spatial variation (Sect. \ref{sec:eq:source}).

\end{itemize}

In this subsection we derive the equations to investigate this.
We first discuss the case that the central illuminating source is a star.
Substituting Eq. (\ref{eq:F_star}) to (\ref{eq:alpha}), and also correcting foreground extinction for $(PI)_{\mathrm{obs};H}$, we derived:
%
\begin{equation}
\alpha(x,y) \sim \frac{r^2 \mathrm{cos}~i}{d^2 \omega_H P_2(H)} \frac{(PI)_{\mathrm{obs};H}(x,y)}{F_{\mathrm{obs};H}}.
\label{eq:alpha:star}
\end{equation}
%
As defined in Eq. (\ref{eq:alpha:definition}), $\alpha(x,y)$ must range between 0 and 1.

We derived the inclination of the disk surface $\delta(x,y)$ in the radial direction in terms of the disk midplane, and its height from the disk midplane as:
\begin{eqnarray}
\delta(x,y) &=& \mathrm{tan}^{-1} [z_\mathrm{disk}(x,y)/r] + \beta(x,y) \nonumber \\
&=& \mathrm{tan}^{-1}[z_\mathrm{disk}(x,y)/r]+\mathrm{sin}^{-1}\alpha(x,y), 
\label{eq:delta} \\
z_\mathrm{disk} (x,y) &=& \int_{0}^{r} \mathrm{tan}~\delta(x,y) ~dr.
\label{eq:z_disk1}
\end{eqnarray}
%

The integrations in Eq. (\ref{eq:z_disk1}) would be made from the central source toward all the position angles. In practice, the central illuminating source is extremely bright compared with the extended emission, making the observations of the PI distribution close to the central source unreliable \citep[][see Sect. \ref{sec:application}]{Liu16,Takami18,Laws20}.
Therefore, we alternatively executed the integration as follows:
%
\begin{equation}
z_\mathrm{disk} (x,y) = z_{\mathrm{disk}; r_\mathrm{min}}(x,y) + \int_{r_\mathrm{min}}^{r} \mathrm{tan}~\delta(x,y) ~dr,
\label{eq:z_disk}
\end{equation}
%
where $z_{\mathrm{disk}; r_\mathrm{min}}(x,y)$ is the height of the disk surface at the outer edge of the software aperture mask at the central source, which is a free parameter. We cannot constrain this parameter from the observations \citep{Takami14}. 
The adopted $z_{\mathrm{disk}; r_\mathrm{min}}(x,y)$ and calculated $z_\mathrm{disk} (x,y)$ must be approximately consistent with $\overline{\gamma}$ defined and used in Sect. \ref{sec:eq:source}.
In Sect. \ref{sec:application:consistencies}, we adjusted $z_{\mathrm{disk}; r_\mathrm{min}}(x,y)$ to investigate whether self-consistent solutions exist.

If the central illuminating source is a compact self-luminous disk, we replaced Eq. (\ref{eq:alpha:star}) with the equation below, using Eq. (\ref{eq:F_disk}) instead of (\ref{eq:F_star}):
%
\begin{equation}
\alpha(x,y) \sim \frac{r^2 \mathrm{cos}~i}{d^2 \omega_H P_2(H)}
\left( \frac{\overline{\gamma}}{\mathrm{cos}~i}   \right)^{-1}
\frac{(PI)_{\mathrm{obs};H}(x,y)}{F_{\mathrm{obs};H}}.
\label{eq:alpha:disk}
\end{equation}
%
We calculated $z_\mathrm{disk} (x,y)$ and $\delta_\mathrm{disk} (x,y)$ by substituting this equation to Eq. (\ref{eq:delta}) and execute numerical integrations with Eq. (\ref{eq:z_disk}).

For some cases, the calculated values of $\alpha(x,y)$ and $\delta(x,y)$ exceed 1 and 90$^\circ$, respectively (Sect. \ref{sec:application:consistencies}). 
In these cases, we conclude that the given combination of the SED, the central source and the dust model is not consistent with the PI distributions in $H$ band.

\subsubsection{Optical thickness of the extended envelope} \label{sec:eq:check:env}

We checked optical thicknesses for the following two directions: the radial direction
($\tau_{\mathrm{r}; \lambda}$); and the line of sight to the observer, that is:
%
\begin{equation}
\tau_{l; \lambda} (x,y) = \frac{\kappa_{\mathrm{ext};\lambda} \Sigma_\mathrm{env}(x,y)}{\mathrm{cos}~i}.
\label{eq:tau_l:definition}
\end{equation}
%
Substituting Eq. (\ref{eq:Sigma})(\ref{eq:F_star})(\ref{eq:F_disk}), and also correcting foreground extinction for $(PI)_{\mathrm{obs};H}$, we derived the following equations in the cases that the central illuminating source is a star and a flat compact disk, respectively:

%
\begin{eqnarray}
\tau_{l; \lambda} (x,y) 
&\sim&
\frac{\kappa_{\mathrm{ext};\lambda}}{\kappa_{\mathrm{ext};H}}
\frac{r^2}{d^2 \omega_H P_2(H)} 
\frac{(PI)_{\mathrm{obs};H}(x,y)}{F_{\mathrm{obs};H}} ,
\label{eq:tau_l:star}
\nonumber \\ \\
\tau_{l; \lambda} (x,y) 
&\sim&
\frac{\kappa_{\mathrm{ext};\lambda}}{\kappa_{\mathrm{ext};H}}
\frac{r^2}{d^2 \omega_H P_2(H)} 
\left( \frac{\overline{\gamma}}{\mathrm{cos}~i} \right)^{-1}
\frac{(PI)_{\mathrm{obs};H}(x,y)}{F_{\mathrm{obs};H}}.
\label{eq:tau_l:disk}
\nonumber \\
\end{eqnarray}
%


Suppose that, at each of the (x,y) positions, the dust grains are vertically and uniformly distributed up to a height twice the typical value $z_\mathrm{env}(x,y)$.
We would then be able to estimate $\tau_{\mathrm{r;\lambda}}$ as:
%
\begin{equation}
\tau_{\mathrm{r}; \lambda} (x,y) 
\sim
\int_{r_\mathrm{min}}^{r_\mathrm{max}} \frac{\tau_{l;\lambda}(x,y)}{4 z_\mathrm{env}(x,y)} dr.
\label{eq:tau_r_tmp}
\end{equation}
%
As $z_\mathrm{env}(x,y)$$\sim$$ \overline{\gamma} r$ (Eq. \ref{eq:gamma}), we revised this equation as:
%
\begin{equation}
\tau_{\mathrm{r}; \lambda} (x,y) 
\sim
\int_{r_\mathrm{min}}^{r_\mathrm{max}} \frac{\tau_{l;\lambda}(x,y)}{4 \overline{\gamma} r} dr.
\label{eq:tau_r}
\end{equation}
%
We regard this value as a typical optical thickness in the radial direction.
As for Eq. (\ref{eq:z_disk}), the integrations were made for all the position angles.
The equation yields a lower limit of the actual optical thickness as we cannot include the opacities within $r_\mathrm{min}$.


\section{Application} \label{sec:application}

\begin{figure}[ht!]
\centering
\includegraphics[width=9cm]{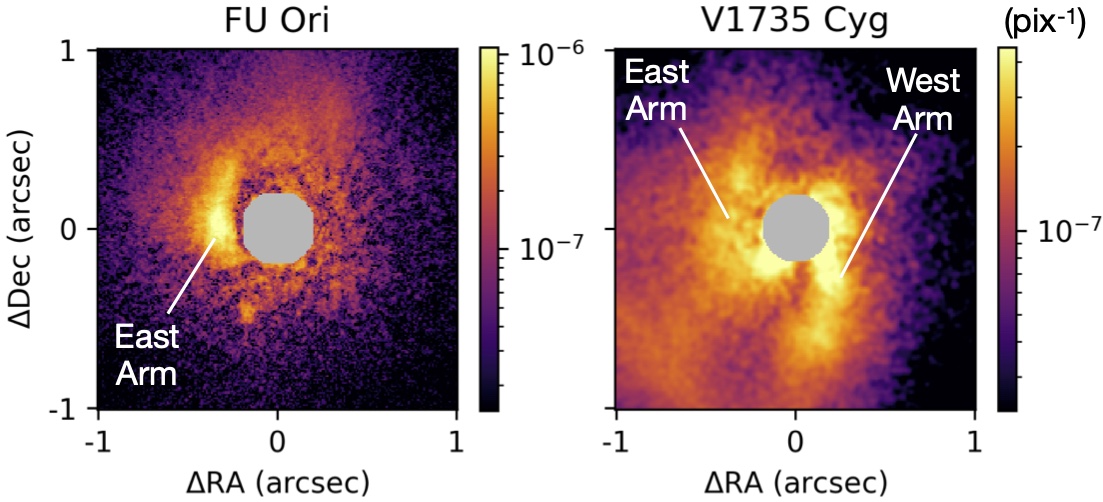}
\caption{PI distribution in $H$ band, $PI_{\mathrm{obs};H}$, for FU Ori and V1735 Cyg \citep{Takami18}, with a pixel scale of 9.5 mas, normalized to the Stokes $I$ flux of the central source. North is up.
In each image, the central region is masked as we were not able to make reliable measurement due to the central source being significantly brighter than the extended emission.
Features such as spiral arms are labeled (see the text for details).
\label{fig:PI_H}
}
\end{figure}

In this section, we apply the equations in Sect. \ref{sec:eq} to two FUors: FU Ori and V1735 Cyg.
We summarize the properties of these objects in Table \ref{tbl:targets}.
For $H$ band imaging polarimetry, we used the images obtained using Subaru-HiCIAO by \citet{Takami18}.
As shown in Fig. \ref{fig:PI_H}, each object is associated with (1) a bright arm-like feature or two; and (2) surrounding diffuse extended emission. As discussed in Sect. \ref{sec:intro}, we regard the former as the features of our major interest for understanding the accretion process onto FUors.

\begin{table}
\caption{Targets. \label{tbl:targets}}
\begin{tabular}{lccc}
\hline\hline
Target  & Distance$^{a}$        & $A_V$ & $L_\mathrm{source}$$^{b}$      \\
                & (pc)                                  &               & ($L_\sun$)
\\ \hline
FU Ori          & 408$\pm$3     & 1.5$^{c}$             & 1.0$\times$10$^2$     \\
V1735 Cyg       & 690$\pm$40    & 7$^{d,c}$             & 43                                    \\
\hline
\end{tabular} \\
$^{a}${Based on the Gaia DR3 parallax measurements \citep{Bailer21}.}\\
$^{b}${The luminosity of the central source (Sect. \ref{sec:application:ill}).}\\
$^{c}${\citet{Gramajo14}.}\\
$^{d}${\citet{Quanz07}.}
\end{table}

In Sect. \ref{sec:application:ill}, we describe radiation from the central source.
In Sect. \ref{sec:application:consistencies}, we check self-consistencies for the individual cases using the equations derived in Sect. \ref{sec:eq:check}.
In Sect. \ref{sec:application:I_indiv}, we investigate the individual emission components, and show that the intensity distribution is dominated by the single scattering process for all the cases.
In Sect. \ref{sec:application:I_a}, we present the distributions for the single-scattering emission for all the cases, compare their intensities, and further discuss their self-consistencies.

We executed calculations for $\overline{\gamma}$=0.1, 0.2 and 0.4. In Sects. \ref{sec:application:consistencies}-\ref{sec:application:I_a} we mainly present the results for $\overline{\gamma}$=0.2, and briefly summarize the results for the cases with $\overline{\gamma}$=0.1 and 0.4. We present the detailed results for the latter cases in Appendix \ref{appendix}.

We executed all the calculations using \texttt{numpy} and \texttt{scipy} for a viewing angle $i$=0$^\circ$ (that is the face-on view).
If the extended emission is due to a disk with an intermediate viewing angle ($i$$\sim$45$^\circ$), we would expect scattered emission from not only the front side of the surface, but also the edge of the opposite side with a dark lane in between \citep[e.g.,][]{Watson07_PPV,Liu16,Dong16}. The $H$ band images at a large field of view do not show evidence for the latter emission component \citep{Liu16,Takami18,Laws20}, supporting the idea of a small viewing angle. For the case that the extended emission is due to an optically thin envelope, we will discuss in Sect. \ref{sec:application:i} how the use of an intermediate viewing angle affects the results.


\subsection{Illumination by the central source} \label{sec:application:ill}
 
We derived the SEDs for these targets using fluxes collected from the literature by \citet{Gramajo14}. The observations were made  using the Infrared Astronomical Satellite (IRAS), the Infrared Space Observatory (ISO), the Midcourse Space Experiment (MSX), the Kuiper Airborne Observatory (KAO), the Balloon-borne Large Aperture Submillimeter Telescope (BLAST), the James Clerk Maxwell Telescope (JCMT) and various ground-based telescopes for optical to mid-infrared wavelengths. These data cover the wavelength ranges of 0.55-850 $\mu$m~and  1.24-500 $\mu$m~for FU Ori and V1735 Cyg, respectively, at angular resolutions up to $\sim$3'. For V1735 Cyg, we added the optical fluxes measured by the Sloan Digital Sky Survey (SDSS) project for the $g$ (0.47 $\mu$m, 20.7 mag.), $r$ (0.62 $\mu$m, 17.4 mag.), $i$ (0.75 $\mu$m, 15.2 mag.), and $z$ bands  (0.89 $\mu$m, 13.4 mag.).
 
\begin{figure}[ht!]
\centering
\includegraphics[width=8.5cm]{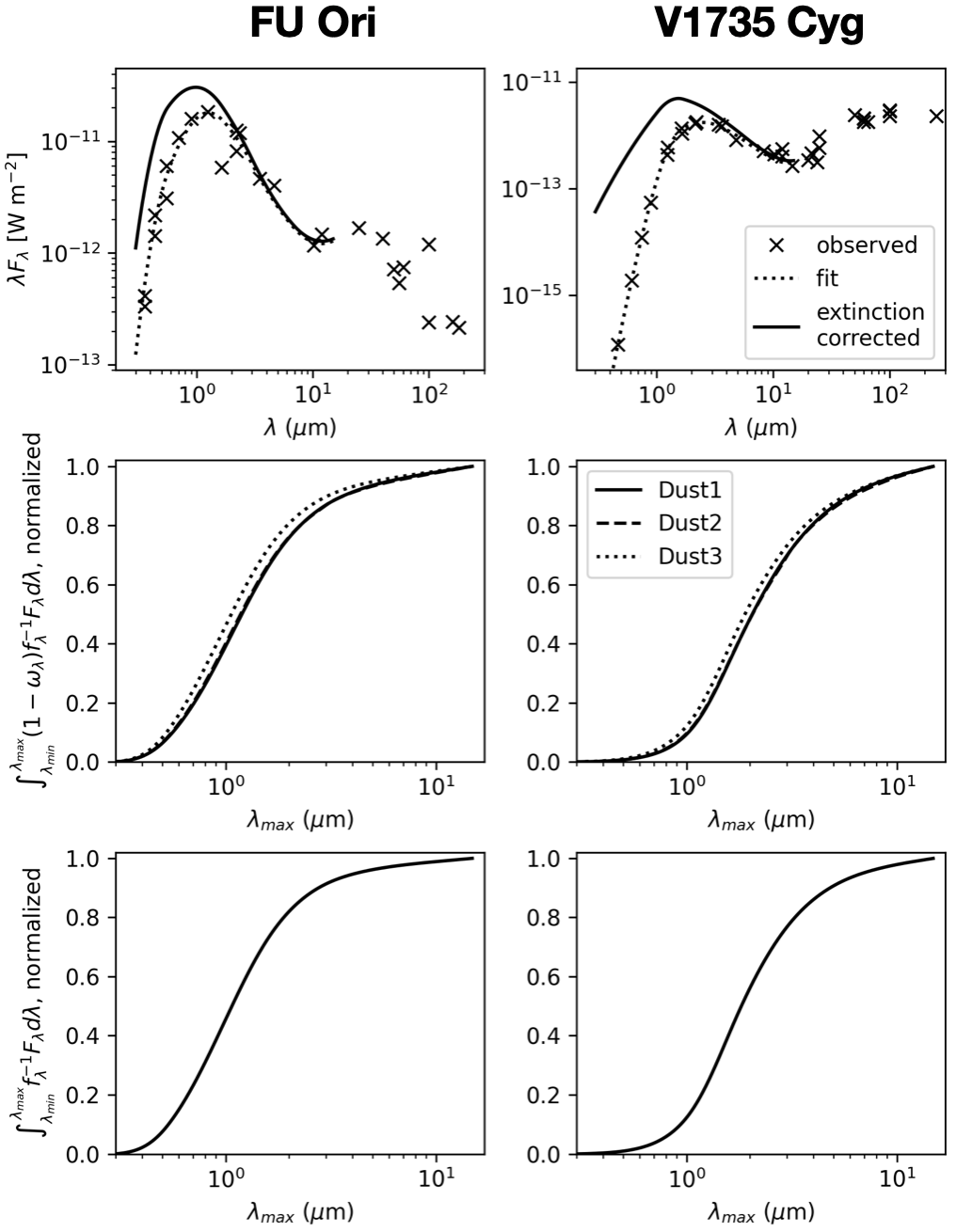}
\caption{
Radiation from the central sources. ($top$) The spectral energy distributions. The crosses show the observations; the dotted curves show fittings; and the solid curves show those corrected for foreground extinction. ($middle$) Cumulative fraction of the integrations used in Eqs. (\ref{eq:T_s:star})(\ref{eq:T_e:star})(\ref{eq:T_e:disk}). The horizontal axis is the maximum wavelength for integration up to $\lambda$=15 $\mu$m.
($bottom$) Same as the middle panels but for the integrations used in Eq. (\ref{eq:T_d:star}).
\label{fig:SEDs}
}
\end{figure}

In Fig. \ref{fig:SEDs}, we show the observed SEDs, the fitting curves obtained using \texttt{scipy.interpolate.interp1d}, and those curves after correcting for the foreground extinction tabulated in Table \ref{tbl:targets}. The observed SEDs show a remarkable far-infrared excess at $\lambda$$>$15 $\mu$m, which is due an envelope, not the central illuminating source \citep[e.g.,][]{Gramajo14}. We therefore obtained the fitting curves at up to $\lambda$=15 $\mu$m~for the integrations with the central illuminating source in Eqs. (\ref{eq:T_s:star})(\ref{eq:T_d:star})(\ref{eq:T_e:star})(\ref{eq:T_e:disk}). For the shortest wavelengths, we extrapolated the curves down to $\lambda$=0.3 $\mu$m~using the same python command. In Table \ref{tbl:targets}, we show the source luminosities obtained using these curves.

In Fig. \ref{fig:SEDs}, we also show the cumulative fraction for the integrations at the right side of Eqs. (\ref{eq:T_s:star})(\ref{eq:T_d:star})(\ref{eq:T_e:star})(\ref{eq:T_e:disk}) as a function of the maximum wavelength. 
The curves in the middle and bottom panels of the figure indicate that the flux at $\lambda$$<$3 $\mu$m is responsible for $\sim$90 \% of radiative heating by the central illuminating source for FU Ori, and 70-80 \% for V1735 Cyg, agreeing with the calculations we made in Sect. \ref{sec:eq:source}.


\subsection{Self-consistencies} \label{sec:application:consistencies}

\subsubsection{Cases illuminated by a star} \label{sec:application:consistencies:star}

\begin{figure*}[ht!]
\centering
\includegraphics[width=16cm]{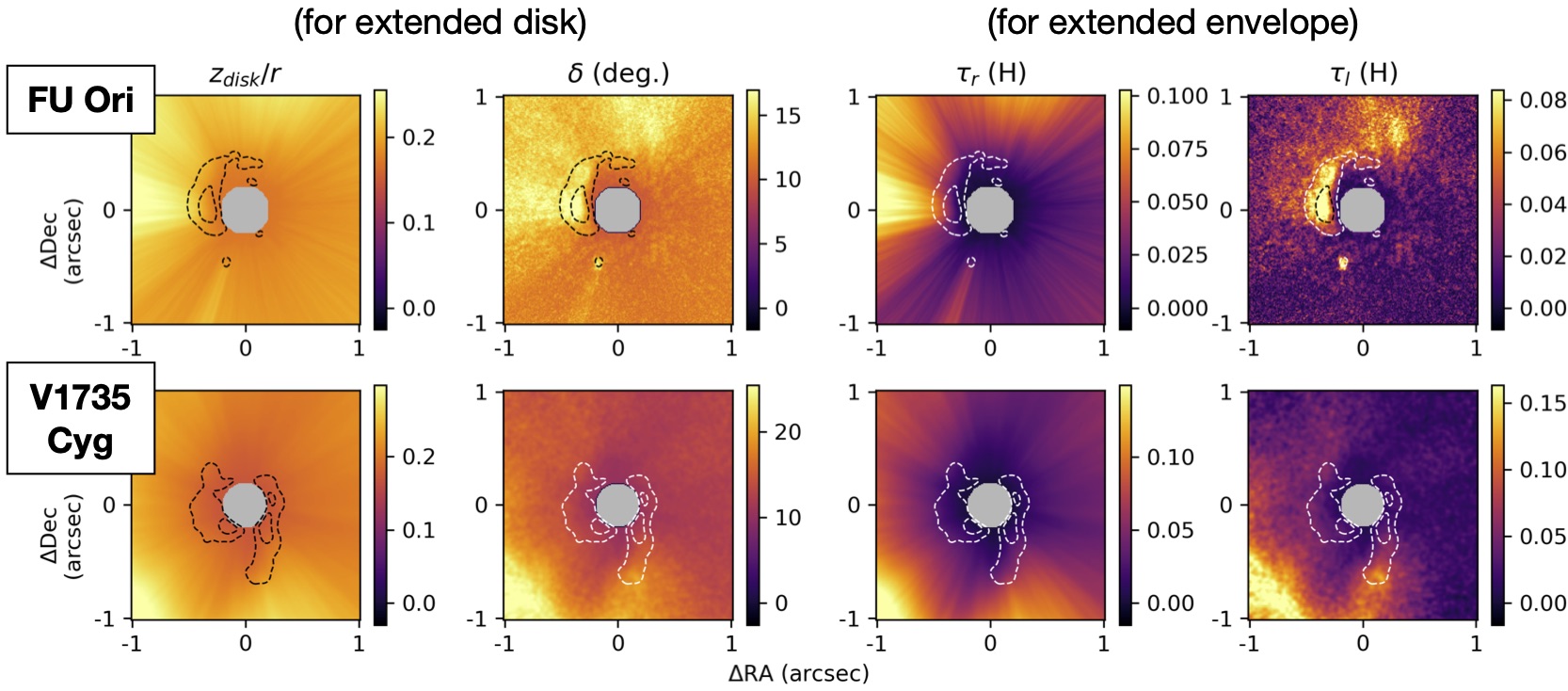}
\caption{
Parameters for disks and envelopes derived from the $H$ band images.
(\textit{From left to right}) the aspect ratio for the surface of the extended disk; the radial inclination angle for the surface of the extended disk; the optical thickness of the extended envelope in $H$ band ($\lambda$=1.65 $\mu$m) from the central star in the radial direction; the same but for line of sight of the observations.
The calculations are made for $\overline{\gamma}$=0.2 with the `Dust1' model.
See the text for the other parameters used to derive the aspect ratio and the radial inclination of the disk surface.
The contours show the brightest regions for the PI distribution in $H$ band, for which we arbitrarily selected the contour levels to indicate the locations of the arm-like structures.
The gray circle in each panel shows the area where we were not able to execute reliable calculations because of the bright central source in the PI image.
\label{fig:check_images_star:gamma02}
}
\end{figure*}

Table \ref{tbl:zr0:gamma02} shows the disk surface aspect ratio $z_\mathrm{disk}/r$ at $r$=$r_\mathrm{min}$ (that is at the edges of the software mask as the center) used for the integration of Eq. (\ref{eq:z_disk}). These values were adjusted to minimize deviation of $\gamma_\mathrm{disk} (x,y)$ from the representative constant $\overline{\gamma}$=0.2  for most of the region. 
Fig. \ref{fig:check_images_star:gamma02} shows the aspect ratio and the radial inclination angle of the disk surface, derived assuming that the observed extended emission is associated with a disk; and the optical thickness for $H$ band in the radial direction and line of sight, derived assuming that the observed extended emission is associated with an envelope. The `Dust1' model is used to derive all the images.
\begin{table}
\caption{Disk surface aspect ratio $z_\mathrm{disk}/r$ at the minimum radius $r$=$r_\mathrm{min}$$^a$. \label{tbl:zr0:gamma02}}
\begin{tabular}{llc}
\hline\hline
Target  & Dust          & Value
\\ \hline
FU Ori          & Dust1 & 0.17  \\
                        & Dust2 & 0.17  \\
                        & Dust3 & 0.16  \\
V1735 Cyg       & Dust1 & 0.17  \\
                        & Dust2 & 0.17  \\
                        & Dust3 & 0.15  \\
\hline
\end{tabular} \\
$^a$Assuming that the central illuminating source is a star, and $\overline{\gamma}$=0.2.
\end{table}

All the calculated images show that the values are small, agreeing with the assumptions and approximations used in Sect. \ref{sec:eq}.
In Fig. \ref{fig:check_images_star:gamma02}, the east arm from FU Ori and the tip of the west arm from V1735 Cyg significantly increase $z_\mathrm{disk}$, $\delta$ and $\tau_{\mathrm{r};H}$ in the outer regions. The figure also show relatively large $\tau_{\mathrm{l};H}$ for these bright features seen in the PI distribution in $H$ band.

\begin{figure}[ht!]
\centering
\includegraphics[width=9cm]{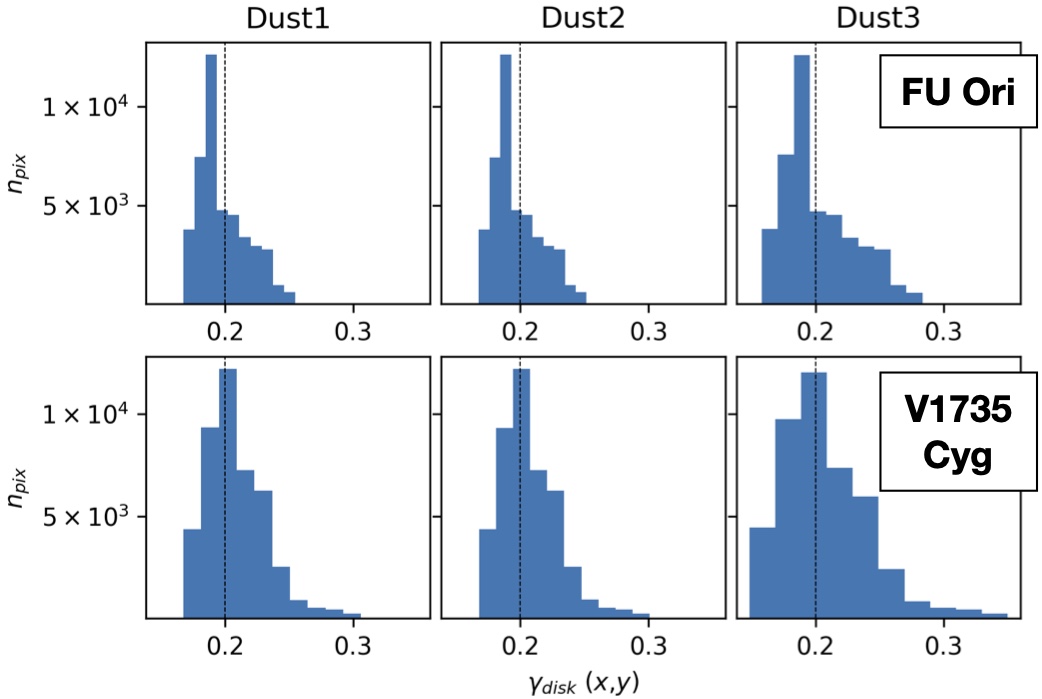}
\caption{
The histograms for $\gamma_\mathrm{disk}$ for the case with a star as the central illuminating source.
The vertical axis shows the number of pixels for each bin.
The vertical dashed line in each panel shows the representative constant $\overline{\gamma}$=0.2.
\label{fig:check_hist_star:gamma02}
}
\end{figure}

Fig. \ref{fig:check_hist_star:gamma02} shows the distributions of pixel values for $\gamma_\mathrm{disk}(x,y)$ for the individual cases.
These show relatively small ranges around the representative constant $\overline{\gamma}$=0.2, within 25 \% for most of the regions. This implies that the amplitudes of the variation of the disk surface is significantly smaller than the height of the disk surface, as shown by analysis and numerical simulations for some protoplanetary disks \citep{Takami14,Zhu15_disk}.

These trends are the same for different targets and dust models. In Table \ref{tbl:maxstd:gamma02} we tabulate the maximum values for the parameters shown in Fig. \ref{fig:check_images_star:gamma02}, and the standard deviation for $\gamma_\mathrm{disk}$ from the representative constant $\overline{\gamma}$=0.2. We find the same trends, that support the self-consistency, for $\overline{\gamma}$=0.1 are 0.4 as well (Appendix \ref{appendix:star}).

\begin{table}
\caption{Parameters for surface geometry of disks and optical thicknesses of envelopes$^a$ \label{tbl:maxstd:gamma02}}
\begin{tabular}{llccccc}
\hline\hline
Target  & Dust          & \multicolumn{4}{c}{Maximum Value}                                                     & $\gamma_\mathrm{disk}$\\ \cline{3-6}
                &               & $z_\mathrm{disk}/r$ & $\delta$ (deg.) & $\tau_{r,H}$ & $\tau_{r,H}$     & r.m.s
\\ \hline
FU Ori          & Dust1 & 0.26  & 19.2  & 0.11  & 0.14  & 0.019 \\
                        & Dust2 & 0.26  & 18.9  & 0.11  & 0.13  & 0.019 \\
                        & Dust3 & 0.30  & 22.8  & 0.16  & 0.20  & 0.028 \\
V1735 Cyg       & Dust1 & 0.32  & 26.9  & 0.18  & 0.18  & 0.025 \\
                        & Dust2 & 0.32  & 26.3  & 0.17  & 0.17  & 0.024 \\
                        & Dust3 & 0.37  & 33.5  & 0.25  & 0.26  & 0.034 \\
\hline
\end{tabular}\\
$^a$Assuming that the central illuminating source is a star, and $\overline{\gamma}$=0.2.
\end{table}

\subsubsection{Cases illuminated by a flat compact disk} \label{sec:application:consistencies:disk}

\begin{figure*}[ht!]
\centering
\includegraphics[width=16cm]{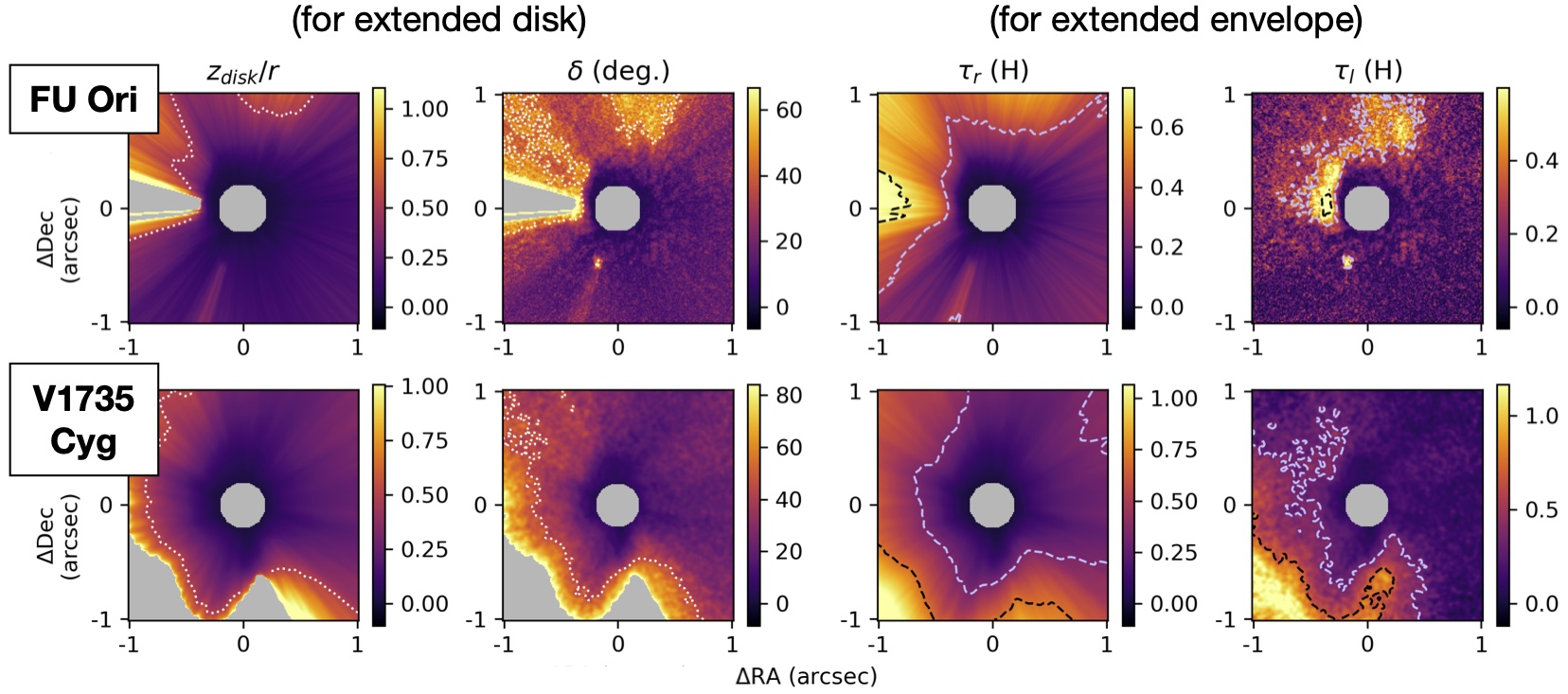}
\caption{
Same as Fig. \ref{fig:check_images_star:gamma02} but with a flat compact disk as the central illuminating source and the `Dust3' model. The white dotted contours in the left two panels indicate the disk surface aspect ratio $z_\mathrm{disk}/r$=0.5 and the radial inclination of the disk surface $\delta$=30$^\circ$, respectively, for the extended disk. The blue and black contours in the right two panels indicate the optical thicknesses $\tau$= 0.35 and 0.7, respectively, in the extended envelope.
The gray circle in each panel shows the area where we were not able to execute reliable calculations because of the bright central source in the PI image.
The other gray areas indicate the regions where we have not been above to execute self-consistent calculations (see the text).
\label{fig:check_images_disk:gamma02}
}
\end{figure*}

\begin{figure}[ht!]
\centering
\includegraphics[width=9cm]{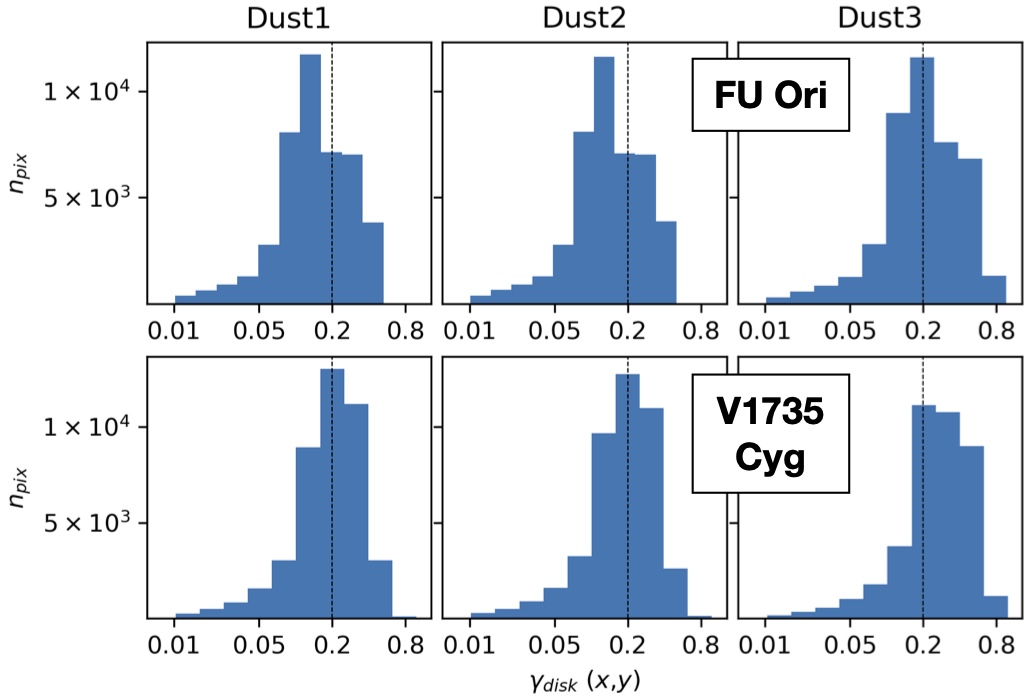}
\caption{
Same as Fig. \ref{fig:check_hist_star:gamma02} but with a flat compact disk as the central illuminating source, and with a logarithmic scale for the horizontal axis.
\label{fig:check_hist_disk:gamma02}
}
\end{figure}

Figs. \ref{fig:check_images_disk:gamma02} and \ref{fig:check_hist_disk:gamma02} are the same as Figs. \ref{fig:check_images_star:gamma02} and \ref{fig:check_hist_star:gamma02} but with a flat compact disk as the illuminating source, and $\overline{\gamma}$=0.2. We derived the parameters for the extended disks in these figures assuming $z_\mathrm{disk}/r$=0.01 at $r$=$r_\mathrm{min}$. In Fig. \ref{fig:check_images_disk:gamma02}, the images for $z_\mathrm{disk}/r$ and $\delta$ have variations significantly larger than the case with a star as the central source shown in Fig. \ref{fig:check_images_star:gamma02}. Fig. \ref{fig:check_hist_disk:gamma02} shows that $\gamma_\mathrm{disk}(x,y)$ also has large variation from $\overline{\gamma}$=0.2, by a factor of more than 2 in 21-35 \% of the region. 


During the integration using Eqs. (\ref{eq:delta}) and (\ref{eq:z_disk}), $\delta(x,y)$ becomes 90$^\circ$~or larger in the left side of FU Ori and the bottom-left corner of the images for  V1735 Cyg, making the integration in the outer regions impossible. As discussed in Sect. \ref{sec:eq:check:disk}, this implies that the given combinations of the central source and the dust model cannot explain the observed PI distribution in these regions. In some of these regions, $\alpha(x,y)$ also exceeds 1, inconsistent with its definition with Eq. (\ref{eq:alpha:definition}).

Fig. \ref{fig:check_images_disk:gamma02} also shows that the optical thickness of the extended envelope is significantly larger than the case with a star as the central source shown in Figs. \ref{fig:check_images_star:gamma02}, by the factor $\overline{\gamma}^{-1}$ as shown in Eq. (\ref{eq:tau_l:disk}). In some regions, the optical thickness exceeds 0.7, implying that self-absorption in the envelope reduces the emission from the central source or toward the observer by a factor of $\sim$2.

In Sect. \ref{sec:application:I_a}, we further investigate the self-consistencies of the calculations for different cases for $\overline{\gamma}$=0.2. 
For $\overline{\gamma}$=0.4, we were able to obtain self-consistent results for all the cases. For $\overline{\gamma}$=0.1, we were able to obtain self-consistent results for the bright part of the east and west arms for V1735 Cyg, for the cases that the extended emission is due to an envelope. See Appendix \ref{appendix:disk} for details.


\subsection{Individual emission components} \label{sec:application:I_indiv}

\begin{figure*}[ht!]
\centering
\includegraphics[width=18cm]{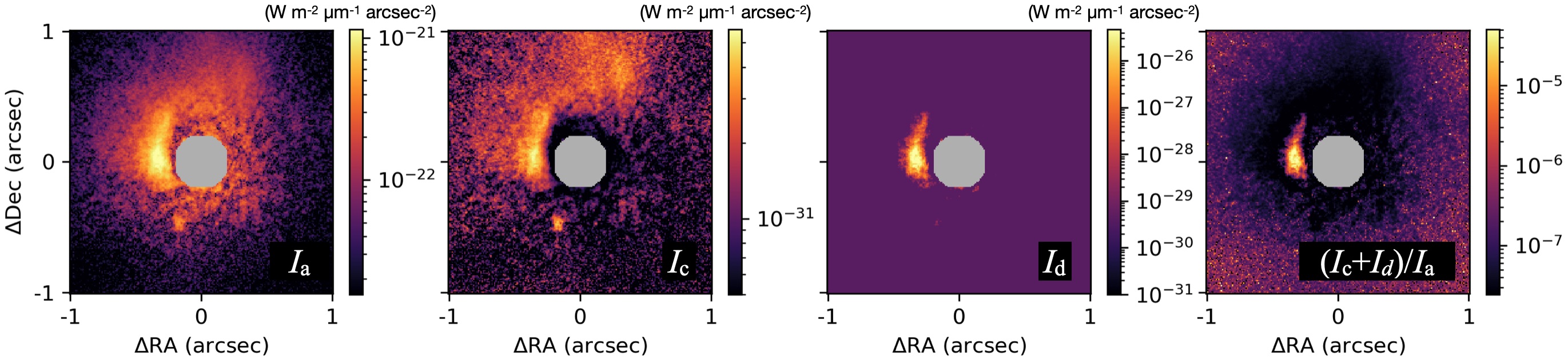}
\caption{
Images for the individual emission components. 
(\textit{From left to right}) the single-scattering component ($I_\mathrm{a}$); thermal emission from the surface layer ($I_\mathrm{c}$) and the disk interior ($I_\mathrm{d}$); and the intensity ratio for the thermal emission ($I_\mathrm{c}$+$I_\mathrm{d}$) per the single scattering emission for FU Ori at $\lambda$=12 $\mu$m, assuming that the extended emission is associated with a disk, and $\overline{\gamma}$=0.2. A combination of a star as the central illuminating source and the `Dust1' model are used.
\label{fig:I_indiv_disk}
}
\end{figure*}

\begin{figure*}[ht!]
\centering
\includegraphics[width=14cm]{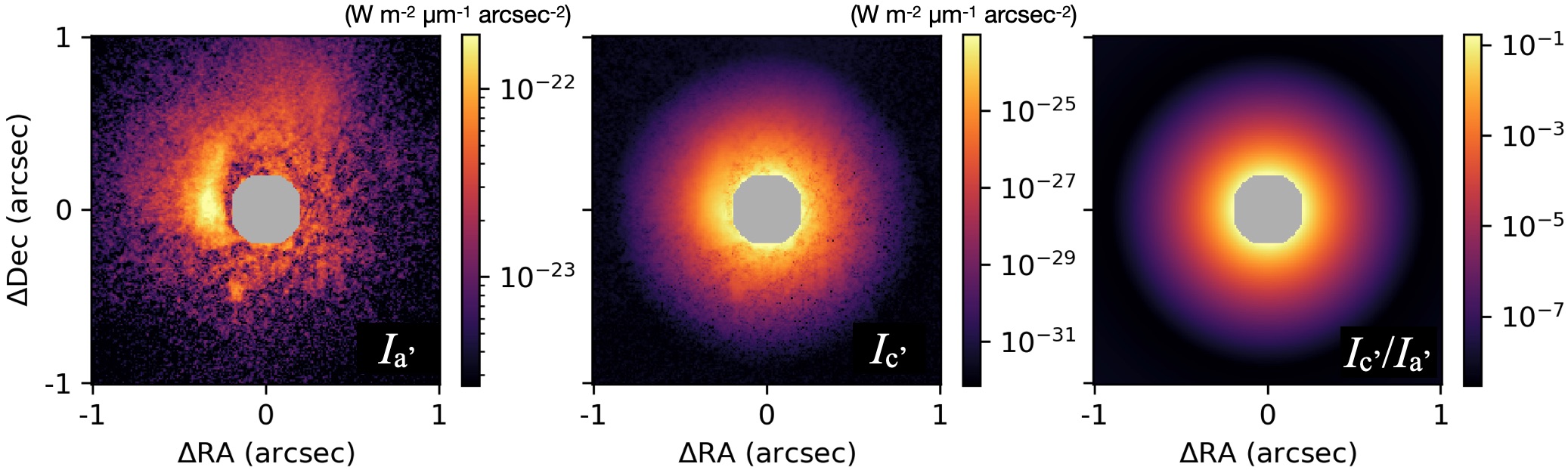}
\caption{
Same as as Fig. \ref{fig:I_indiv_disk} but assuming that the extended emission is associated with an envelope. (\textit{from left to right}) Intensity distributions for the scattered emission ($I_\mathrm{a}'$); thermal emission ($I_\mathrm{c}'$) and their intensity ratio.
\label{fig:I_indiv_env}
}
\end{figure*}

Figs. \ref{fig:I_indiv_disk} and \ref{fig:I_indiv_env} show the individual emission components for FU Ori at $\lambda$=12 $\mu$m, assuming that the extended emission is due to a disk and an envelope, respectively. 
The calculations were made for $\overline{\gamma}$=0.2 with a combination of a star as the central illuminating source and the `Dust1' model.
For thermal emission, we set the minimum temperature of the disk and the envelope to be 30 K.
The fraction of the thermal emission to the total emission is largest for the selected wavelength (that is the longest wavelength).

Under the given approximation, the intensity distributions for scattered emission $I_a(x,y)$, $I_b(x,y)$ and $I_{a'}(x,y)$ are identical to that of the PI distribution (Sects. \ref{sec:eq:thick:I_b} and \ref{sec:eq:source:neweqs}). In Table \ref{tbl:Ib_per_Ia} we show the $I_b(x,y)$/$I_a(x,y)$ intensity ratios calculated using Eq. (\ref{eq:thick:I_b}). As shown in the table, $I_b(x,y)$ is fainter than $I_a(x,y)$ by a factor of 5 or larger. As discussed in Sect. \ref{sec:eq:thick:I_b}, the calculated $I_b(x,y)$ may be overestimated, therefore the actual contribution of the double scattering emission to the total intensity may be even smaller.

\begin{table}
\centering
\caption{Upper limit intensity ratios for the double-scattering emission per single-scattering emission. \label{tbl:Ib_per_Ia}}
\begin{tabular}{clll}
\hline\hline
Wavelength ($\mu$m)     & Dust1 & Dust2 & Dust3
\\ \hline
3.5     & 0.12  & 0.18  & 0.22  \\
4.8     & 0.086 & 0.15  & 0.18  \\
12      & 0.0041        & 0.0082        & 0.053 \\
\hline
\end{tabular} \\
\end{table}

Figs. \ref{fig:I_indiv_disk} and \ref{fig:I_indiv_env} show that the thermal emission components $I_\mathrm{c}$, $I_\mathrm{d}$, and $I_\mathrm{c'}$ are significantly fainter than the single scattering  emission, at least by a factor of 100 for most of the regions. This is due to the low temperatures ($T$$\lesssim$70 K) derived in these regions. In Table \ref{tbl:max_I_th}, we list the maximum intensity ratios for thermal emission divided by the single scattering emission for a variety of cases. The table shows a maximum value of 0.25 for V1735 Cyg at $\lambda$=12 $\mu$m, with the `Dust1' model and a star as the illuminating source, and for the case where the extended emission is due to an envelope. Even in this case, the ratio is below 0.03 for more than 98 \% of the region. For the remaining cases, the tabulated maximum intensity ratios are 0.08. 

\begin{table*}
\centering
\caption{Maximum ratios for thermal to single-scattering emission$^a$. \label{tbl:max_I_th}}
\begin{tabular}{lcllllllll}
\hline\hline
Extended                &       Wavelength      & Target        & \multicolumn{3}{c}{Illuminated by Star} && \multicolumn{3}{c}{Illuminated by Flat Compact Disk}        \\ \cline{4-6} \cline{8-10}
Emission is:    & ($\mu$m)      &               & 
\multicolumn{1}{c}{Dust1} & 
\multicolumn{1}{c}{Dust2} & 
\multicolumn{1}{c}{Dust3} && 
\multicolumn{1}{c}{Dust1} & 
\multicolumn{1}{c}{Dust2} & 
\multicolumn{1}{c}{Dust3}
\\ \hline
Disk    & 3.5   & FU Ori                        & 2.1$\times$10$^{-36}$ &5.8$\times$10$^{-37}$  &7.8$\times$10$^{-33}$  && 4.1$\times$10$^{-37}$   &1.2$\times$10$^{-37}$  &1.6$\times$10$^{-33}$ \\
        &       & V1735 Cyg             & 1.4$\times$10$^{-47}$ &9.8$\times$10$^{-48}$  &7.3$\times$10$^{-48}$  && 2.8$\times$10$^{-48}$   &2.0$\times$10$^{-48}$  &1.5$\times$10$^{-48}$ \vspace{0.1cm} \\

        & 4.8   & FU Ori                        & 1.8$\times$10$^{-24}$ & 5.6$\times$10$^{-25}$   & 4.6$\times$10$^{-22}$ && 3.6$\times$10$^{-25}$        & 1.1$\times$10$^{-25}$   & 9.2$\times$10$^{-23}$ \\
        &       & V1735 Cyg             & 1.2$\times$10$^{-31}$ & 7.5$\times$10$^{-32}$ & 4.8$\times$10$^{-32}$   && 2.5$\times$10$^{-32}$        & 1.5$\times$10$^{-32}$ & 9.7$\times$10$^{-33}$ \vspace{0.1cm} \\

        & 12 & FU Ori                   & 1.4$\times$10$^{-3}$  &6.9$\times$10$^{-4}$   &7.4$\times$10$^{-5}$   && 2.7$\times$10$^{-4}$    &1.4$\times$10$^{-4}$   &1.5$\times$10$^{-5}$ \\
        &       & V1735 Cyg             & 2.3$\times$10$^{-5}$  &1.2$\times$10$^{-5}$   &1.3$\times$10$^{-6}$   && 4.7$\times$10$^{-6}$    &2.4$\times$10$^{-6}$   &2.5$\times$10$^{-7}$ \vspace{0.1cm} \\

Envelope        &3.5    & FU Ori                & 4.7$\times$10$^{-21}$ &5.0$\times$10$^{-22}$  &1.0$\times$10$^{-20}$  && 7.9$\times$10$^{-35}$   &3.7$\times$10$^{-36}$  &2.8$\times$10$^{-34}$ \\
                &       & V1735 Cyg     & 2.2$\times$10$^{-35}$ &7.1$\times$10$^{-37}$  &3.1$\times$10$^{-36}$  && 1.1$\times$10$^{-50}$   &6.2$\times$10$^{-51}$  &6.8$\times$10$^{-51}$ \vspace{0.1cm} \\

                &4.8    & FU Ori                & 1.2$\times$10$^{-13}$ & 1.8$\times$10$^{-14}$   & 1.4$\times$10$^{-13}$ && 1.1$\times$10$^{-23}$        & 9.1$\times$10$^{-25}$   & 1.8$\times$10$^{-23}$ \\
                &       & V1735 Cyg     & 5.1$\times$10$^{-24}$ & 3.2$\times$10$^{-25}$ & 8.0$\times$10$^{-25}$   && 1.1$\times$10$^{-34}$        & 5.3$\times$10$^{-35}$ & 4.8$\times$10$^{-35}$\vspace{0.1cm} \\

                &12 & FU Ori            & 2.5$\times$10$^{-1}$  &7.6$\times$10$^{-2}$   &2.6$\times$10$^{-2}$   && 2.4$\times$10$^{-5}$    &5.8$\times$10$^{-6}$   &2.8$\times$10$^{-6}$ \\
                &       & V1735 Cyg     & 4.0$\times$10$^{-5}$  &8.6$\times$10$^{-6}$   &1.8$\times$10$^{-6}$   && 2.4$\times$10$^{-8}$    &1.2$\times$10$^{-8}$   &1.7$\times$10$^{-9}$ \\
\hline
\end{tabular}\\
$^a$For $\overline{\gamma}$=0.2.
\end{table*}

Table \ref{tbl:min_Ia} shows the minimum fraction of the single scattering emission to the total intensity.
The table shows that the single scattering emission is responsible for more than 80 \% of the total intensity for all cases and positions.
This emission component does not significantly suffer from the possible large systematic errors (A)(B)(C) discussed in Sect. \ref{sec:eq:check:disk} for the extended disks, as explained below. For (A), the equations for single-scattering emission (\ref{eq:thick:I_a:star}, \ref{eq:thick:I_a:disk}) do not include cos~$i$ that can cause large systematic errors. The cos~$i$ terms with $\overline{\gamma}$ in the former equation can yield a systematic error only up to $\sim$30 \% (Sects. \ref{sec:eq:thick:I_a}, \ref{sec:eq:source:overview}). The term for (B) has been canceled out in Eqs. (\ref{eq:thick:I_a:star}) and (\ref{eq:thick:I_a:disk}) as this effect equally affect the observed PI distribution in $H$ band and modeled mid-infrared emission. For (C), $\overline{\gamma}$ is included in Eq. (\ref{eq:thick:I_a:star}) for which the illuminating source is a star. Therefore, the spatial variation of  $\gamma_\mathrm{disk} (x,y)$ from the representative constant $\overline{\gamma}$ is relatively small ($<$25 \% for most of the pixels; Sect. \ref{sec:application:consistencies:star}).

\begin{table*}
\centering
\caption{Minimum fraction of single-scattering emission to total intensity$^a$. \label{tbl:min_Ia}}
\begin{tabular}{lclcccccccc}
\hline\hline
Extended                &       Wavelength      & Target        & \multicolumn{3}{c}{Illuminated by Star} && \multicolumn{3}{c}{Illuminated by Compact Disk}     \\ \cline{4-6} \cline{8-10}
Emission is:    & ($\mu$m)      &               & Dust1 & Dust2 & Dust3 && Dust1 & Dust2 & Dust3
\\ \hline
Disk    & 3.5           & FU Ori                & 0.88 & 0.84 & 0.81 && 0.88 & 0.84 & 0.81  \\
        &               & V1735 Cyg     & 0.88 & 0.84 & 0.81 && 0.88 & 0.84 & 0.81 \vspace{0.1cm} \\

        & 4.8           & FU Ori                & 0.92 & 0.87 & 0.85 && 0.92 & 0.87 & 0.85 \\
        &               & V1735 Cyg     & 0.92 & 0.87 & 0.85 && 0.92 & 0.87 & 0.85\vspace{0.1cm} \\

        & 12            & FU Ori                & $>$0.99 & 0.99 & 0.95 && $>$0.99 & 0.99 & 0.95  \\
        &               & V1735 Cyg     & $>$0.99 & 0.99 & 0.95 && $>$0.99 & 0.99 & 0.95 \vspace{0.1cm} \\

Envelope        & 3.5   &        FU Ori          & $>$0.99 & $>$0.99 & $>$0.99 && $>$0.99 & $>$0.99 & $>$0.99 \\
                &               & V1735 Cyg      & $>$0.99 & $>$0.99 & $>$0.99 && $>$0.99 & $>$0.99 & $>$0.99\vspace{0.1cm} \\

                &4.8            & FU Ori                 & $>$0.99 & $>$0.99 & $>$0.99 && $>$0.99 & $>$0.99 & $>$0.99 \\
                &               & V1735 Cyg      & $>$0.99 & $>$0.99 & $>$0.99 && $>$0.99 & $>$0.99 & $>$0.99\vspace{0.1cm} \\

                &12             & FU Ori                 & 0.89 & 0.96 & 0.99 && $>$0.99 & $>$0.99 & $>$0.99 \\
                &               & V1735 Cyg      & $>$0.99 & $>$0.99 & $>$0.99 && $>$0.99 & $>$0.99 & $>$0.99 \\
\hline
\end{tabular}\\
$^a$For $\overline{\gamma}$=0.2.
\end{table*}

For the case that the extended emission is due to an envelope, some regions suffer from large optical thickness ($\tau$$\gtrsim$0.7, corresponding to a degradation in the flux or intensity of a factor of $\gtrsim$2), which we did not include when deriving the equations (Sect. \ref{sec:application:consistencies:disk}). Fortunately, these are still small for the arm-like features (Sect. \ref{sec:application:I_a}), that is the features of our major interest for investigating the mass accretion process (Sect. \ref{sec:intro}).

As described in the beginning of this subsection, all of the above calculations are made for 
$\overline{\gamma}$=0.2. 
The single-scattering emission dominates the total intensity distributions for $\overline{\gamma}$=0.1 and 0.4 as well (Appendix).
We note that all the images for single scattering emission still suffer from an internal systematic error by a factor of up to $\sim$2, based on the simplification $a_\lambda$ (Sect. \ref{sec:eq:thick:I_a}) made to avoid the complexity of calculations with a variety of scattering angles.


\subsection{Single-scattering emission} \label{sec:application:I_a}

\begin{figure*}[ht!]
\centering
\includegraphics[width=16cm]{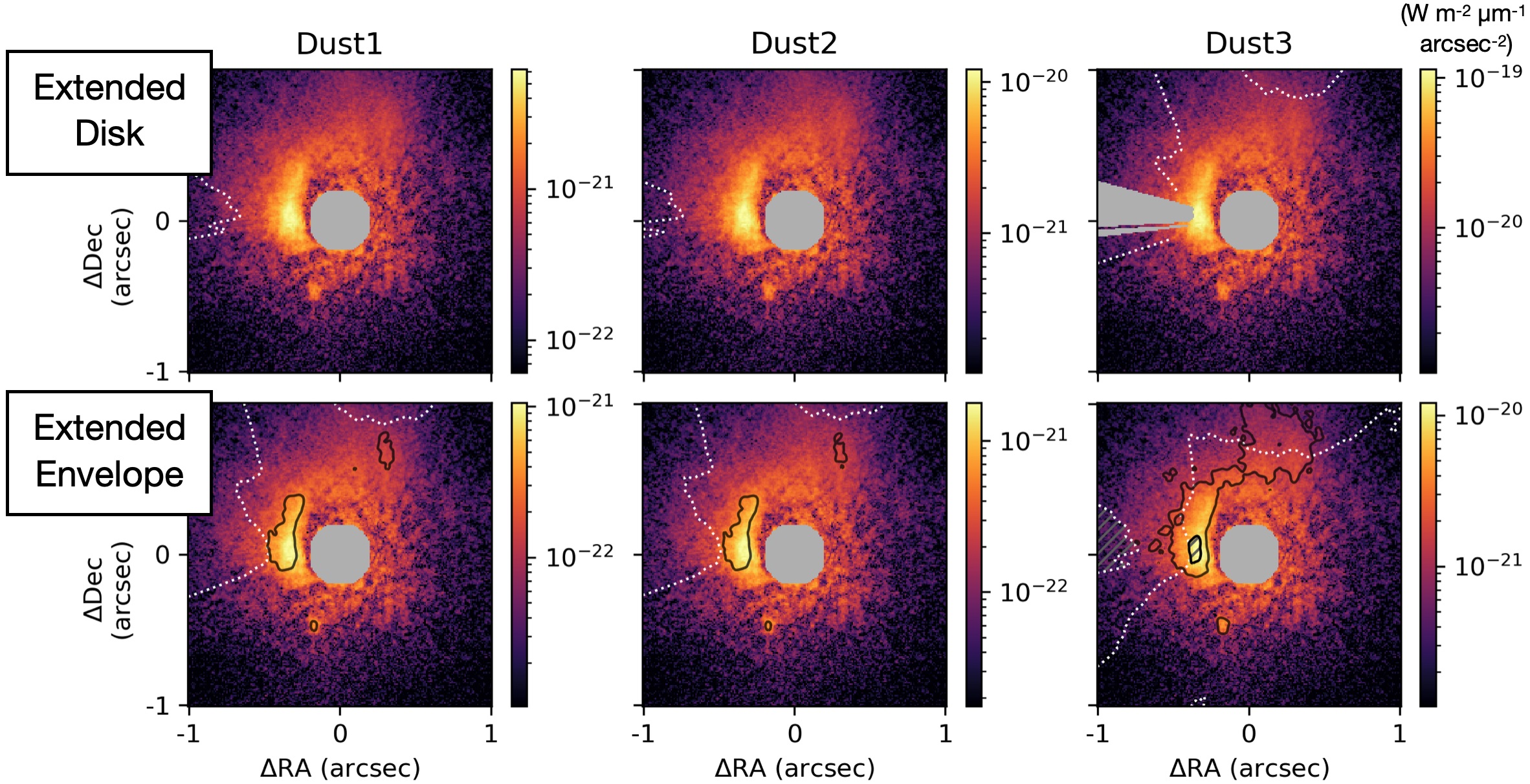}
\caption{
The intensities for the single-scattering component for FU Ori at $\lambda$=12 $\mu$m, for the cases where the extended emission is due to a disk ($top$) and an envelope ($bottom$), and for different dust models. The central illuminating source is assumed to be a flat compact disk. The white dotted contours in the upper panels are for the disk surface aspect ratio $z_\mathrm{disk}/r$=0.5; the white dotted contours in the lower panels are for the $H$ band optical thickness in the radial direction $\tau_r$=0.35 and 0.7; and the black solid contours in the lower panels are for the $H$ band optical thickness along line of sight $\tau_l$=0.35 and 0.7. The regions with $\tau_r$$>$0.7 and $\tau_l$$>$0.7 are hatched in the bottom-right panel. In the top-left panel, the left (east) side of the east arm is masked as the calculations are not self-consistent (see the text).
\label{fig:Ia_FU_gamma02}
}
\end{figure*}

\begin{figure*}[ht!]
\centering
\includegraphics[width=16cm]{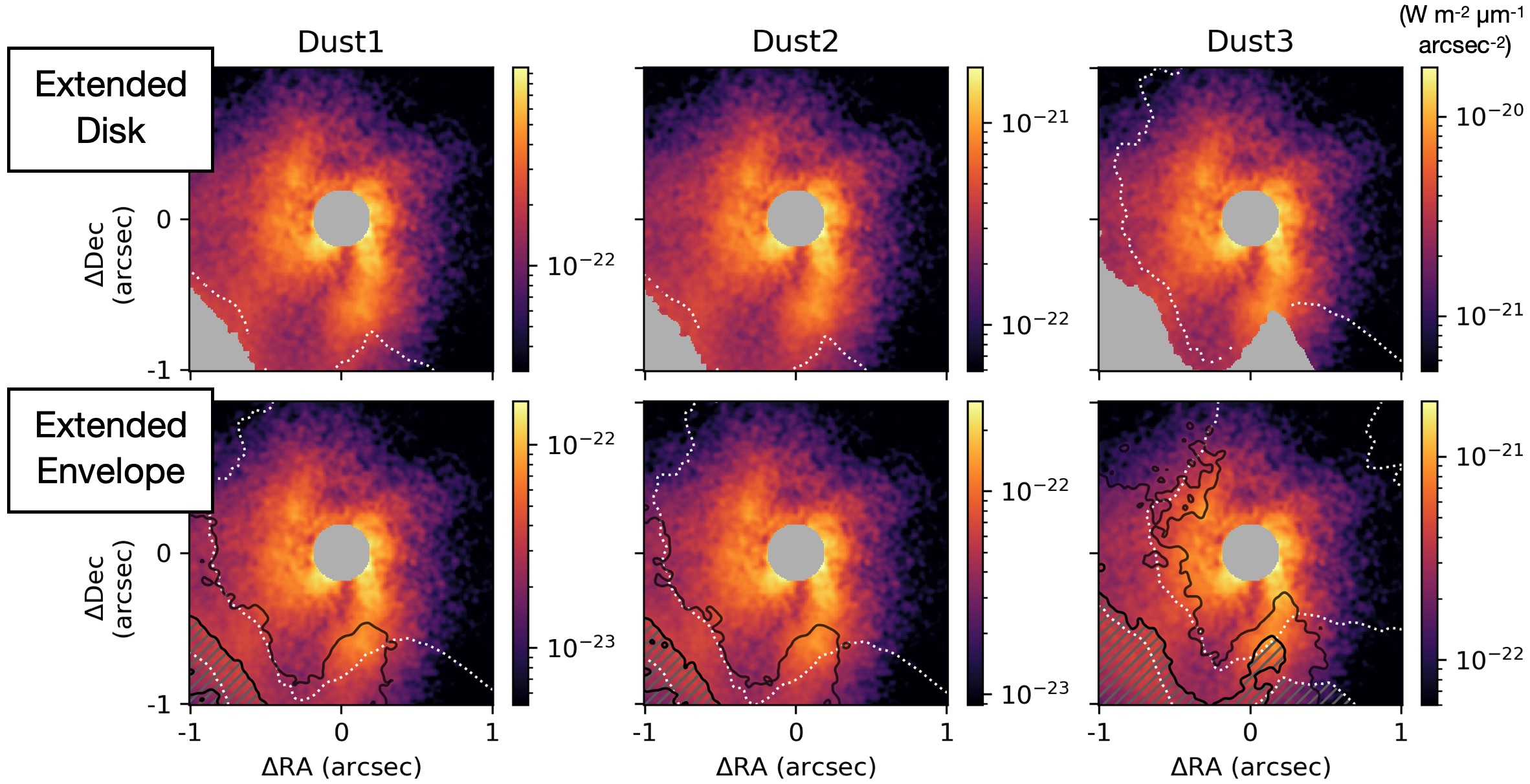}
\caption{
Same as Fig. \ref{fig:Ia_FU_gamma02} but for V1735 Cyg.
\label{fig:Ia_V17_gamma02}
}
\end{figure*}

Figs. \ref{fig:Ia_FU_gamma02} and \ref{fig:Ia_V17_gamma02} show the intensity distributions for single-scattering emission at $\lambda$=12 $\mu$m for FU Ori and V1735 Cyg, respectively, for all the cases which assumed that the central illuminating source is a flat compact disk, and $\overline{\gamma}$=0.2. 
As discussed above, all the images are identical to the PI distribution in $H$ band but with different intensity levels. As shown in Eqs. (\ref{eq:thick:I_a:disk}) and (\ref{eq:thin:I_a:disk}), the intensity is independent of the assumed $\overline{\gamma}$ for these cases.


In Fig. \ref{fig:Ia_FU_gamma02}, the left (east) side of the east arm is masked for the case in which the extended emission is due to a disk with the `Dust3' model, as the radial inclination of the disk surface $\delta$ exceeded 90$^\circ$ during the integration. This is caused by the emission of the bright part of the east arm. Following our discussion in Sect. \ref{sec:application:consistencies:disk}, we conclude that the given conditions for this image are not consistent with the presence of this east arm, that is the major feature for this object.
In the upper panels of Fig. \ref{fig:Ia_V17_gamma02}, the masked regions for $\delta$$\ge$90$^\circ$ are  located in the left corner, and also the bottom edge of the image for the `Dust3' model. This implies that the emission in these regions cannot be explained with an extended disk. It is still possible to attribute the emission in the inner region (including two arms) to an extended disk, and the outer region to an extended envelope.

In the bottom panels of Figs. \ref{fig:Ia_FU_gamma02} and \ref{fig:Ia_V17_gamma02}, the optical thickness from the center or along the line of sight exceeds 0.7 (corresponding a decrease in the flux or intensity by a factor of $\gtrsim$2) in $H$ band in the hatched areas. When we derived the equations in Sect. \ref{sec:eq}, we assumed that the extended envelope is optically thin, and did not include the decreased intensities due to optical thickness. Therefore, the derived intensities are not very accurate in and near these regions, with a systematic error of a factor of $\gtrsim$2. Fortunately, most of the emission associated with the arms does not suffer from this optical thickness issue.

\begin{figure}[ht!]
\centering
\includegraphics[width=9cm]{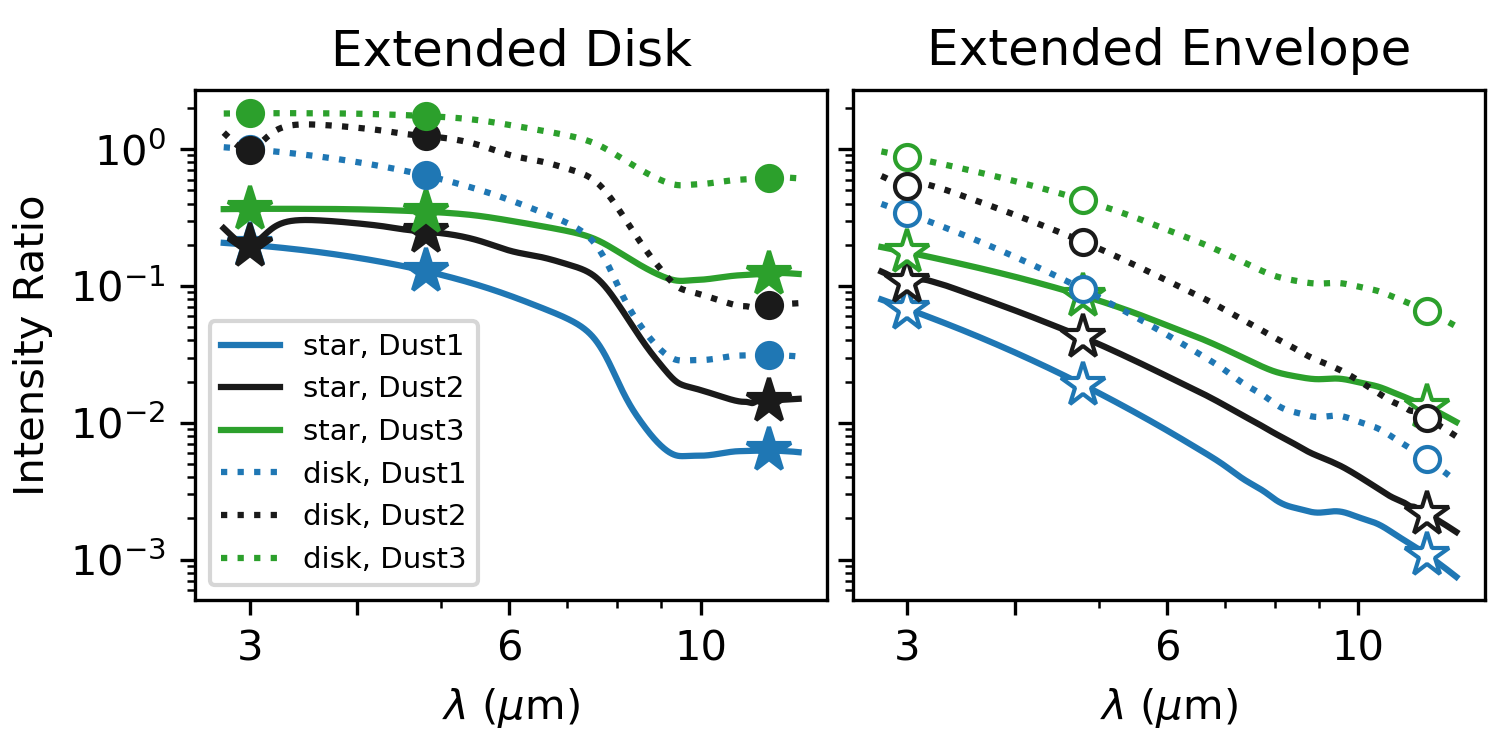}
\caption{
Relative intensities for single scattering emission for various cases. All the values are normalized to that of an extended disk with the `Dust1' model, with a flat compact disk as the illuminating source, and $\lambda$=3.5 $\mu$m. The left and right panels are for the case where we assumed the extended emission is associated with a disk and an envelope, respectively, For each panel, we show the intensity ratios for different illuminating sources and dust models.
\label{fig:I_ratios}
}
\end{figure}

If the central source is a star, the intensities are fainter by a factor of $\overline{\gamma}$ as shown in Eqs. (\ref{eq:thick:I_a:star})(\ref{eq:thin:I_a:star})(\ref{eq:thick:I_a:disk})(\ref{eq:thin:I_a:disk}). All the results are self-consistent for these cases as discussed in Sect. \ref{sec:application:consistencies:star}.
In the areas where the approximation works well, the intensity ratios are derived as follows using Eqs. (\ref{eq:thick:I_a:star})(\ref{eq:thin:I_a:star})(\ref{eq:thick:I_a:disk})(\ref{eq:thin:I_a:disk}):
%
\begin{eqnarray}
\frac{I_{a,\lambda_2}}{I_{a,\lambda_1}} & = & \frac{a_{\lambda_2} \omega_{\lambda_2}}{a_{\lambda_1} \omega_{\lambda_1}}, \label{eq:ratio1}\\
\frac{I_{a',\lambda}}{I_{a,\lambda}} & = & \frac{\kappa_{\mathrm{ext};\lambda}}{\kappa_{\mathrm{ext};H}}  \label{eq:ratio2}.
\end{eqnarray}
%
Fig. \ref{fig:I_ratios} shows the intensities normalized to that of an extended disk, a flat compact disk as the illuminating source and the `Dust1' model. These intensity ratios are independent of the target object (FU Ori or V1735 Cyg). For each wavelength, the intensities range over a factor of 20-800, depending on the central illuminating source, whether the extended emission is due to a disk or an envelope, the dust models and the observing wavelengths ($\lambda$=3--13 $\mu$m). The ranges are smaller for shorter wavelengths due to the smaller differences in dust properties from those for the $H$ band.

The intensity ratio given by Eq. (\ref{eq:ratio2}) is smaller than 1 at the target wavelengths (see Fig. \ref{fig:dust_general}), implying that the extended emission is fainter if it is due to an envelope. Therefore, the intensities in this case would be lower limits. In reality, the optically thinner nature for longer wavelengths may allow us to reveal the presence of an embedded disk even in the case that the emission observed in $H$ band is dominated by an envelope, as described in Section \ref{sec:intro}.


\subsection{Dependence on viewing angle} \label{sec:application:i}

We made calculations in Sects. \ref{sec:application:consistencies}-\ref{sec:application:I_a} for the viewing angle $i$=0$^\circ$.
As discussed in the beginning of the section, the viewing angle appears to be close to zero if the extended emission is due to a disk, however, it could be intermediate ($i$$\sim$45$^\circ$) if the extended emission is due to an envelope. As discussed in Sect. \ref{sec:application:I_indiv}, this does not significantly affect the calculated single scattering emission, which dominates the total intensity distribution, as the terms with the viewing angle have been canceled out in Eqs. (\ref{eq:thin:I_a:star}) and  (\ref{eq:thin:I_a:disk}).

In contrast, an intermediate viewing angle would yield a large $r$ and smaller $F_\lambda(x,y)$ with Eqs. (\ref{eq:F_star}) and (\ref{eq:F_disk}), and therefore a larger optical thicknesses for the extended envelope, for example, by a factor of 2-3 for $i$$\sim$45$^\circ$ (see Eqs. \ref{eq:tau_l:star}-\ref{eq:tau_r}). This would increase the areas where the calculated intensity distribution for the single scattering emission is not reliable. In Figs. \ref{fig:Ia_FU_gamma02} and \ref{fig:Ia_V17_gamma02}, the figures obtained for $i$=0$^\circ$, the regions with $\tau_{l;H}$$>$0.35 (therefore $\tau_{l;H}$$>$0.7-1 for $i$=45$^\circ$) cover significant fractions of the arm features for FU Ori with all the dust models, and V1735 Cyg with the `Dust3' models. One would regard the calculated emission in these regions as not reliable for such a viewing angle. We emphasize that, however, the discussion in this subsection does not exclude a possibility that the disk or the envelope associated with these targets have a significantly smaller viewing angle, close to $i$=0$^\circ$.


\section{Benchmark calculations with monochromatic radiative transfer simulations} \label{sec:sim}

One may be concerned that the approximations used to calculate the disk emission (Section \ref{sec:eq:thick}) may significantly degrade the accuracy for our calculations. In this section, we perform monochromatic Monte-Carlo radiative transfer simulations using the Sprout code \citep{Takami13}\footnote{https://github.com/JenniferKarr/Sprout\_Code} and demonstrate how well the semi-analytic calculations using $H$ band calculations reproduce the simulated images.
We explain the method and show the results in Sect. \ref{sec:sim:mono}. In Sect. \ref{sec:sim:full} we briefly discuss what we need for full radiative transfer simulations to verify the calculations for the thermal emission components.

\subsection{Method and results}\label{sec:sim:mono}

As is common in some full radiative transfer codes \citep{Whitney03a,Robitaille11}, we used the following equation for a flared disk \citep{Shakura73,Lynden-Bell74}:
\begin{equation}
\rho (r,z) = \rho_0 \left[1-\sqrt{\frac{R_0}{r}} \right] \left(\frac{R_0}{r} \right)^\alpha ~ \mathrm{exp}~ \left\{- \frac{1}{2} \left[\frac{\it z}{h}\rm \right]^2 \right\},
\label{eq:ssdisk}
\end{equation} 
where $\rho_0$ is a constant to scale the density; $R_0$ is a reference radius for $\rho_0$; $\alpha$ is the radial density exponent; and $h$ is the disk scale height. The scale height $h$ increases with radius as $h = h_0 (r/R_0)^\beta$, where $h_0$ is the disk scale height at the reference radius; $\beta$ is the flaring power. 
As is also common in some models \citep{Robitaille06,Robitaille07,Dong12b,Takami13}, we assumed $\alpha = \beta+1$, which yields the surface density distribution $\Sigma (r) \propto r^{-1}$, approximately agreeing with that inferred from millimeter interferometry for disks associated with many low-mass pre-main sequence stars \citep{Andrews09,Andrews10b}. Use of an independent $\alpha$ is beyond the scope of this work.

Table \ref{tbl:disk_params} summarizes the parameter set we used for simulations. We used a disk mass of 0.1 $M_\sun$ based on recent mini-survey observations of FUor disks using ALMA \citep{Kospal21}. Similar disk masses were also estimated by fitting the observed infrared spectral distributions for FUors using full radiative transfer simulations \citep{Gramajo14}.
The outer disk radius of 500 au was set based on the $H$ band imaging polarimetry shown in Section \ref{sec:application}. The inner disk radius was chosen to not let it block the light from the central source to the outer disk surface. Under this condition, this parameter does not affect the results significantly as long as it is located within the central mask for the $H$ band image (Sect. \ref{sec:application}). The selected $\beta$ of 1.3 is comparable to that used for protoplanetary disks associated with normal YSOs \citep[e.g.,][]{Dong12b,Takami13,Follette15}. The parameters $\rho_0$ and $h_0$ were chosen to satisfy the given disk mass and $\overline{\gamma}$$\sim$0.2.

\begin{table}
\caption{Disk parameters for the monochromatic Monte-Carlo radiative transfer simulations \label{tbl:disk_params}}
\begin{tabular}{ll}
\hline\hline
Parameter 
& Value \\ \hline

Disk mass 
& 0.1 $M_\sun$ \\

Minimum disk radius
& 1 au \\

Maximum disk radius
& 500 au \\

Power index for scale height $\beta$
& 1.3  \\

Reference disk radius $R_0$
& 13.3 $R_\sun$\\

Density at the reference position $\rho_0$
& 3.55$\times$10$^{-7}$ g cm$^{-3}$ \\

Scale height at the reference disk radius $h_0$
& 5.56$\times$10$^{-3}$ $R_0$\\

\hline \\
\end{tabular} \\
\end{table}

In this section we used the `Dust3' model, which consists of astronomical silicate and graphite. The optical constants for the target wavelengths are obtained from \citet{Draine84}.
We note that different authors use different types of carbon dust, either graphite or amorphous carbon \citep{Cotera01,Wood02b}. While graphite has been extensively used \citep[e.g.,][]{Draine84,Laor93,Kim94,Whitney03a,Robitaille06,Dong12b,Dong12a}, far-infrared SEDs of YSOs and evolved stars suggest the absence of graphite and presence of amorphous carbon in circumstellar dust \citep[][and references therein]{Jager98}. We used graphite for consistency with the dust parameter files used in earlier sections.

Following Sect. \ref{sec:eq:source},
we executed Monte-Carlo scattering simulations using this disk model for the following cases: (1) $\lambda$=1.65 $\mu$m for the case that the central illuminating source is a star; and (2) $\lambda$=1.65, 3.5, 4.8 and 12 $\mu$m for the case that the central source is a flat compact self-luminous disk. We used 5$\times$10$^8$ photons for each case.
Each photon was initially ejected from the central source, which is approximated as a point source either for a star or a flat self-luminous compact disk. The photons were isotropically ejected from the central source for both cases. Each photon has a `weight' corresponding to its Stokes $I$. The weights for all the initial photons are the same if the central source is a star. If the central source is a flat compact disk, we scaled the weight of the photons based on its grazing angle.

The light path for the next scattering position was calculated for the opacity distribution determined by Equation (\ref{eq:ssdisk}), Table \ref{tbl:disk_params}, and the dust opacity for the wavelength of the calculations. The scattering angle and the resultant Stokes parameters were calculated based on the Mie theory. The former was calculated using the `table look-up method', which yields accuracies better than the Henyey-Greenstein approximation used in Section \ref{sec:eq} \citep{Fischer94,Takami13}.
We then scaled the photon weight by the scattering albedo, and continued the calculation until each photons escapes from the region of the interest. To save time for calculations, photons with more than ten scatterings were removed because of small photon weights, and therefore small contribution to the simulated images.
The photons escaping out from the region of our interest were collected with ten image detectors for different viewing angles.
Each detector consists of 101$\times$101 pixels to sample the image of the entire disk.
The partition of the viewing angles for the detectors was made to receive equal number of photons in the case that the radiation is isotropic.
For face-on and intermediate viewing angles, we used the results for the photons integrated over the angles of 0$^\circ$--26$^\circ$ and
37$^\circ$--46$^\circ$, respectively. See \citet{Takami13} for other details of the simulations.

\begin{figure}[ht!]
\centering
\includegraphics[width=9cm]{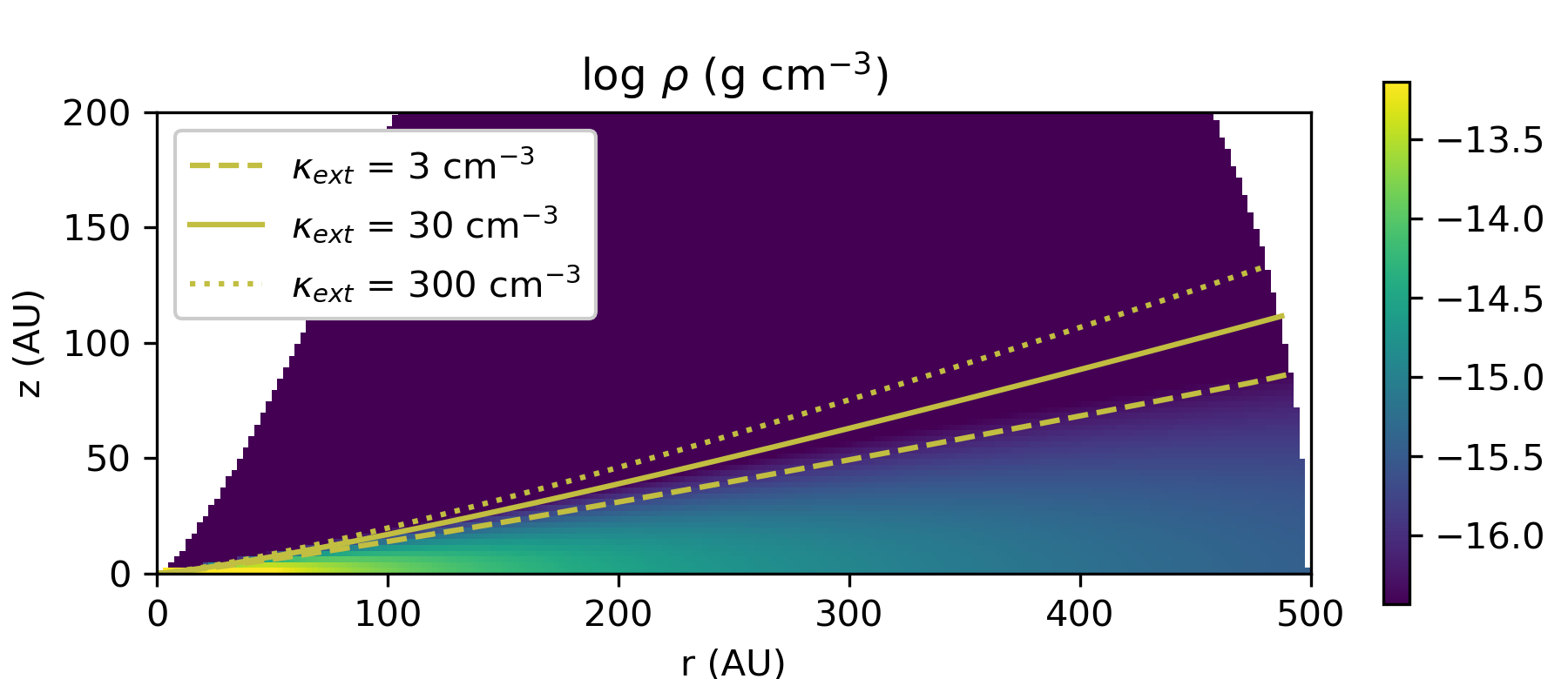}
\caption{
Density distribution used for the monochromatic Monte-Carlo radiative transfer simulations. The green curves show the $\tau$=1 disk surfaces for $\kappa_\mathrm{ext}$=3, 30, and 300 cm$^2$ g$^{-1}$.
\label{fig:sprout_tau_1}
}
\end{figure}

\begin{figure*}[ht!]
\centering
\includegraphics[width=18.5cm]{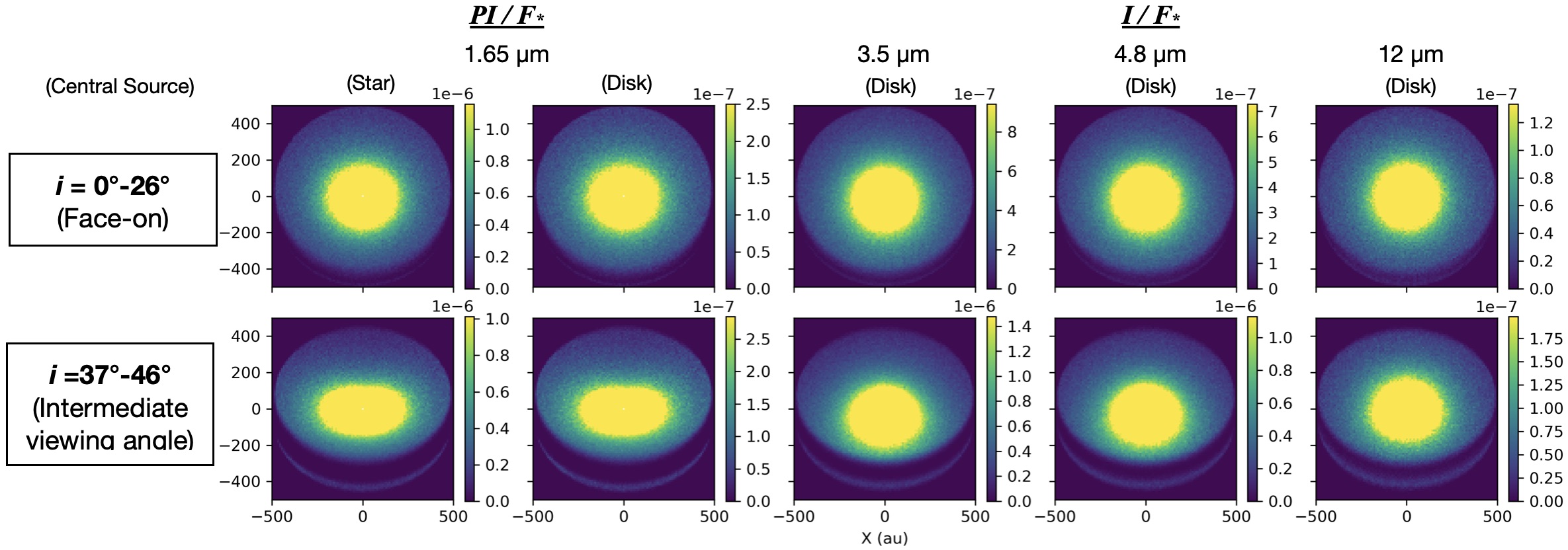}
\caption{
Images obtained using monochromatic Monte-Carlo radiative transfer simulations. See text for details.
\label{fig:sprout_all}
}
\end{figure*}

Fig. \ref{fig:sprout_tau_1} shows the $\tau$=1 disk surfaces for $\kappa_\mathrm{ext}$=3, 30, and 300 cm$^2$ g$^{-1}$. 
To derive the equations in Section \ref{sec:eq:thick}, we have approximated that the location of the disk surface is the same for all the wavelengths of our interest shown in Fig. \ref{fig:dust_general}. Small differences in the locations of the disk surfaces in Fig. \ref{fig:sprout_tau_1} justify this approximation.
Fig. \ref{fig:sprout_all} shows all the simulated images for the face-on ($i$=0$^\circ$--26$^\circ$) and the intermediate viewing angle (37$^\circ$--46$^\circ$). While most of the emission is associated with the front side of the disk, we also find the emission from the edge of the back side of the disk in the bottom of each image, and a dark lane which corresponds to the outer disk edge. These features are prominent in the images for the intermediate viewing angle.

\begin{figure*}[ht!]
\centering
\includegraphics[width=18.5cm]{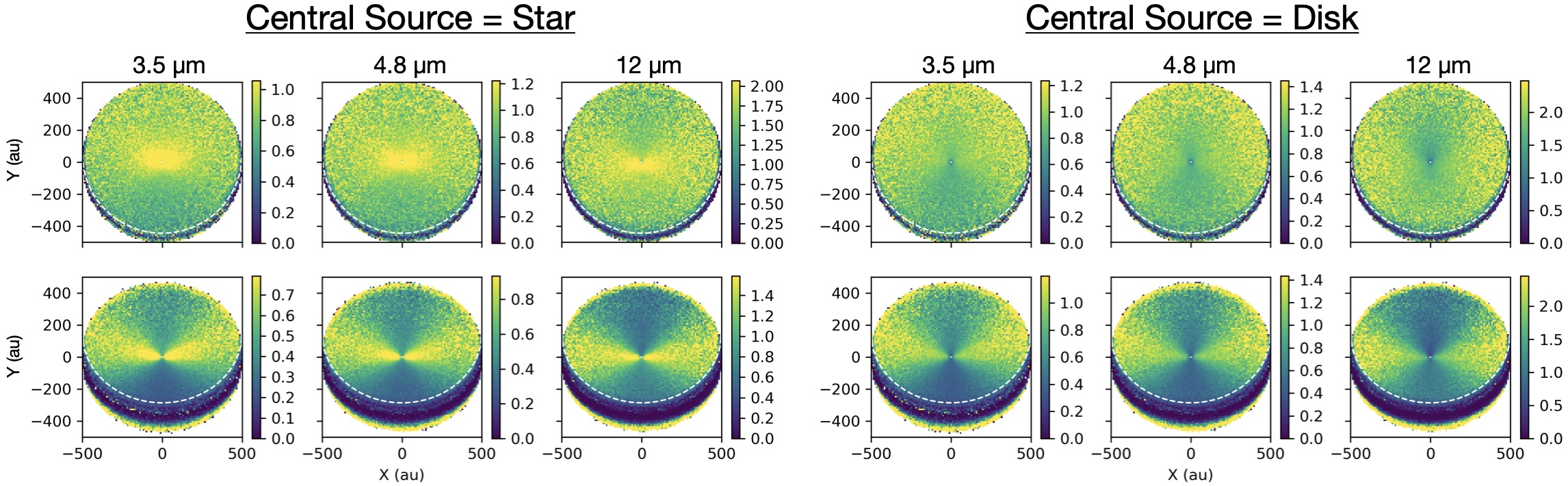}
\caption{
Intensities derived using the new semi-analytic method, divided by those directly derived using monochromatic radiative transfer simulations. 
As for Fig. \ref{fig:sprout_all}, the top and bottom panels show images for the face-on view ($i$=0$^\circ$--26$^\circ$) and the intermediate viewing angle ($i$=37$^\circ$--46$^\circ$), respectively.
In each panel, the region below the white dashed curve, that are associated with the emission from the outer disk edge, show large variations. These regions were not included in our analysis to verify the calculations (see text for details).
\label{fig:sprout_ratio_maps}
}
\end{figure*}

\begin{figure*}[ht!]
\centering
\includegraphics[width=18.5cm]{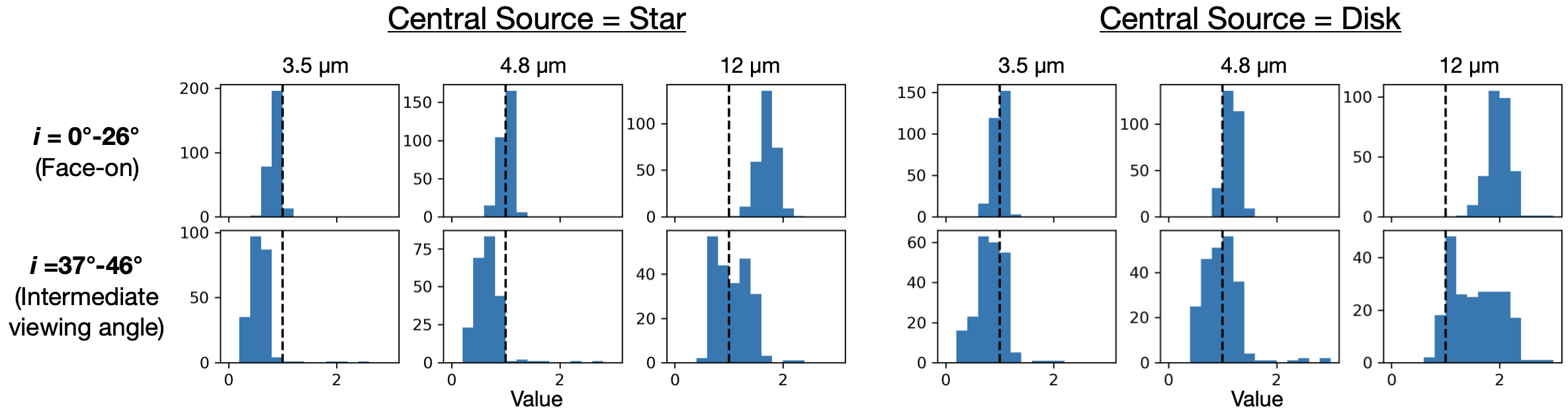}
\caption{
Histograms for intensity ratios for Fig. \ref{fig:sprout_ratio_maps} after applying 5$\times$5 binning. The dashed vertical panel indicate an intensity ratio of 1.
See text for details.
\label{fig:sprout_ratio_hist}
}
\end{figure*}

Using the simulated $H$ band polarimetry at $\lambda$=1.65 $\mu$m, we derived the images at 3.5, 4.8 and 12 $\mu$m using Equations (\ref{eq:thick:I_a:star}), (\ref{eq:thick:I_a:disk}), and (\ref{eq:thick:I_b}). In Fig. \ref{fig:sprout_ratio_maps}, we show these images divided by those directly obtained by monochromatic radiative transfer simulations (Fig. \ref{fig:sprout_all}).
Fig. \ref{fig:sprout_ratio_hist} shows the histograms of the intensity ratios for the individual cases. These histograms were made as follows. From the simulated images we first removed the emission associated with the dark lane and the back side of the disk. This is because the intensity ratios are significantly different from the front side of the disk surface, as shown in Fig. \ref{fig:sprout_ratio_maps}, but the actual observations in Fig. \ref{fig:PI_H} does not clearly show the presence of such an outer disk edge. We then applied 5$\times$5-pixel binning to increase the signal-to-noise before taking the intensity ratios.

In Figs. \ref{fig:sprout_ratio_maps} and \ref{fig:sprout_ratio_hist}, the measured intensity ratios range between $\sim$0.5 and $\sim$2, implying that our new semi-analytic method yields intensities at an accuracy within a factor of $\sim$2. Large values are observed for the longest wavelengths ($\lambda$=12 $\mu$m), as the actual disk surface is below that for $\lambda$=1.65 $\mu$m used for the semi-analytic calculations, making the numerical values smaller. Larger dispersions of the ratios for the intermediate viewing angle is due to a combination of a  large variation of the scattering angles at the disk surface, and the use of a representative $a_\lambda$ defined with Eq. (\ref{eq:a}).

\subsection{Toward full radiative transfer simulations}\label{sec:sim:full}

To better check our semi-analytic calculations, one may wish to see full radiative transfer simulations including radiation and viscous heating of the disk, to accurately calculate thermal emission. Such simulation codes include HO-CHUNK, HOCHUNK-3D \citep{Whitney03a,Whitney13} and Hyperion \citep{Robitaille11}. 
These codes can only place a star as the central source, not a compact self-luminous disk.
The presence of such an optically thick ionized disk for FUors has been extensively discussed over many years \citep[e.g.,][for reviews]{Hartmann96,Audard14}.
As discussed in Sect. \ref{sec:eq:overview}, the absence of such a disk would also yield the following problem. The inner edge of the dust disk is exposed to stellar radiation, yielding temperatures in the disk significantly higher than calculated using our semi-analytic method. Therefore, we need to revise the existing codes for the above issues to perform comparisons with our semi-analytic calculations.

The typical disk model used in Sect. \ref{sec:sim:mono} yields a very large optical thickness in the inner radii. It therefore demands significant computational time due to the number of scatterings, absorptions and reemissions. Our trials with Hyperion, which includes algorithms to efficiently solve this problem, suggest the need of a cluster computer with several tens or hundred cores to obtain results with different mass accretion rates within a few days.


\section{Summary and discussion} \label{sec:sum}

We have developed a semi-analytic method to estimate the intensities of mid-infrared extended emission associated with FUors from the observed PI distributions in the $H$ band ($\lambda$=1.65 $\mu$m). 
The extended 
disk or envelope responsible for the infrared extended emission is illuminated by the central source, which is either a star or a flat compact self-luminous disk. For our calculations, we assumed that the disk or the envelope is radiatively heated. The disks associated with YSOs, including FUors, must be optically thick at optical to mid-infrared wavelengths. We approximated that an extended disk consists of a surface layer with an optical thickness of  approximately one along the line of sight of the radiation from the central source, and an optically thick disk interior. We then calculated emission from extended disks for single scattering, double scattering, and thermal emission from the surface layer and the disk interior. For the envelope, we assumed that it is optically thin, and we derived the equations for single-scattering emission and thermal emission. We also derived equations to check consistencies of the calculations for the assumptions and the approximations we used.

We carried out the calculations for two FUors, FU Ori and V1735 Cyg, for three wavelengths ($\lambda$=3.5, 4.8, and 12 $\mu$m), using three dust models and (1) $\overline{\gamma}$ ($\sim$$z/r$)=0.1, 0.2, and 0.4, and (2) a viewing angle of zero (face-on). The near-infrared emission observed toward these targets supports the idea of the face-on view if the extended emission is due to a disk. The viewing angle may be intermediate (for example, $i$$\sim$45$^\circ$) if the extended emission is due to an envelope, but this does not affect the trends summarized below.

Our calculations suggest that the mid-infrared extended emission at the above wavelengths is dominated by the single scattering process. The contribution from thermal emission is negligible unless we add an additional heating mechanism such as adiabatic heating in spiral structures and/or fragments. The uncertain nature of the central illuminating source (a star or a compact disk), the nature of the extended emission (a disk or an envelope), and optical properties of dust grains yield uncertainties in the modeled intensities on orders of magnitude, for example, 20--800 for $\overline{\gamma}$=0.2 at $\lambda$=3--13 $\mu$m. The calculated infrared emission is significantly fainter if the central illuminating source is a star rather than a compact disk, approximately by a factor of the aspect ratio for the disk surface or the envelope.

The new method we developed is useful for estimating the detectability of the extended mid-infrared emission. Such observations would lead us to, if successful, better understanding the hierarchy of the mass accretion processes onto FUors (the envelope to the disk, and the disk to the star), which holds keys for understanding the mechanism of their accretion outbursts and the growth of low-mass YSOs
in general.
This method is complementary to full radiative transfer simulations, which offer more accurate calculations but only with specific dynamical models and significant computational time.

Throughout the calculations for the extended disks, we assumed that the radiation heating through the inner edge of the dust disk is negligible, because of the obscuration by an optically thick ionized disk. This assumption may have to be further investigated in the future using realistic simulations for such disks.


\begin{acknowledgements}
We thank Professor Oliver Absil for useful discussion and the anonymous referee for useful comments.
M.T. is supported by the National Science and Technology Council (NSTC) of Taiwan (grant No. 109-2112-M-001-019, 110-2112-M-001-044-, 111-2112-M-001-059-).
M.T. and S.-Y. L. are supported by NSTC Taiwan 108-2923-M-001-006-MY3 for the Taiwanese-Russian collaboration project.
P.-G. G. is supported by NSTC Taiwan 111-2112-M-001-037.
This work has made use of data from the European Space Agency (ESA) mission Gaia (https://www.cosmos.esa.int/gaia), processed by the Gaia Data Processing and Analysis Consortium (DPAC, https://www.cosmos.esa.int/web/gaia/dpac/consortium). Funding for the DPAC has been provided by national institutions, in particular the institutions participating in the Gaia Multilateral Agreement.
This research made use of the Simbad database operated at CDS, Strasbourg, France, and the NASA's Astrophysics Data System Abstract Service.
\end{acknowledgements}


\bibliographystyle{aa} 
\bibliography{astro} 

\begin{thebibliography}{65}
\expandafter\ifx\csname natexlab\endcsname\relax\def\natexlab#1{#1}\fi

\bibitem[{{Andrews} {et~al.}(2009){Andrews}, {Wilner}, {Hughes}, {Qi}, \&
  {Dullemond}}]{Andrews09}
{Andrews}, S.~M., {Wilner}, D.~J., {Hughes}, A.~M., {Qi}, C., \& {Dullemond},
  C.~P. 2009, \apj, 700, 1502

\bibitem[{{Andrews} {et~al.}(2010){Andrews}, {Wilner}, {Hughes}, {Qi}, \&
  {Dullemond}}]{Andrews10b}
{Andrews}, S.~M., {Wilner}, D.~J., {Hughes}, A.~M., {Qi}, C., \& {Dullemond},
  C.~P. 2010, \apj, 723, 1241

\bibitem[{{Audard} {et~al.}(2014){Audard}, {{\'A}brah{\'a}m}, {Dunham},
  {Green}, {Grosso}, {Hamaguchi}, {Kastner}, {K{\'o}sp{\'a}l}, {Lodato},
  {Romanova}, {Skinner}, {Vorobyov}, \& {Zhu}}]{Audard14}
{Audard}, M., {{\'A}brah{\'a}m}, P., {Dunham}, M.~M., {et~al.} 2014, Protostars
  and Planets VI, 387

\bibitem[{{Bailer-Jones} {et~al.}(2021){Bailer-Jones}, {Rybizki}, {Fouesneau},
  {Demleitner}, \& {Andrae}}]{Bailer21}
{Bailer-Jones}, C.~A.~L., {Rybizki}, J., {Fouesneau}, M., {Demleitner}, M., \&
  {Andrae}, R. 2021, VizieR Online Data Catalog, I/352

\bibitem[{{Boss}(2003)}]{Boss03}
{Boss}, A.~P. 2003, \apj, 599, 577

\bibitem[{{Caratti o Garatti} {et~al.}(2017){Caratti o Garatti}, {Stecklum},
  {Garcia Lopez}, {Eisl{\"o}ffel}, {Ray}, {Sanna}, {Cesaroni}, {Walmsley},
  {Oudmaijer}, {de Wit}, {Moscadelli}, {Greiner}, {Krabbe}, {Fischer}, {Klein},
  \& {Iba{\~n}ez}}]{Caratti17}
{Caratti o Garatti}, A., {Stecklum}, B., {Garcia Lopez}, R., {et~al.} 2017,
  Nature Physics, 13, 276

\bibitem[{{Cardelli} {et~al.}(1989){Cardelli}, {Clayton}, \&
  {Mathis}}]{Cardelli89}
{Cardelli}, J.~A., {Clayton}, G.~C., \& {Mathis}, J.~S. 1989, \apj, 345, 245

\bibitem[{{Chiang} \& {Goldreich}(1997)}]{Chiang97}
{Chiang}, E.~I. \& {Goldreich}, P. 1997, \apj, 490, 368

\bibitem[{{Chiang} \& {Goldreich}(1999)}]{Chiang99}
{Chiang}, E.~I. \& {Goldreich}, P. 1999, \apj, 519, 279

\bibitem[{{Chiang} {et~al.}(2001){Chiang}, {Joung}, {Creech-Eakman}, {Qi},
  {Kessler}, {Blake}, \& {van Dishoeck}}]{Chiang01}
{Chiang}, E.~I., {Joung}, M.~K., {Creech-Eakman}, M.~J., {et~al.} 2001, \apj,
  547, 1077

\bibitem[{{Cieza} {et~al.}(2018){Cieza}, {Ru{\'{\i}}z-Rodr{\'{\i}}guez},
  {Perez}, {Casassus}, {Williams}, {Zurlo}, {Principe}, {Hales}, {Prieto},
  {Tobin}, {Zhu}, \& {Marino}}]{Cieza18}
{Cieza}, L.~A., {Ru{\'{\i}}z-Rodr{\'{\i}}guez}, D., {Perez}, S., {et~al.} 2018,
  \mnras, 474, 4347

\bibitem[{{Cotera} {et~al.}(2001){Cotera}, {Whitney}, {Young}, {Wolff}, {Wood},
  {Povich}, {Schneider}, {Rieke}, \& {Thompson}}]{Cotera01}
{Cotera}, A.~S., {Whitney}, B.~A., {Young}, E., {et~al.} 2001, \apj, 556, 958

\bibitem[{{Dong} {et~al.}(2012{\natexlab{a}}){Dong}, {Hashimoto}, {Rafikov},
  {Zhu}, {Whitney}, {Kudo}, {Muto}, {Brandt}, {McClure}, {Wisniewski}, {Abe},
  {Brandner}, {Carson}, {Egner}, {Feldt}, {Goto}, {Grady}, {Guyon}, {Hayano},
  {Hayashi}, {Hayashi}, {Henning}, {Hodapp}, {Ishii}, {Iye}, {Janson},
  {Kandori}, {Knapp}, {Kusakabe}, {Kuzuhara}, {Kwon}, {Matsuo}, {McElwain},
  {Miyama}, {Morino}, {Moro-Martin}, {Nishimura}, {Pyo}, {Serabyn}, {Suto},
  {Suzuki}, {Takami}, {Takato}, {Terada}, {Thalmann}, {Tomono}, {Turner},
  {Watanabe}, {Yamada}, {Takami}, {Usuda}, \& {Tamura}}]{Dong12b}
{Dong}, R., {Hashimoto}, J., {Rafikov}, R., {et~al.} 2012{\natexlab{a}}, \apj,
  760, 111

\bibitem[{{Dong} {et~al.}(2012{\natexlab{b}}){Dong}, {Rafikov}, {Zhu},
  {Hartmann}, {Whitney}, {Brandt}, {Muto}, {Hashimoto}, {Grady}, {Follette},
  {Kuzuhara}, {Tanii}, {Itoh}, {Thalmann}, {Wisniewski}, {Mayama}, {Janson},
  {Abe}, {Brandner}, {Carson}, {Egner}, {Feldt}, {Goto}, {Guyon}, {Hayano},
  {Hayashi}, {Hayashi}, {Henning}, {Hodapp}, {Honda}, {Inutsuka}, {Ishii},
  {Iye}, {Kandori}, {Knapp}, {Kudo}, {Kusakabe}, {Matsuo}, {McElwain},
  {Miyama}, {Morino}, {Moro-Martin}, {Nishimura}, {Pyo}, {Suto}, {Suzuki},
  {Takami}, {Takato}, {Terada}, {Tomono}, {Turner}, {Watanabe}, {Yamada},
  {Takami}, {Usuda}, \& {Tamura}}]{Dong12a}
{Dong}, R., {Rafikov}, R., {Zhu}, Z., {et~al.} 2012{\natexlab{b}}, \apj, 750,
  161

\bibitem[{{Dong} {et~al.}(2016){Dong}, {Vorobyov}, {Pavlyuchenkov}, {Chiang},
  \& {Liu}}]{Dong16}
{Dong}, R., {Vorobyov}, E., {Pavlyuchenkov}, Y., {Chiang}, E., \& {Liu}, H.~B.
  2016, \apj, 823, 141

\bibitem[{{Draine} \& {Lee}(1984)}]{Draine84}
{Draine}, B.~T. \& {Lee}, H.~M. 1984, \apj, 285, 89

\bibitem[{{Dullemond} {et~al.}(2007){Dullemond}, {Hollenbach}, {Kamp}, \&
  {D'Alessio}}]{Dullemond07_PPV}
{Dullemond}, C.~P., {Hollenbach}, D., {Kamp}, I., \& {D'Alessio}, P. 2007,
  Protostars and Planets V, 555

\bibitem[{{Elbakyan} {et~al.}(2019){Elbakyan}, {Vorobyov}, {Rab}, {Meyer},
  {G{\"u}del}, {Hosokawa}, \& {Yorke}}]{Elbakyan19}
{Elbakyan}, V.~G., {Vorobyov}, E.~I., {Rab}, C., {et~al.} 2019, \mnras, 484,
  146

\bibitem[{{Fischer} {et~al.}(1994){Fischer}, {Henning}, \& {Yorke}}]{Fischer94}
{Fischer}, O., {Henning}, T., \& {Yorke}, H.~W. 1994, \aap, 284, 187

\bibitem[{{Follette} {et~al.}(2015){Follette}, {Grady}, {Swearingen}, {Sitko},
  {Champney}, {van der Marel}, {Takami}, {Kuchner}, {Close}, {Muto}, {Mayama},
  {McElwain}, {Fukagawa}, {Maaskant}, {Min}, {Russell}, {Kudo}, {Kusakabe},
  {Hashimoto}, {Abe}, {Akiyama}, {Brandner}, {Brandt}, {Carson}, {Currie},
  {Egner}, {Feldt}, {Goto}, {Guyon}, {Hayano}, {Hayashi}, {Hayashi}, {Henning},
  {Hodapp}, {Ishii}, {Iye}, {Janson}, {Kandori}, {Knapp}, {Kuzuhara}, {Kwon},
  {Matsuo}, {Miyama}, {Morino}, {Moro-Martin}, {Nishimura}, {Pyo}, {Serabyn},
  {Suenaga}, {Suto}, {Suzuki}, {Takahashi}, {Takato}, {Terada}, {Thalmann},
  {Tomono}, {Turner}, {Watanabe}, {Wisniewski}, {Yamada}, {Takami}, {Usuda}, \&
  {Tamura}}]{Follette15}
{Follette}, K.~B., {Grady}, C.~A., {Swearingen}, J.~R., {et~al.} 2015, \apj,
  798, 132

\bibitem[{{Gramajo} {et~al.}(2014){Gramajo}, {Rod{\'o}n}, \&
  {G{\'o}mez}}]{Gramajo14}
{Gramajo}, L.~V., {Rod{\'o}n}, J.~A., \& {G{\'o}mez}, M. 2014, \aj, 147, 140

\bibitem[{{Greenberg} \& {Li}(1996)}]{Greenberg96}
{Greenberg}, J.~M. \& {Li}, A. 1996, \aap, 309, 258

\bibitem[{{Grosso} {et~al.}(2003){Grosso}, {Alves}, {Wood}, {Neuh{\"a}user},
  {Montmerle}, \& {Bjorkman}}]{Grosso03}
{Grosso}, N., {Alves}, J., {Wood}, K., {et~al.} 2003, \apj, 586, 296

\bibitem[{{Hartmann} \& {Kenyon}(1996)}]{Hartmann96}
{Hartmann}, L. \& {Kenyon}, S.~J. 1996, \araa, 34, 207

\bibitem[{{Henning} \& {Meeus}(2009)}]{Henning09}
{Henning}, T. \& {Meeus}, G. 2009, arXiv e-prints, arXiv:0911.1010

\bibitem[{{Herbig} {et~al.}(2003){Herbig}, {Petrov}, \& {Duemmler}}]{Herbig03}
{Herbig}, G.~H., {Petrov}, P.~P., \& {Duemmler}, R. 2003, \apj, 595, 384

\bibitem[{{Ilee} {et~al.}(2011){Ilee}, {Boley}, {Caselli}, {Durisen},
  {Hartquist}, \& {Rawlings}}]{Ilee11}
{Ilee}, J.~D., {Boley}, A.~C., {Caselli}, P., {et~al.} 2011, \mnras, 417, 2950

\bibitem[{{J\"ager} {et~al.}(1998){J\"ager}, {Mutschke}, \&
  {Henning}}]{Jager98}
{J\"ager}, C., {Mutschke}, H., \& {Henning}, T. 1998, \aap, 332, 291

\bibitem[{{Kim} {et~al.}(1994){Kim}, {Martin}, \& {Hendry}}]{Kim94}
{Kim}, S., {Martin}, P.~G., \& {Hendry}, P.~D. 1994, \apj, 422, 164

\bibitem[{{K{\'o}sp{\'a}l} {et~al.}(2008){K{\'o}sp{\'a}l}, {{\'A}brah{\'a}m},
  {Apai}, {Ardila}, {Grady}, {Henning}, {Juh{\'a}sz}, {Miller}, \&
  {Mo{\'o}r}}]{Kospal08}
{K{\'o}sp{\'a}l}, {\'A}., {{\'A}brah{\'a}m}, P., {Apai}, D., {et~al.} 2008,
  \mnras, 383, 1015

\bibitem[{{K{\'o}sp{\'a}l} {et~al.}(2021){K{\'o}sp{\'a}l}, {Cruz-S{\'a}enz de
  Miera}, {White}, {{\'A}brah{\'a}m}, {Chen}, {Csengeri}, {Dong}, {Dunham},
  {Feh{\'e}r}, {Green}, {Hashimoto}, {Henning}, {Hogerheijde}, {Kudo}, {Liu},
  {Takami}, \& {Vorobyov}}]{Kospal21}
{K{\'o}sp{\'a}l}, {\'A}., {Cruz-S{\'a}enz de Miera}, F., {White}, J.~A.,
  {et~al.} 2021, \apjs, 256, 30

\bibitem[{{Laor} \& {Draine}(1993)}]{Laor93}
{Laor}, A. \& {Draine}, B.~T. 1993, \apj, 402, 441

\bibitem[{{Laws} {et~al.}(2020){Laws}, {Harries}, {Setterholm}, {Monnier},
  {Rich}, {Aarnio}, {Adams}, {Andrews}, {Bae}, {Calvet}, {Espaillat},
  {Hartmann}, {Hinkley}, {Isella}, {Kraus}, {Wilner}, \& {Zhu}}]{Laws20}
{Laws}, A. S.~E., {Harries}, T.~J., {Setterholm}, B.~R., {et~al.} 2020, \apj,
  888, 7

\bibitem[{{Liu} {et~al.}(2016){Liu}, {Takami}, {Kudo}, {Hashimoto}, {Dong},
  {Vorobyov}, {Pyo}, {Fukagawa}, {Tamura}, {Henning}, {Dunham}, {Karr},
  {Kusakabe}, \& {Tsuribe}}]{Liu16}
{Liu}, H.~B., {Takami}, M., {Kudo}, T., {et~al.} 2016, Science Advances, 2,
  e1500875

\bibitem[{{Long} {et~al.}(2018){Long}, {Pinilla}, {Herczeg}, {Harsono},
  {Dipierro}, {Pascucci}, {Hendler}, {Tazzari}, {Ragusa}, {Salyk}, {Edwards},
  {Lodato}, {van de Plas}, {Johnstone}, {Liu}, {Boehler}, {Cabrit}, {Manara},
  {Menard}, {Mulders}, {Nisini}, {Fischer}, {Rigliaco}, {Banzatti}, {Avenhaus},
  \& {Gully-Santiago}}]{Long18}
{Long}, F., {Pinilla}, P., {Herczeg}, G.~J., {et~al.} 2018, \apj, 869, 17

\bibitem[{{Lynden-Bell} \& {Pringle}(1974)}]{Lynden-Bell74}
{Lynden-Bell}, D. \& {Pringle}, J.~E. 1974, \mnras, 168, 603

\bibitem[{{Nayakshin}(2010)}]{Nayaksin10}
{Nayakshin}, S. 2010, \mnras, 408, L36

\bibitem[{{Padgett} {et~al.}(1999){Padgett}, {Brandner}, {Stapelfeldt},
  {Strom}, {Terebey}, \& {Koerner}}]{Padgett99}
{Padgett}, D.~L., {Brandner}, W., {Stapelfeldt}, K.~R., {et~al.} 1999, \aj,
  117, 1490

\bibitem[{{Quanz} {et~al.}(2007){Quanz}, {Henning}, {Bouwman}, {van Boekel},
  {Juh{\'a}sz}, {Linz}, {Pontoppidan}, \& {Lahuis}}]{Quanz07}
{Quanz}, S.~P., {Henning}, T., {Bouwman}, J., {et~al.} 2007, \apj, 668, 359

\bibitem[{{Robitaille}(2011)}]{Robitaille11}
{Robitaille}, T.~P. 2011, \aap, 536, A79

\bibitem[{{Robitaille} {et~al.}(2007){Robitaille}, {Whitney}, {Indebetouw}, \&
  {Wood}}]{Robitaille07}
{Robitaille}, T.~P., {Whitney}, B.~A., {Indebetouw}, R., \& {Wood}, K. 2007,
  \apjs, 169, 328

\bibitem[{{Robitaille} {et~al.}(2006){Robitaille}, {Whitney}, {Indebetouw},
  {Wood}, \& {Denzmore}}]{Robitaille06}
{Robitaille}, T.~P., {Whitney}, B.~A., {Indebetouw}, R., {Wood}, K., \&
  {Denzmore}, P. 2006, \apjs, 167, 256

\bibitem[{{Sandell} \& {Weintraub}(2001)}]{Sandell01}
{Sandell}, G. \& {Weintraub}, D.~A. 2001, \apjs, 134, 115

\bibitem[{{Shakura} \& {Sunyaev}(1973)}]{Shakura73}
{Shakura}, N.~I. \& {Sunyaev}, R.~A. 1973, \aap, 24, 337

\bibitem[{{Spitzer}(1978)}]{Spitzer78}
{Spitzer}, L. 1978, {Physical processes in the interstellar medium}
  ({Wiley-InterScience})

\bibitem[{{Stamatellos} \& {Herczeg}(2015)}]{Stamatellos15}
{Stamatellos}, D. \& {Herczeg}, G.~J. 2015, \mnras, 449, 3432

\bibitem[{{Takami} {et~al.}(2019){Takami}, {Chen}, {Liu}, {Hirano},
  {K{\'o}sp{\'a}l}, {{\'A}brah{\'a}m}, {Vorobyov}, {Cruz-S{\'a}enz de Miera},
  {Csengeri}, {Green}, {Hogerheijde}, {Hsieh}, {Karr}, {Dong}, {Trejo}, \&
  {Chen}}]{Takami19}
{Takami}, M., {Chen}, T.-S., {Liu}, H.~B., {et~al.} 2019, \apj, 884, 146

\bibitem[{{Takami} {et~al.}(2018){Takami}, {Fu}, {Liu}, {Karr}, {Hashimoto},
  {Kudo}, {Vorobyov}, {K{\'o}sp{\'a}l}, {Scicluna}, {Dong}, {Tamura}, {Pyo},
  {Fukagawa}, {Tsuribe}, {Dunham}, {Henning}, \& {de Leon}}]{Takami18}
{Takami}, M., {Fu}, G., {Liu}, H.~B., {et~al.} 2018, \apj, 864, 20

\bibitem[{{Takami} {et~al.}(2014){Takami}, {Hasegawa}, {Muto}, {Gu}, {Dong},
  {Karr}, {Hashimoto}, {Kusakabe}, {Chapillon}, {Tang}, {Itoh}, {Carson},
  {Follette}, {Mayama}, {Sitko}, {Janson}, {Grady}, {Kudo}, {Akiyama}, {Kwon},
  {Takahashi}, {Suenaga}, {Abe}, {Brandner}, {Brandt}, {Currie}, {Egner},
  {Feldt}, {Guyon}, {Hayano}, {Hayashi}, {Hayashi}, {Henning}, {Hodapp},
  {Honda}, {Ishii}, {Iye}, {Kandori}, {Knapp}, {Kuzuhara}, {McElwain},
  {Matsuo}, {Miyama}, {Morino}, {Moro-Martin}, {Nishimura}, {Pyo}, {Serabyn},
  {Suto}, {Suzuki}, {Takato}, {Terada}, {Thalmann}, {Tomono}, {Turner},
  {Wisniewski}, {Watanabe}, {Yamada}, {Takami}, {Usuda}, \&
  {Tamura}}]{Takami14}
{Takami}, M., {Hasegawa}, Y., {Muto}, T., {et~al.} 2014, \apj, 795, 71

\bibitem[{{Takami} {et~al.}(2013){Takami}, {Karr}, {Hashimoto}, {Kim},
  {Wisniewski}, {Henning}, {Grady}, {Kandori}, {Hodapp}, {Kudo}, {Kusakabe},
  {Chou}, {Itoh}, {Momose}, {Mayama}, {Currie}, {Follette}, {Kwon}, {Abe},
  {Brandner}, {Brandt}, {Carson}, {Egner}, {Feldt}, {Guyon}, {Hayano},
  {Hayashi}, {Hayashi}, {Ishii}, {Iye}, {Janson}, {Knapp}, {Kuzuhara},
  {McElwain}, {Matsuo}, {Miyama}, {Morino}, {Moro-Martin}, {Nishimura}, {Pyo},
  {Serabyn}, {Suto}, {Suzuki}, {Takato}, {Terada}, {Thalmann}, {Tomono},
  {Turner}, {Watanabe}, {Yamada}, {Takami}, {Usuda}, \& {Tamura}}]{Takami13}
{Takami}, M., {Karr}, J.~L., {Hashimoto}, J., {et~al.} 2013, \apj, 772, 145

\bibitem[{{Tazaki} {et~al.}(2016){Tazaki}, {Tanaka}, {Okuzumi}, {Kataoka}, \&
  {Nomura}}]{Tazaki16}
{Tazaki}, R., {Tanaka}, H., {Okuzumi}, S., {Kataoka}, A., \& {Nomura}, H. 2016,
  \apj, 823, 70

\bibitem[{{Vorobyov}(2013)}]{Vorobyov13}
{Vorobyov}, E.~I. 2013, \aap, 552, A129

\bibitem[{{Vorobyov} {et~al.}(2020){Vorobyov}, {Matsukoba}, {Omukai}, \&
  {Guedel}}]{Vorobyov20}
{Vorobyov}, E.~I., {Matsukoba}, R., {Omukai}, K., \& {Guedel}, M. 2020, \aap,
  638, A102

\bibitem[{{Vorobyov} {et~al.}(2022){Vorobyov}, {Skliarevskii}, {Molyarova},
  {Akimkin}, {Pavlyuchenkov}, {K{\'o}sp{\'a}l}, {Liu}, {Takami}, \&
  {Topchieva}}]{Vorobyov22}
{Vorobyov}, E.~I., {Skliarevskii}, A.~M., {Molyarova}, T., {et~al.} 2022, \aap,
  658, A191

\bibitem[{{Watson} {et~al.}(2007){Watson}, {Stapelfeldt}, {Wood}, \&
  {M{\'e}nard}}]{Watson07_PPV}
{Watson}, A.~M., {Stapelfeldt}, K.~R., {Wood}, K., \& {M{\'e}nard}, F. 2007,
  Protostars and Planets V, 523

\bibitem[{{Whitney} {et~al.}(2004){Whitney}, {Indebetouw}, {Bjorkman}, \&
  {Wood}}]{Whitney04}
{Whitney}, B.~A., {Indebetouw}, R., {Bjorkman}, J.~E., \& {Wood}, K. 2004,
  \apj, 617, 1177

\bibitem[{{Whitney} {et~al.}(2013){Whitney}, {Robitaille}, {Bjorkman}, {Dong},
  {Wolff}, {Wood}, \& {Honor}}]{Whitney13}
{Whitney}, B.~A., {Robitaille}, T.~P., {Bjorkman}, J.~E., {et~al.} 2013, \apjs,
  207, 30

\bibitem[{{Whitney} {et~al.}(2003{\natexlab{a}}){Whitney}, {Wood}, {Bjorkman},
  \& {Cohen}}]{Whitney03b}
{Whitney}, B.~A., {Wood}, K., {Bjorkman}, J.~E., \& {Cohen}, M.
  2003{\natexlab{a}}, \apj, 598, 1079

\bibitem[{{Whitney} {et~al.}(2003{\natexlab{b}}){Whitney}, {Wood}, {Bjorkman},
  \& {Wolff}}]{Whitney03a}
{Whitney}, B.~A., {Wood}, K., {Bjorkman}, J.~E., \& {Wolff}, M.~J.
  2003{\natexlab{b}}, \apj, 591, 1049

\bibitem[{{Whittet} {et~al.}(2001){Whittet}, {Gerakines}, {Hough}, \&
  {Shenoy}}]{Whittet01}
{Whittet}, D.~C.~B., {Gerakines}, P.~A., {Hough}, J.~H., \& {Shenoy}, S.~S.
  2001, \apj, 547, 872

\bibitem[{{Wood} {et~al.}(2002){Wood}, {Wolff}, {Bjorkman}, \&
  {Whitney}}]{Wood02b}
{Wood}, K., {Wolff}, M.~J., {Bjorkman}, J.~E., \& {Whitney}, B. 2002, \apj,
  564, 887

\bibitem[{{Yen} {et~al.}(2019){Yen}, {Gu}, {Hirano}, {Koch}, {Lee}, {Liu}, \&
  {Takakuwa}}]{Yen19}
{Yen}, H.-W., {Gu}, P.-G., {Hirano}, N., {et~al.} 2019, \apj, 880, 69

\bibitem[{{Yen} {et~al.}(2014){Yen}, {Takakuwa}, {Ohashi}, {Aikawa}, {Aso},
  {Koyamatsu}, {Machida}, {Saigo}, {Saito}, {Tomida}, \& {Tomisaka}}]{Yen14}
{Yen}, H.-W., {Takakuwa}, S., {Ohashi}, N., {et~al.} 2014, \apj, 793, 1

\bibitem[{{Zhu} {et~al.}(2015){Zhu}, {Dong}, {Stone}, \&
  {Rafikov}}]{Zhu15_disk}
{Zhu}, Z., {Dong}, R., {Stone}, J.~M., \& {Rafikov}, R.~R. 2015, \apj, 813, 88

\bibitem[{{Zhu} {et~al.}(2008){Zhu}, {Hartmann}, {Calvet}, {Hernandez},
  {Tannirkulam}, \& {D'Alessio}}]{Zhu08}
{Zhu}, Z., {Hartmann}, L., {Calvet}, N., {et~al.} 2008, \apj, 684, 1281

\end{thebibliography}



\begin{appendix} 

\section{Results for $\overline{\gamma}$=0.1 and 0.4}\label{appendix}

In Sects. \ref{appendix:star} and \ref{appendix:disk}, we present consistency checks for the cases where the extended disk or envelope is illuminated by a star and a flat compact disk, respectively. In Sect. \ref{appendix:I}, we show that the single-scattering emission dominates the total intensity distribution, as for $\overline{\gamma}$=0.2 (Sect. \ref{sec:application:I_indiv}), and comment on their intensities.

\subsection{Consistency check for cases illuminated by a star} \label{appendix:star}

\begin{figure*}[ht!]
\centering
\includegraphics[width=18cm]{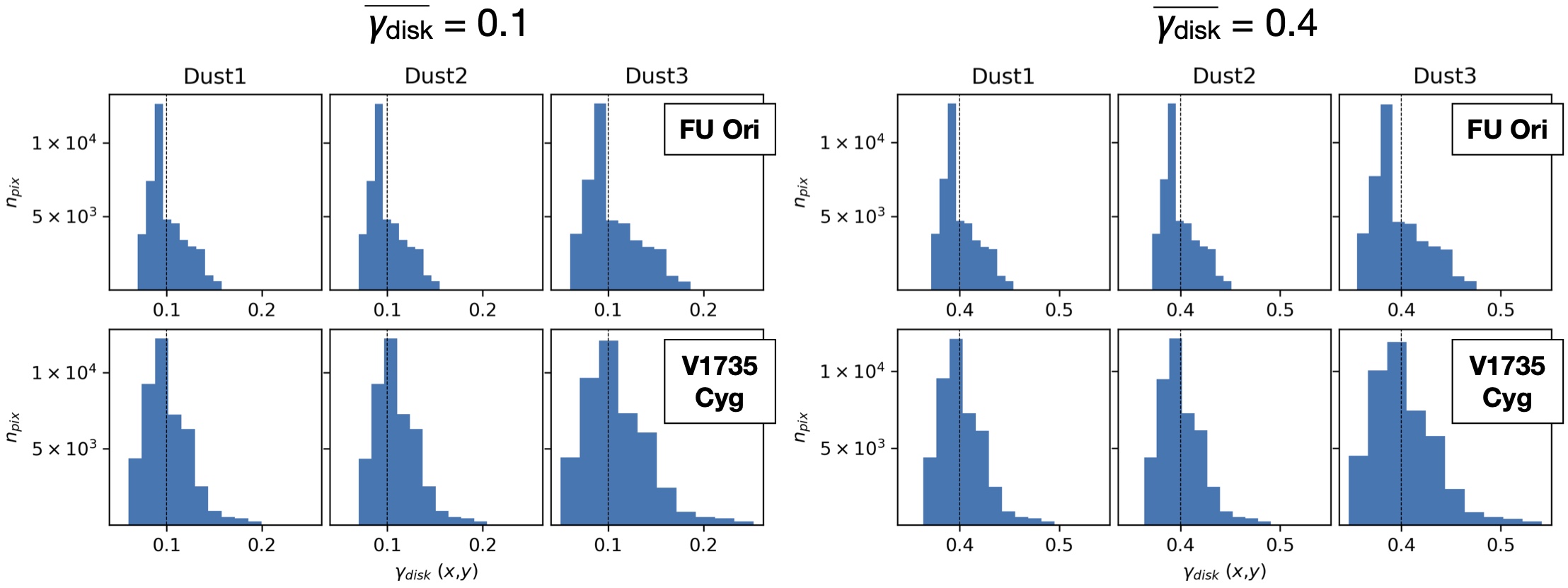}
\caption{
Same as Fig. \ref{fig:check_hist_star:gamma02} but for $\overline{\gamma}$=0.1 ($left$) and 0.4 ($right$).
\label{fig:check_hist_star:others}
}
\end{figure*}

Table \ref{tbl:zr0:others} shows $z_\mathrm{disk}/r$ at $r$=$r_\mathrm{min}$ used for the integration of Eq. (\ref{eq:z_disk}). These values were adjusted to minimize the deviation of $\gamma_\mathrm{disk} (x,y)$ from $\overline{\gamma}$=0.1 and 0.4  for most of the region. 
Fig. \ref{fig:check_hist_star:others} shows the distributions of pixel values for $\gamma_\mathrm{disk}(x,y)$ for the individual cases.
These show relatively small ranges around $\overline{\gamma}$, within 50 \% for most of the regions. 

\FloatBarrier

\begin{table}
\caption{Disk surface aspect ratio $z_\mathrm{disk}/r$ at the minimum radius $r$=$r_\mathrm{min}$$^a$.\label{tbl:zr0:others}}
\begin{tabular}{clclc}
\hline\hline
$\overline{\gamma}$     & Target        & Central & Dust        & Value \\
                                                        &               & Source
\\ \hline
0.1     & FU Ori                & star  & Dust1 & 0.07  \\
        &                       & star  & Dust2 & 0.07  \\
        &                       & star  & Dust3 & 0.06  \\
        & V1735 Cyg     & star  & Dust1 & 0.06  \\
        &                       & star  & Dust2 & 0.07  \\
        &                       & star  & Dust3 & 0.05  \\
0.4     & FU Ori                & star  & Dust1 & 0.40  \\
        &                       & star  & Dust2 & 0.40  \\
        &                       & star  & Dust3 & 0.38  \\
        & V1735 Cyg     & star  & Dust1 & 0.39  \\
        &                       & star  & Dust2 & 0.39  \\
        &                       & star  & Dust3 & 0.37  \\
        & FU Ori                & disk  & Dust1 & 0.34  \\
        &                       & disk  & Dust2 & 0.34  \\
        &                       & disk  & Dust3 & 0.30  \\
        & V1735 Cyg     & disk  & Dust1 & 0.33  \\
        &                       & disk  & Dust2 & 0.33  \\
        &                       & disk  & Dust3 & 0.28  \\
\hline
\end{tabular} \\
$^a$For $\overline{\gamma}$=0.1 and 0.4.
\end{table}

In Table \ref{tbl:maxstd:others}, we tabulate the maximum values for $z_\mathrm{disk}/r$, $\delta$, $\tau_{r,H}$, and $\tau_{r,H}$, and the standard deviations for $\gamma_\mathrm{disk}$ from $\overline{\gamma}$. The optical thicknesses of the envelopes are $\sim$0.2 or less, agreeing with the assumption when we derived the equations in Sect. \ref{sec:eq}.
As discussed in Sect. \ref{sec:application:I_indiv} for $\overline{\gamma}$=0.2, the parameters $z_\mathrm{disk}/r$ and $\delta$ do not significant affect the accuracies of the calculations as long as the intensity distributions are dominated by single-scattering emission (Sect. \ref{appendix:I}).

\begin{table*}
\caption{Parameters for surface geometry of disks and optical thicknesses of envelopes. \label{tbl:maxstd:others}}
\begin{tabular}{lllccccc}
\hline\hline
$\overline{\gamma}$ & Target    & Dust          & \multicolumn{4}{c}{80\%-ile/Maximum Values} & $\gamma_\mathrm{disk}$\\ \cline{4-7}
&               &               & $z_\mathrm{disk}/r$ & $\delta$ (deg.) & $\tau_{r,H}$ & $\tau_{l,H}$     & r.m.s
\\ \hline
0.1     & FU Ori                & Dust1 & 0.12/0.16     & 9.1/13.5              & 0.12/0.22       & 0.04/0.13     & 0.019 \\
        &                       & Dust2 & 0.12/0.16     &  8.9/13.2     & 0.12/0.22       & 0.04/0.13     & 0.019 \\
        &                       & Dust3 & 0.13/0.19     & 10.7/17.1     & 0.18/0.32       & 0.06/0.19     & 0.028 \\
        & V1735 Cyg     & Dust1 & 0.12/0.20     & 10.3/20.6     & 0.15/0.35     & 0.07/0.18       & 0.024 \\
        &                       & Dust2 & 0.13/0.21     & 10.7/20.5     & 0.15/0.34       & 0.06/0.17     & 0.025 \\
        &                       & Dust3 & 0.14/0.26     & 12.8/27.6     & 0.22/0.50       & 0.09/0.26     & 0.035 \\
0.4     & FU Ori                & Dust1 & 0.46/0.51     & 26.9/31.4     & 0.03/0.06       & 0.04/0.13     & 0.018 \\
        &                       & Dust2 & 0.46/0.51     & 26.7/31.0     & 0.03/0.05       & 0.04/0.13     & 0.017 \\
        &                       & Dust3 & 0.46/0.54     & 28.1/34.6     & 0.04/0.08       & 0.06/0.19     & 0.026 \\
        & V1735 Cyg     & Dust1 & 0.46/0.57     & 28.3/38.7     & 0.04/0.09     & 0.07/0.18       & 0.022 \\
        &                       & Dust2 & 0.46/0.56     & 28.0/38.1     & 0.04/0.08       & 0.06/0.17     & 0.021 \\
        &                       & Dust3 & 0.47/0.64     & 30.3/45.6     & 0.05/0.13       & 0.09/0.26     & 0.033 \\
\hline
\end{tabular}\\
\end{table*}

\subsection{Consistency check for cases illuminated by a compact disk} \label{appendix:disk}

\begin{figure*}[ht!]
\centering
\includegraphics[width=18cm]{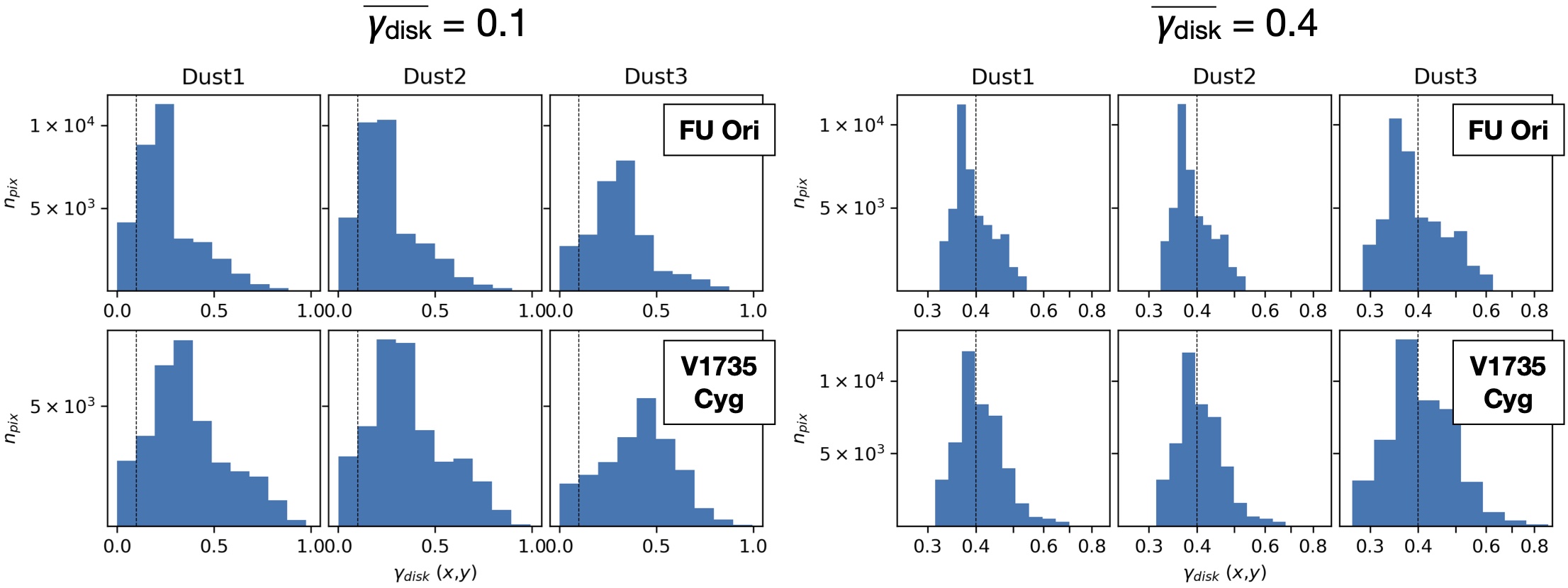}
\caption{
Same as Fig. \ref{fig:check_hist_disk:gamma02} but for $\overline{\gamma}$=0.1 ($left$) and 0.4  ($right$).
\label{fig:check_hist_disk:others}
}
\end{figure*}

Fig. \ref{fig:check_hist_disk:others} shows the distributions of pixel values for $\gamma_\mathrm{disk}(x,y)$ for the individual cases.
For $\overline{\gamma}$=0.1, the values significantly exceed $\overline{\gamma}$ for most of the regions, despite the fact that we set $z_\mathrm{disk}/r$=0 at $r$=$r_\mathrm{min}$ to minimize them. This implies that the calculations are inconsistent, and therefore $\overline{\gamma}$ cannot be 0.1 for the case that the extended emission is due to a disk, and with a flat compact disk as the illuminating source.

Unlike the cases where $\overline{\gamma}$=0.1 and 0.2, the dispersion of $\gamma_\mathrm{disk}(x,y)$ is small for $\overline{\gamma}$=0.4. As a result, we needed to adjust $z_\mathrm{disk}/r$ at $r$=$r_\mathrm{min}$ to obtain approximately self-consistent results, as for the case that the central illuminating source is a star. These values are tabulated in Table \ref{tbl:zr0:others}.
In Fig. \ref{fig:check_hist_disk:others}, $\gamma_\mathrm{disk}$ shows relatively small ranges around $\overline{\gamma}$=0.4, within 30 \% for most of the regions. As discussed in Sect. \ref{sec:application:I_indiv}, this dispersion does not affect the total intensity distribution as long as it is dominated by single-scattering emission.

\begin{figure*}[ht!]
\centering
\includegraphics[width=16cm]{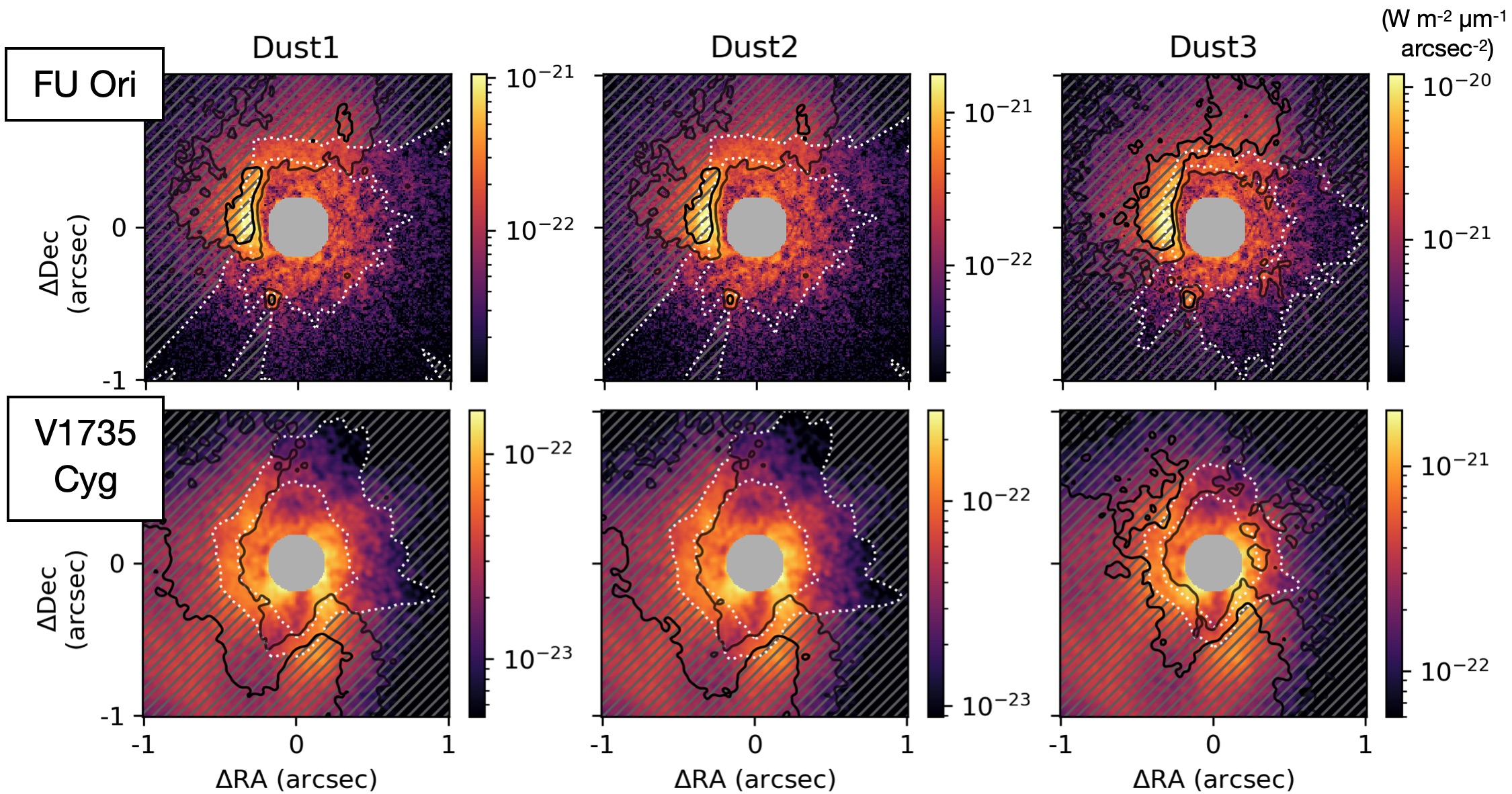}
\caption{
Same as the bottom panles of Figs. \ref{fig:Ia_FU_gamma02} and \ref{fig:Ia_V17_gamma02} but for $\overline{\gamma}$=0.1.
As for these figures, the white dotted contours in the upper panels are for $z_\mathrm{disk}/r$=0.5; the white dotted contours in the lower panels are for $\tau_r$=0.35 and 0.7; and the black solid contours in the lower panels are for $\tau_l$=0.35 and 0.7. The regions with $\tau_r$$>$0.7 and $\tau_l$$>$0.7 are hatched in the bottom panels.
\label{fig:Ia_gamma01}
}
\end{figure*}

Fig. \ref{fig:Ia_gamma01} shows the intensity distribution for the single scattering emission for $\overline{\gamma}$=0.1 in the case that the extended emission is due to an envelope. The contours shows that, unlike the case for $\overline{\gamma}$=0.2 (Sect. \ref{sec:application:I_a}), the optical thickness exceeds 0.7 in significant fractions of the emission regions. In particular, these cover a significant fraction of the east arm associated FU Ori. In these regions, systematic errors for the intensity would exceed a factor of 2. We therefore conclude that the calculations for this star are not self-consistent. For V1735 Cyg, the optical thicknesses are below 0.7 for the bright parts of the east and west arms. 

\begin{figure*}[ht!]
\centering
\includegraphics[width=16cm]{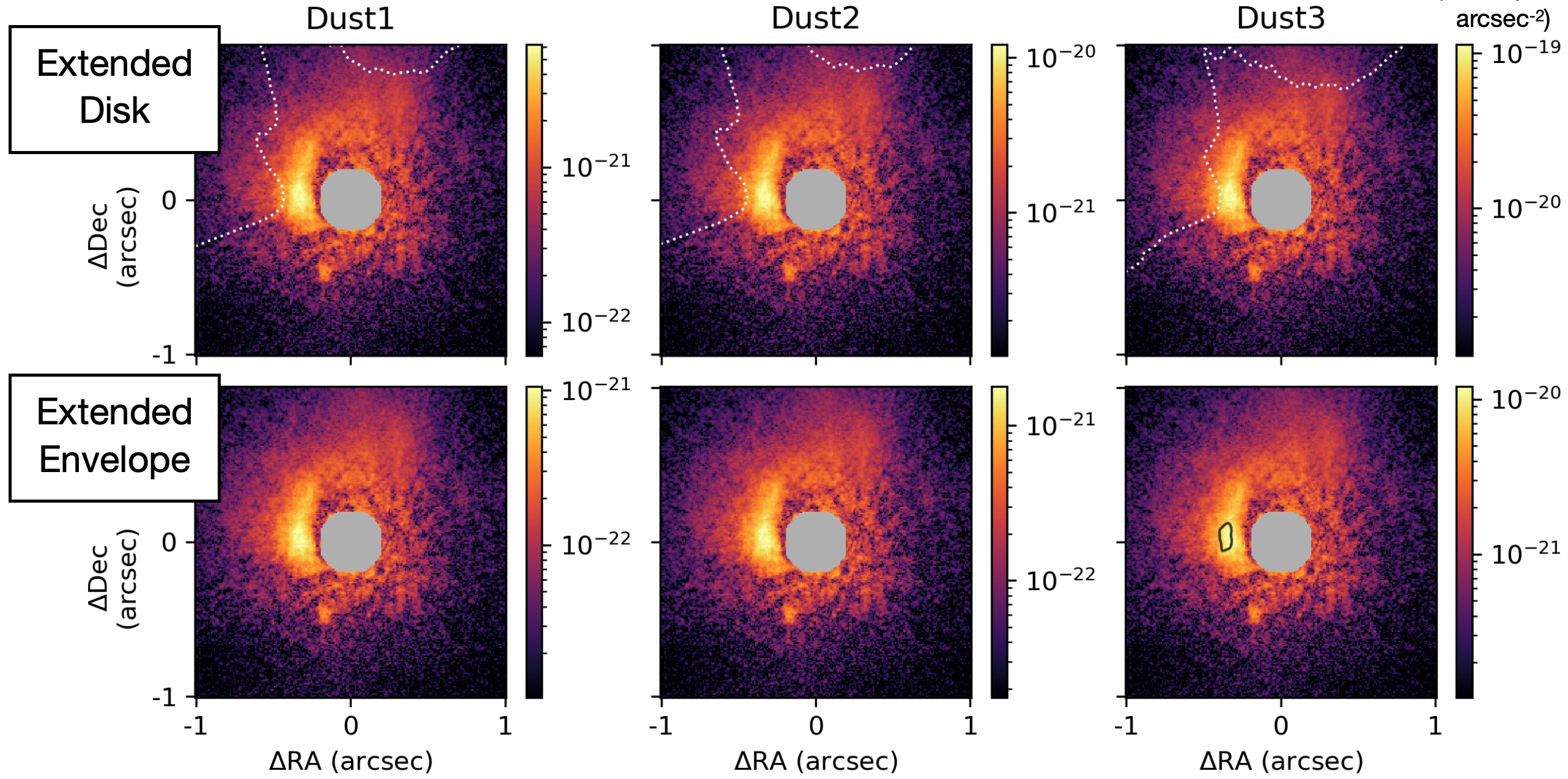}
\caption{
Same as Fig. \ref{fig:Ia_FU_gamma02} but for $\overline{\gamma}$=0.4.
The white dotted contours in the upper panels are for $z_\mathrm{disk}/r$=0.5; and the black solid contours in the bottom-right panels are for $\tau_l$=0.35.
\label{fig:Ia_FU_gamma04}
}
\end{figure*}

\begin{figure*}[ht!]
\centering
\includegraphics[width=16cm]{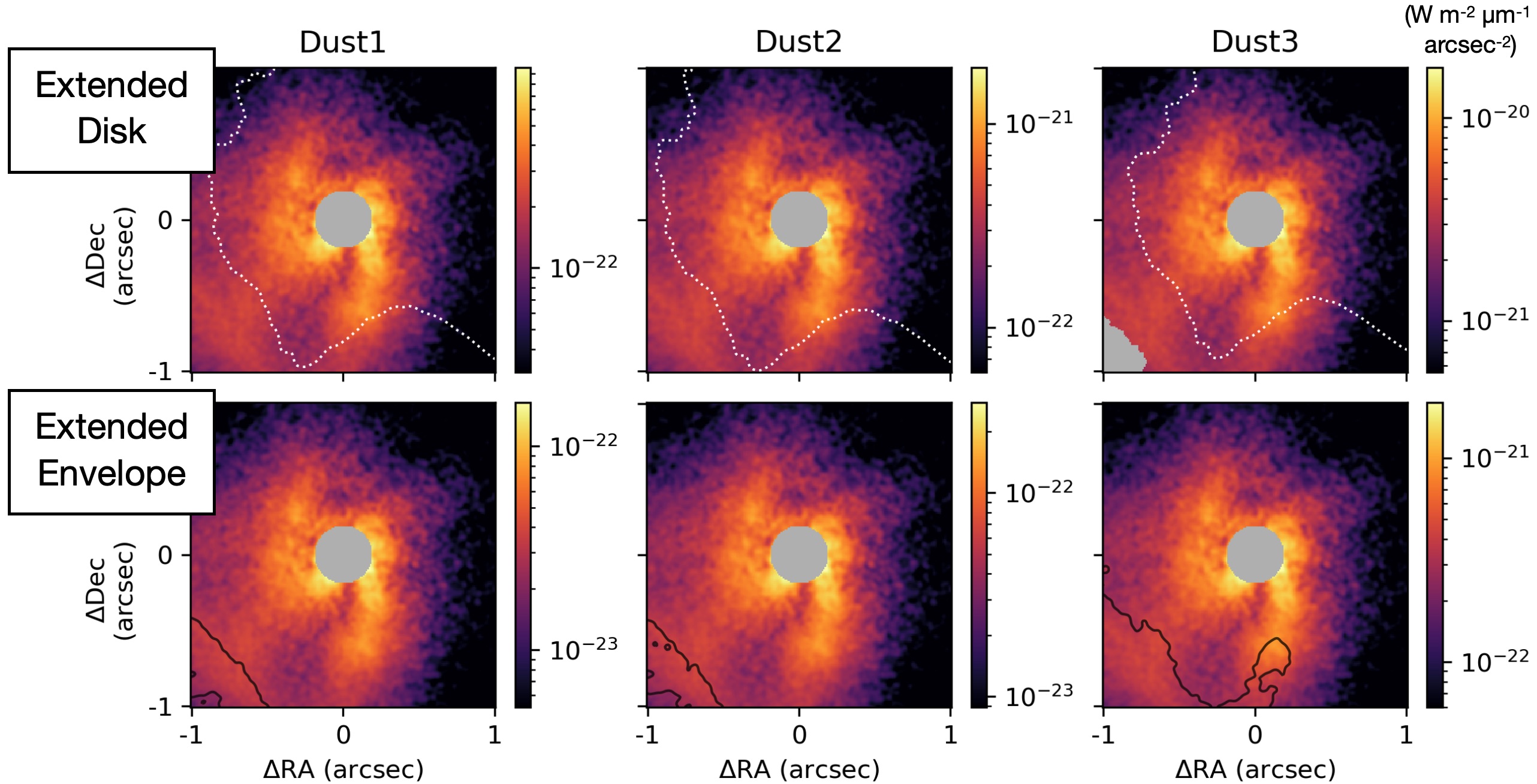}
\caption{
Same as Fig. \ref{fig:Ia_FU_gamma04} but for V1735 Cyg.
\label{fig:Ia_V17_gamma04}
}
\end{figure*}

Figs. \ref{fig:Ia_FU_gamma04} and \ref{fig:Ia_V17_gamma04} show the intensity distributions for single scattering emission for FU Ori and V1735 Cyg, respectively, for $\overline{\gamma}$=0.4, with contours for optical thicknesses and $z_\mathrm{disk}/r$. The optical thicknesses in the envelopes exceed 0.35 only at the brightness peak in the east arm for FU Ori, and a part of the outer diffuse emission for V1735 Cyg, but their optical thicknesses are still below 0.7. Therefore, the optically thin assumption used to derive the equations in Sect. \ref{sec:eq} is valid. The disk surface aspect ratios $z_\mathrm{disk}/r$ exceed 0.5 in some regions, however, it does not affect the accuracy of the calculations (Sect. \ref{sec:application:I_indiv}).


\subsection{Intensities} \label{appendix:I}

Table \ref{tbl:min_Ia:others} shows the minimum fraction of single-scattering emission to the total intensity for various cases. These are nearly identical to the values for the cases with $\overline{\gamma}$=0.2 but for FU Ori at $\lambda$=12 $\mu$m, the extended emission is due to an envelope illuminated by a star. As for the cases with $\overline{\gamma}$=0.2, the single-scattering emission is responsible for more than 80 \% of the total intensity for all the cases.

\begin{table*}
\centering
\caption{Minimum fraction of single-scattering emission to total intensity.\label{tbl:min_Ia:others}}
\begin{tabular}{clclcccccccc}
\hline\hline
$\overline{\gamma}$ & Extended          &       Wavelength      & Target         & \multicolumn{3}{c}{Illuminated by Star} && \multicolumn{3}{c}{Illuminated by Compact Disk}        \\ \cline{5-7} \cline{9-11}
&Emission       & ($\mu$m)      &               & Dust1 & Dust2 & Dust3 && Dust1 & Dust2 & Dust3
\\ \hline
0.1     & Disk  & 3.5   & FU Ori                & 0.89 & 0.85 & 0.82 && --- & --- & --- \\
        &       &       & V1735 Cyg     & 0.89 & 0.85 & 0.82 && --- & --- & --- \vspace{0.1cm} \\
        &       & 4.8   & FU Ori                & 0.92 & 0.87 & 0.85 && --- & --- & --- \\
        &       &       & V1735 Cyg     & 0.92 & 0.87 & 0.85 && --- & --- & --- \vspace{0.1cm} \\
        &       & 12 & FU Ori           & $>$0.99 & 0.99 & 0.95 && --- & --- & ---  \\
        &       &       & V1735 Cyg     & $>$0.99 & 0.99 & 0.95 && --- & --- & ---  \vspace{0.1cm} \\
        & Envelope      & 3.5   & FU Ori                 & $>$0.99 & $>$0.99 & $>$0.99 && --- & --- & --- \\
        &               &       & V1735 Cyg      & $>$0.99 & $>$0.99 & $>$0.99 && $>$0.99 & $>$0.99 & $>$0.99 \vspace{0.1cm} \\
        &               &4.8    & FU Ori                 & $>$0.99 & $>$0.99 & $>$0.99 && --- & --- & --- \\
        &               &       & V1735 Cyg      & $>$0.99 & $>$0.99 & $>$0.99 && $>$0.99 & $>$0.99 & $>$0.99 \vspace{0.1cm} \\
        &               &12 & FU Ori             & 0.83 & 0.94 & 0.98 && --- & --- & --- \\
        &               &       & V1735 Cyg      & $>$0.99 & $>$0.99 & $>$0.99 && $>$0.99 & $>$0.99 & $>$0.99 \vspace{0.1cm} \\

0.4     & Disk  & 3.5   & FU Ori                & 0.89 & 0.85 & 0.82 && 0.89 & 0.85 & 0.82 \\
        &       &       & V1735 Cyg     & 0.89 & 0.85 & 0.82 && 0.89 & 0.85 & 0.82 \vspace{0.1cm} \\
        &       & 4.8   & FU Ori                & 0.92 & 0.87 & 0.85 && 0.92 & 0.87 & 0.85 \\
        &       &       & V1735 Cyg     & 0.92 & 0.87 & 0.85 && 0.92 & 0.87 & 0.85 \vspace{0.1cm} \\
        &       & 12 & FU Ori           & $>$0.99 & 0.99 & 0.95 && $>$0.99 & 0.99 & 0.95  \\
        &       &       & V1735 Cyg     & $>$0.99 & 0.99 & 0.95 && $>$0.99 & 0.99 & 0.95  \vspace{0.1cm} \\
        & Envelope      & 3.5   & FU Ori                 & $>$0.99 & $>$0.99 & $>$0.99 && $>$0.99 & $>$0.99 & $>$0.99 \\
        &               &       & V1735 Cyg      & $>$0.99 & $>$0.99 & $>$0.99 && $>$0.99 & $>$0.99 & $>$0.99 \vspace{0.1cm} \\
        &               &4.8    & FU Ori                 & $>$0.99 & $>$0.99 & $>$0.99 && $>$0.99 & $>$0.99 & $>$0.99 \\
        &               &       & V1735 Cyg      & $>$0.99 & $>$0.99 & $>$0.99 && $>$0.99 & $>$0.99 & $>$0.99 \vspace{0.1cm} \\
        &               &12 & FU Ori             & 0.95 & 0.99 & 0.99 && $>$0.99 & $>$0.99 & $>$0.99 \\
        &               &       & V1735 Cyg      & $>$0.99 & $>$0.99 & $>$0.99 && $>$0.99 & $>$0.99 & $>$0.99 \\
\hline
\end{tabular}\\
\end{table*}

As discussed in Sect. \ref{sec:application:I_a}, the intensity distributions are identical to those for $\overline{\gamma}$=0.2 for the cases where the extended disk or the envelope is illuminated by a flat compact disk. The intensity distributions are lower by a factor of $\overline{\gamma}$ for the cases where the extended disk or envelope is illuminated by a star. As a result, the uncertainty of the intensities increases to a factor of 100-800 with an uncertainty of $\overline{\gamma}$ of 0.1-0.4.

\end{appendix}

\end{document}